\documentclass[]{aa} 
\usepackage{txfonts}
\usepackage{natbib}
\usepackage{graphicx, xspace}
\usepackage{multirow}

\usepackage{ amssymb }

\usepackage[colorlinks=true,
			linkcolor=blue,
			citecolor=blue, 
			filecolor=blue, 
			urlcolor=blue]{hyperref}

\newcommand{\vsini}{$v \sin i$}
\newcommand{\logg}{$\log g$} 
\newcommand{\Teff}{$T_{\rm eff}$}
\newcommand{\cd}{\,c/d\xspace}

\newcommand{\days} {\,days\xspace}
\newcommand{\muHz}{\,$\mu$Hz\xspace}

\newcommand{\nm}{\,nm\xspace}
\newcommand{\kms} {\,km/s\xspace}
\newcommand{\num} {$\nu_{\rm max}$\xspace}
\newcommand{\dnu} {$\Delta\nu$\xspace}

\newcommand{\deltaf}{$\delta f$\xspace}
\newcommand{\dpi}{$\Delta \Pi_1$\xspace}
\newcommand{\Dpi}{$\Delta \Pi_1$\xspace}

\newcommand{\sindex}{$\mathcal{S}$-index\xspace}
\newcommand{\Sindex}{$\mathcal{S}$-index\xspace}
\newcommand{\Ssymbol}{$\mathcal{S}$\xspace}

\newcommand{\Ro}{$\mathcal{R}_{\rm O}$\xspace}

\newcommand{\Kepler} {\textit{Kepler}\xspace}
\newcommand{\kepler} {\textit{Kepler}\xspace}
\newcommand{\Hermes} {\textsc{Hermes}\xspace}
\newcommand{\hermes} {\textsc{Hermes}\xspace}
\newcommand{\Mercator} {\textsc{Mercator}\xspace}

\newcommand{\KIC}[1]{{KIC\,#1\xspace}}

\newcommand{\Asterix}{KIC\,9163796\xspace}

\newcommand{\ttau}{$\theta^1$\,Tau\xspace}

\newcommand{\spd}{{\sc spd}\xspace}

\newcommand{\gssp}{{\sc gssp}\xspace}

\newcommand{\Figure}[1]{Figure\,\ref{#1}\xspace}
\newcommand{\Table}[1]{Table\,\ref{#1}\xspace}

\newcommand{\Section}[1]{Section\,\ref{#1}\xspace}

\makeatletter

\newcommand{\Rmnum}[1]{\expandafter\@slowromancap\romannumeral #1@}
\makeatother

\usepackage[]{color}

\newcommand{\new}[1]{\blu{\bf #1}}

\renewcommand{\new}[1]{#1}
\renewcommand{\new}[1]{#1}

\def\teff{$T_{\mathrm{eff}}$}

\def\dn1{$\delta\nu_{01}$}
\def\dn2{$\delta\nu_{02}$}
\def\sun{\hbox{$_\odot$}\xspace}

\usepackage[switch]{lineno}
\begin{document}

\title{Seismic probing of the first dredge-up event through the eccentric red-giant\,\&\,red-giant spectroscopic binary \Asterix\thanks{Based on observations made with the \textit{Kepler} space telescope and the \hermes spectrograph mounted on the 1.2\,m Mercator Telescope at the Spanish Observatorio del Roque de los Muchachos of the Instituto de Astrof{\'i}sica de Canarias.}}
\subtitle{How different are red-giant stars with a mass ratio of $\sim$1.015?}

\authorrunning{Beck et al.}
\titlerunning{\Asterix $-$ Seismic probing of the first dredge-up event}

\author{P.\,G.~Beck\inst{1,2,3,4}
\and T.~Kallinger\inst{5}
\and K.~Pavlovski\inst{6}
\and A.~Palacios\inst{7}
\and A.~Tkachenko\inst{8}
\and  S.~Mathis\inst{3,4,9}
\and R.\,A.~Garc{\'i}a\inst{3,4},\\
 E.~Corsaro\inst{1,2,3,4}
\and C.~Johnston\inst{8}
\and B.~Mosser\inst{9}
\and T.~Ceillier\inst{3,4} 
\and {J.-D.~do\,Nascimento~Jr.\inst{10,11}} 
\and G.~Raskin\inst{8} }



\date{Recieved: 30 May 2017 / Accepted: 13 December 2017}

\institute{
 Instituto de Astrof\'{\i}sica de Canarias, E-38200 La Laguna, Tenerife, Spain 
\email{paul.beck@iac.es}
\and Departamento de Astrof\'{\i}sica, Universidad de La Laguna, E-38206 La Laguna, Tenerife, Spain 
\and IRFU, CEA, Universit\'e Paris-Saclay, F-91191 Gif-sur-Yvette, France
\and Universit\'e Paris Diderot, AIM, Sorbonne Paris Cit\'e, CEA, CNRS, F-91191 Gif-sur-Yvette, France
\and Institut f\"ur Astronomie der Universit\"at Wien, T\"urkenschanzstr. 17, 1180 Wien, Austria 
\and Department of Physics, Faculty of Science, University of Zagreb, Croatia 
\and LUPM, UMR5299, Universit\'e de Montpellier, CNRS, F-34095 Montpellier cedex 5, France
\and Instituut voor Sterrenkunde, KU Leuven, 3001 Leuven, Belgium 
\and LESIA, Obs. de Paris, PSL Research Univ., CNRS, Univ.\,Pierre et Marie Curie, Univ.\,Paris Diderot,  92195\,Meudon, France 
\and Harvard-Smithsonian Center for Astrophysics, Cambridge, MA 02138, USA
\and Departamento de F\'isica, Universidade Federal do Rio Grande do Norte, CEP: 59072-970 Natal, RN, Brazil
}

\abstract
{Binaries in double-lined spectroscopic systems (SB2) provide a homogeneous set of stars. Differences of parameters, such as age or initial conditions, which otherwise would have strong impact on the stellar evolution, can be neglected. The observed differences are determined by the difference in stellar mass between the two components. The mass ratio can be determined with much higher accuracy than the actual stellar mass. }
{In this work, we aim to study the eccentric binary system \Asterix, whose two components are very close in mass and both are low-luminosity red-giant stars.}
{We analysed four years of \Kepler space photometry and we obtained high-resolution spectroscopy with the \Hermes instrument. 
The orbital elements and the spectra of both components were determined using spectral disentangling methods. The effective temperatures, and metallicities were extracted from disentangled spectra of the two stars.
Mass and radius of the primary were determined through asteroseismology. The surface rotation period of the primary is determined from the \Kepler light curve.  From representative theoretical models of the star, we derived the internal rotational gradient, while for a grid of models, the measured lithium abundance is compared with theoretical predictions.}
{From seismology the primary of \Asterix is a star of 1.39$\pm$0.06\,M\sun, while the spectroscopic mass ratio between both components can be determined with much higher precision by spectral disentangling to be 1.015$\pm$0.005. With such mass and a difference in effective temperature of 600\,K from spectroscopy, the secondary and primary are, \new{respectively,} in the early and advanced stage of the first dredge-up event on the red-giant branch. The period of the primary's surface rotation resembles the orbital period within ten days. The radial rotational gradient  between the surface and core in \Asterix  is found to be 6.9$^{+2.0}_{-1.0}$. This is a low value but not exceptional if compared to the sample of typical single field stars. The seismic average of the envelope's rotation agrees with the surface rotation rate. The lithium abundance is in agreement with quasi rigidly-rotating~models.}
{The agreement between the surface rotation with the seismic result indicates that the full convective envelope is rotating quasi-rigidly. 
The models of the lithium abundance are compatible with a rigid rotation in the radiative zone during the main sequence. Because of the many constraints offered by oscillating stars in binary systems, such objects are important test beds of stellar evolution.}

\maketitle




\begin{figure*}[th!]
\centering
\includegraphics[width=0.99\textwidth
]{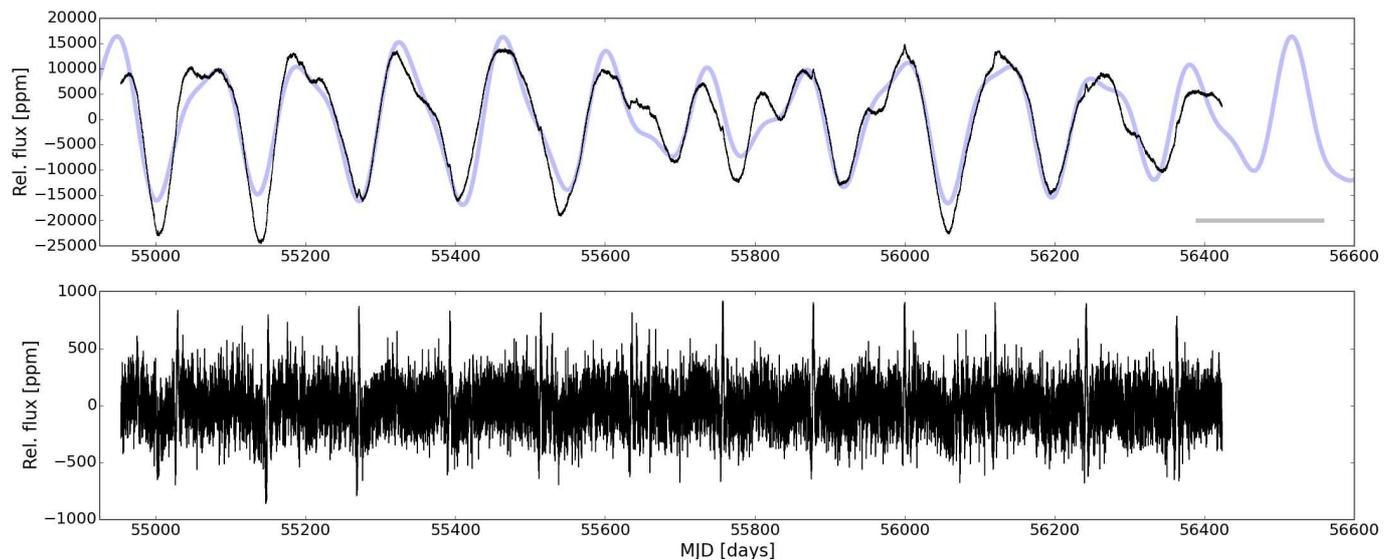}
\caption{Light curve of \Asterix. The top and lower panels show the full light curve (Q0-Q17), smoothed with a filter of 150 and 4 days, respectively. The synthetic light curve, depicted as solid light grey line is calculated from the four significant frequencies from \Table{tab:surfaceModulationFrequencies}. The time base of the spectroscopic monitoring is indicated through the horizontal grey bar.
\label{fig:LCcurve} }
\end{figure*}



\section{Introduction}
Rotation, activity and the surface abundance of lithium in a star are subject to its internal structure and evolution \citep[e.g.][]{Skumanich1972}. The key parameter to understand stellar evolution is the mass of the star. It is, however, challenging to determine stellar masses for objects that are not members of a stellar cluster or a binary system. An overestimated mass leads to an underestimate of the stellar age \citep*[e.g.][and references therein]{kippenhahn2013,Salaris2005}. 
For red-giant stars the current uncertainty is typically better than $\sim$30\% \citep[e.g.][]{Casagrande2016, SilvaAguirre2016}. Recent independent studies by  \cite{Mosser2013}, \cite{Epstein2014} and \cite{Gaulme2016}
gave strong indications that the seismic scaling relations 
for solar-like oscillators \citep[e.g.][]{Kjeldsen1995,Chaplin2011} are overestimating the stellar mass of red giants by about 15\%.
While the current precision of the seismic inferred mass is about 5\%, 
double-lined spectroscopic binaries (SB2) however allow us to constrain the ratio of mass between the two stellar targets with a much higher precision and accuracy than the masses of the two stellar components can be determined. Furthermore, both stars in such systems have the same age \new{and initial composition} (unless they originated from a rare capturing event). Therefore, the differences found between the stellar component of such system can be accounted to the difference in mass. Indeed, this makes such systems attractive targets to study rotation, mixing and stellar evolution in general.

Due to the low small rotation rates and the typically low activity level, surface rotation of red-giant stars is notoriously difficult to measure from photometry. They also often rotate too slowly to be resolved by the rotational broadening of the absorption lines, even with a high-resolution spectrograph.
Only recently, \cite{Ceillier2017} has performed a consistent search amongst $\sim$17000 \Kepler red giants, from which they determined rotation periods for $\sim$360 stars, of which a large fraction could be binaries or mergers. 
Given the observational limitations, the evolution of the surface rotation during the red-giant phase is challenging to be monitored. Ironically, the rotation of the deep interior of red-giant stars is nowadays much more straightforward to measure. By means of seismology, \cite{Beck2012}  constrained the core-to-surface rotation rate of red giants and by analysing the average core rotation in a large set of red giants. \cite{Mosser2012c}  showed that stars in the red clump exhibit slower core rotation than stars on the red-giant branch (RGB). Through inversion techniques \citep{Deheuvels2012,Deheuvels2015,Beck2014a,diMauro2016, Triana2017},  it is in principle possible to determine also the rotation rate of the upper layers. Such a result has, however, has never been confronted with independently determined surface rotation rates for red giants.
On the main sequence, several studies have derived surface rotation periods for several hundred stars \citep{Garcia2014b, McQuillan2014}. 
\cite{Benomar2015} found from asteroseismic inversion that the rotational gradients in solar-type stars are flat and that rotation rates derived from stellar spots give higher values than seismology.

In this context, the abundance of the fragile element lithium (Li) depends on rotation, mixing as well as on the depth of the convective zone and delivers good diagnostics for stellar evolution \new{for red-giant and main-sequence stars} \citep[e.g.][\new{\citealt{Lambert2004,Castro2016,Guiglion2016, Beck2017Li}} and references therein]{Zahn1992,Zahn1994,Charbonnel1994,Talon1998}. \new{These d}etailed studies \new{have shown that for} red-giant stars Li becomes increasingly depleted with decreasing effective temperature.  The main event of lithium-dilution is the first dredge-up (FDU), which occurs at the bottom of the RGB, when the convective envelope of the low-luminosity red-giant starts to deepen into the star \citep[e.g.][]{Charbonnel2000}. This phase is, compared to time scales of stellar evolution extremely short and it is rare to find binaries where both components are undergoing the FDU event at the same time.
However, it is still not fully understood why less than 1\% of red giants exhibit large surface lithium abundances \citep[see][]{Brown1989,Liu2014}. In those cases, the analysis highly benefits from asteroseismology of red-giant stars, by measuring mass as well as identifying evolutionary stage \citep[e.g.][]{SilvaAguirre2014,Jofre2015}.

%

 When studying aspects of stellar evolution, such as the lithium abundance on a star-by-star basis, binary systems provide an enormous analytical advantage. In contrast to comparing two individual field stars, the analysis of components of a double-lines spectroscopic binary system, which are born at the same time from the same cloud \new{\citep[e.g.][and references therein]{Satsuka2017}} allows to eliminate many uncertainty factors as  the two stars  are equal in age, initial composition and starting conditions as well as the distance and interstellar absorption. The analysis of the differences between the two components of a binary system sheds light on processes under the stellar surface, such as rotational mixing or transport by gravity waves \new{and initial conditions \citep[e.g.][and references therein]{Zahn1994, Charbonnel2000,Lebreton2001,Grundahl2008, Appourchaux2015, SchmidAerts2016}}. This is especially interesting, if such systems are in phases of stellar evolution when small differences in mass have a substantial impact. 

Around 50 binary systems with at least one oscillating red-giant component were observed by the NASA \Kepler space telescope \citep{Borucki2010}. Stellar binarity of these objects was indicated either through the presence of stellar eclipses \citep[][]{Gaulme2013,Gaulme2014,Gaulme2016,Beck2015Toulouse} or from the presence of tidally induced flux modulations during the periastron passage \citep{Beck2014a} in the \Kepler light curve.
Although all these objects were characterised through asteroseismology. By now only three binaries with a red-giant component have been analysed in detail using \Kepler photometry and ground-based spectroscopy, 
\KIC{8410637} \citep{Hekker2010,Frandsen2013}, \KIC{5006817} \citep{Beck2014a}, and recently \KIC{9246715} \citep{Rawls2016}. \new{In another, study soon to appear, \cite{Themessl2017} present a reanalysis of KIC\,8410637 from the full four years of \Kepler photometry and extended spectroscopic monitoring and present a detailed analysis of  KIC\,6540750 and KIC\,9540226.}
On the main sequence, several binary stars have been analysed, such as \object{KIC\,7510397} \citep{Appourchaux2015}, KIC\,10124866 \citep[][]{White2017} or the system of \object{16\,Cyg A\&B} \citep{Metcalfe2015,Davies2015} in which both components are oscillating.
The \new{red-giant or sub-giant} primaries of only few binary systems, Procyon, \ttau, \new{$\mu$\,Herculi}, were studied with seismology through the ground-based observations, utilising high-resolution spectroscopy with a metre-per-second accuracy \citep[][\new{respectively}]{Arentoft2008,Beck2015a,Grundahl2017}.

In this paper, we use photometric data from the \Kepler space telescope and  \Hermes ground-based spectrograph (\Section{sec:Observations}). A comprehensive analysis of the light curve (\Section{sec:surfaceRotation}), spectroscopy (\Section{sec:specAnalysis}), and the global asteroseismic parameters seismology for both stellar components (\Section{sec:PSD}) and the seismically determined radial rotational gradient in the primary (\Section{sec:seismoRotation}) of the binary system KIC\,9163796 (TYC\,3557\,2118\,1, V\,=\,9.82\,mag) is presented. 
In \Section{sec:activity} we discuss the effects of stellar activity on the primary. This system yields well-constrained input parameters for the theoretical modelling of the mixing of chemical species, in particular of lithium (\Section{sec:lithium}) and the tidal evolution of the system (\Section{sec:modelingTides}). 
Finally, the conclusions of this paper are summarised in \Section{sec:Conclusions}.


\section{Observations \label{sec:Observations}}

\begin{table}[t!]
\caption{Journal of observations and the radial velocities for both components of the system \KIC{9163796}.}
\centering
\tabcolsep=7pt
\begin{tabular}{crrrr}
\hline\hline
\multicolumn{1}{c}{BJD'} &
\multicolumn{1}{c}{ExpTime} &
\multicolumn{1}{c}{S/N(Mg)} &
\multicolumn{1}{c}{RV1} &
\multicolumn{1}{c}{RV2} \\
\multicolumn{1}{c}{$[$days]} &
\multicolumn{1}{c}{[s]} &
\multicolumn{1}{c}{} &
\multicolumn{1}{c}{[km\,s${-1}$]} &
\multicolumn{1}{c}{[km\,s${-1}$]}  \\
\hline
6388.6218	&750 	& 50  &	-14.70	&	-8.16 	  \\
6392.6194	&1250 	& 65  &	-16.41 	&	-4.34 	 \\
6415.6180	&1200 	& 70  &	-21.06	&	-0.19 	 \\
6432.5633	&1800 	& 80  &	-21.62	&	0.30 		 \\
6438.7131	&600 	& 50  &	-21.37	&	0.18 		 \\
6454.6504	&1800 	& 85  &	-19.12	&	-2.01 	 \\
6457.7038	&400 	& 40  &	-18.23	&	-2.75 	 \\
6463.4188	&600 	& 40  &	-15.76	&	-5.47 	 \\
6469.5667	&1800 	& 80  &	-10.88	&	-10.88 	 \\
6472.4185	&400 	& 30  &	-7.35		&	-12.84 	 \\
6479.5045	&1800 	& 85  &	13.22	&	-34.67 	 \\
6481.4543	&1800 	& 90  &	25.55	&	-47.66	 \\
6483.4543	&1800 	& 60  &	41.33	&	-63.22	 \\
6485.5717	&1800 	& 80  &	48.50	&	-71.13	 \\
6488.4878	&1800 	& 75  &	30.87	&	-53.19	 \\
6506.3947	&600 	& 35  &	-11.92	&	-11.92	 \\
6510.6143	&1200 	& 55  &	-14.93	&	-7.19		 \\
6512.3816	&600 	& 40  &	-15.78	&	-6.00		 \\
6515.5870	&600 	& 40  &	-17.09	&	-3.88		 \\
6524.5240	&600 	& 40  &	-19.54	&	-1.85		 \\
6533.6284	&600 	& 40  &	-20.74	&	-0.44		 \\
6535.4217	&600 	& 45  &	-21.03	&	-0.41		 \\
6542.4943	&600 	& 35  &	-21.54	&	0.06		 \\
6555.4203	&600 	& 50  &	-21.65	&	0.18		 \\
6560.3886	&600 	& 50  &	-21.37	&	0.02		 \\ \hline
7170.5893 	&1800 	& 80  &	-21.03	&	-0.13		 \\
7198.6549 	&1050 	& 60  &	-10.19	&	-10.19		 \\
7239.5499  	&1800 	& 80  &	-15.26 		&      -6.40\\
\hline
\end{tabular}
\tablefoot{The midpoint of the observation in the abbreviated barycentric julian date, (BJD'\,=\,BJD\,-\,2450000.0), exposure times of each individual spectrum, the signal to noise at the  \'echelle order 69, and the derived radial velocities for the high (RV1) and small profile (RV2) in the cross correlation function.  We adopt the 2\,$\sigma$-threshold of 140\,m\,s${-1}$ of the night-to-night stability as the typical uncertainty of the measurement. The horizontal line marks the beginning of \hbox{the observations in 2015}. 
\label{tab:journalOfObservations}}
\vspace{-5mm}
\end{table}%

The binary nature of the system \Asterix was first reported by \citet[][hereafter referred to as BHV14]{Beck2014a} from eight quarters of \Kepler observations (Q, i.e. 90-day data segments). 
BHV14 reported the detection of 18 binary systems, which exhibit an ellipsoidal flux modulation during the phase of closest encounter of the two stellar components (periastron). These stars are colloquially referred to as \textit{Heartbeat stars},  a term coined by \cite{Thompson2012}.  As most of the systems reported by BHV14, \Asterix is not eclipsing, though it exhibits clear ellipsoidal flux modulation every 121.3\days (\Figure{fig:LCcurve}, bottom panel). This feature will be discussed in more detail in \Section{sec:modelingTides}. All systems were monitored with the \Hermes spectrograph in 2013 by BHV14 and found to be eccentric with 0.2\,$\lesssim$\,$e$\,$\lesssim$\,0.8.
From this sample, \Asterix is one of the most interesting systems from a spectroscopic point of view. Visual inspection of the spectroscopic observations and the average line profiles from cross correlation show that this system is an obvious double-lined spectroscopic binary (SB2). Throughout this paper, we will refer to the brighter and more massive component as the \textit{primary} (or for parameters the index 1) and fainter, less massive star as the \textit{secondary} (index 2), respectively.
Although the inclination cannot be determined precisely due to the lack of eclipses, the SB2 nature on the other side allows for the determination of the mass ratio $q$, separation of the individual spectra of the components and therefore detail atmospheric diagnostics.


\begin{figure*}[t!]
\centering
\includegraphics[width=0.99\textwidth]{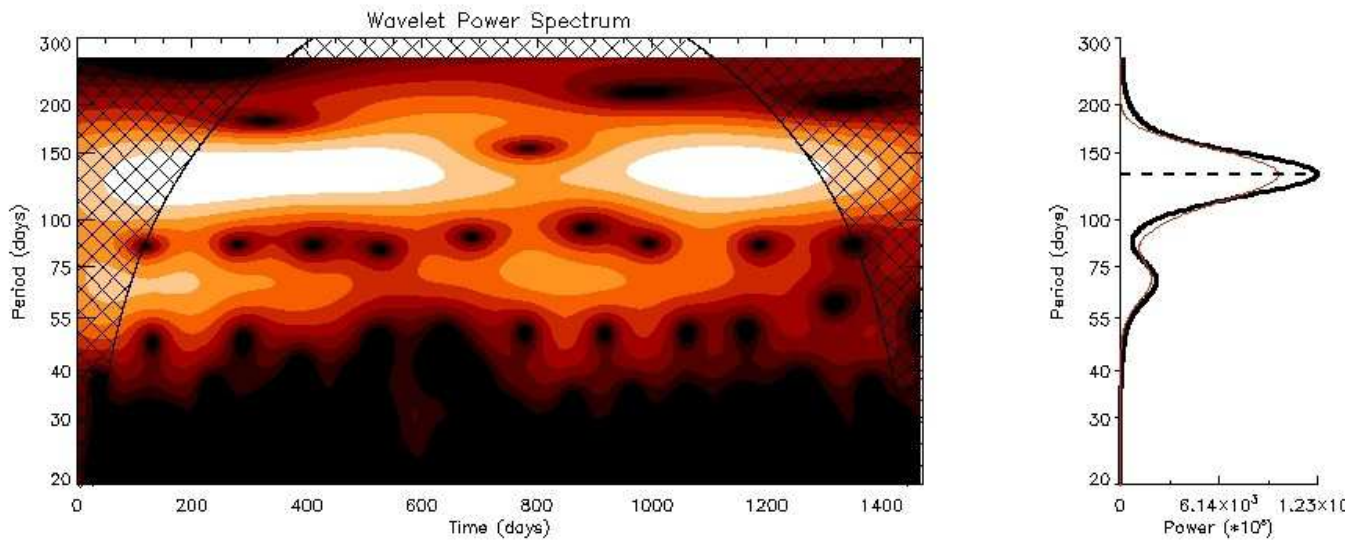}
\caption{Analysis of the long periodic light curve modulation of \Asterix in the period range between 20 and 200 days. The left panel depicts the wavelet analysis of the full time series in period space. Period regions in the wavelet for which the window function is larger than the covered time base are indicated through a grid pattern. The right panel shows  the GWPS. 
\label{fig:surfaceRotation}}
\end{figure*}

\subsection{Space photometry}
The \Kepler light curve of \Asterix, depicted in \Figure{fig:LCcurve}, covers overs 1470\,days (Q0-Q17) with a duty cycle higher than 90\%. The light curves of individual quarters were extracted from the target pixel data following \cite{Bloemen2013} and stacked following \cite{Garcia2011}. Regular gaps originate from dumping the saturated angular momentum of the reaction wheels and the interruptions from data download and spacecraft roll are complicating the spectral-window function. These alias frequencies are  dominating the high-frequent regime of the power spectral density in ppm$^2$/$\mu$Hz (hereafter PSD) and if not corrected for, the low-amplitude oscillation signal is hidden by the window function. To reduce these effects, data gaps of up to two days in the light curve  were filled through an inpainting technique described by \cite{Garcia2014}.

From a visual inspection of the individual \Kepler quarters, a long periodic photometric variation becomes apparent (\Figure{fig:LCcurve}, top panel). 
To preserve this variation in the stacked light curve, the individual quarters were stitched according to their slope 
and smoothed with a running triangular smoothing function, with a  width of 150\,days. This approach allows us to remove the 380-day flux modulation, originating from the \Kepler orbit, but preserves the signature from spots and ellipsoidal modulation as shown in the bottom panel of \Figure{fig:LCcurve}.  To remove the spot signal but preserve the flux modulation, originating from binarity, a second triangular filter with a width of 4 days was used. The resulting light curve containing spot modulations as well as the 'unspotted' version of the light curve, is shown in the top and bottom panel of \Figure{fig:LCcurve}, respectively. For both light curves the PSD is computed.
We note that the amplitude of the oscillation signal is very small, so we can only detect the oscillation signature obtained with the inpainted data owing to the increased signal-to-noise ratio (S/N).

\subsection{Ground-based spectroscopy}
We obtained 25 spectra spread over about 170 days in 2013 (partly published in BHV14), and we revisited the system in 2015 to take three additional spectra during about 70 days. All spectra were obtained with the \textit{High Efficiency and Resolution \Mercator Echelle Spectrograph}  \citep[\Hermes, ][]{Raskin2011,RaskinPhD} mounted on the 1.2\,m \Mercator telescope at the Spanish Observatorio del Roque de los Muchachos of the Instituto de Astrof{\'i}sica de Canarias. The spectrograph has a resolving power of  R\,$\simeq$\,85\,000 and covers a wavelength range between 375 and 900\nm. 

The journal of the observations is given in \Table{tab:journalOfObservations}. Exposure times were typically chosen to range between 600 and 1800 seconds and lead to a S/N of typically better than 40 in the range of the Mg\,\textsc{i} triplet ($\sim$518\nm). The S/N is measured from the gain corrected instrumental flux at the centre of the 69$^{\rm th}$ \'echelle order of \hermes, which corresponds to the region containing the Mg triplet. 

The raw data were reduced with the instrument specific pipeline. The radial velocities were derived through weighted cross-correlation of each spectrum with an Arcturus template. The best results are obtained by using the wavelength range between 478 and 653\nm for the cross correlation. The response of the cross correlation function clearly shows a double-peaked structure with both peaks moving in anti phase, confirming the SB2 nature of the system. For more details of the reduction process as well as the observation of red-stars in binaries with \Hermes, we refer to \cite{Raskin2011} and \cite{Beck2014a,Beck2015a}, respectively.



\begin{table}[t!]
\caption{Long periodic variations found in \Asterix from the wavelet and Fourier analysis.}
\centering
\tabcolsep=3pt
\begin{tabular}{lcccccccc}
\hline\hline
 & H & Frequency&	period	&phase&	\multicolumn{2}{c}{Amplitude}  &	S/N \\
&&[$\mu$Hz] & [day] & [] & [ppm] & [mmag] \smallskip\\
\hline
w$_1$ &$-$&0.088&131$\pm$11 &$-$&$-$&$-$&$-$\\
w$_2$  & $\simeq$w$_1\cdot$2 &0.168& 69$\pm$4 &$-$&$-$&$-$&$-$\smallskip\\
\hline
f$_1$	& $-$ &	0.087	&	133.036	&	0.335	&	12958.4	& 	12.4	&	15.8\\
f$_2$	& $\simeq$f$_1\cdot$2 &	0.163	&	70.827	&	0.316	&	3904.1	& 	3.7	&	4.4\\
f$_3$	& $-$ &	0.070	&	166.381	&	0.801	&	3781.9	& 	3.6	&	4.0\\
f$_4$ &$\simeq$f$_1$	&	0.083	&	140.128	&	0.095	&	3690.5	&	3.5	&	4.5\\
\hline
\end{tabular}
\tablefoot{The number of modulation component from the wavelet analysis ($w_i$) and the Fourier decomposition ($f_i$) is given.  The harmonic relation to other period components, frequency, period, phase with respect to the zero point, the semi-amplitude in parts-per-million and milli-magnitudes are reported. The final column gives the signal-to-noise of the frequency, which was computed for each oscillation mode within a frequency box of 0.1\cd.}\vspace{-5mm}
\label{tab:surfaceModulationFrequencies}
\end{table}%

\section{Long periodic brightness variations \label{sec:surfaceRotation}}

Photometric rotational variability is found only for a small fraction of the red giants observed with the \Kepler space telescope \citep{Ceillier2017}. \Asterix is one of these rare cases and falls into the less-populated long periodic tail of the distribution found by \cite{Ceillier2017}. 
Over the course of a year a star is observed by 4 different CCD detectors with varying systematic properties, because the \Kepler telescope is rotated at the end of each quarter by 90 degrees to keep the solar cells pointed at the Sun.
Therefore, intrinsic variations with periods longer than about 90 days can only be detected reliably if their amplitudes are strong enough. It has been shown by \cite{Ceillier2017} that from \Kepler light curves, rotation periods up to $\sim$175~days can be investigated. 


The frequency analysis, depicted in the left panel of \Figure{fig:surfaceRotation} follows the approaches of \cite{Garcia2014} and \cite{Ceillier2017}, using a wavelet decomposition with a Morlet mother wavelet. By collapsing this decomposition on the periods axis, the so-called Global Wavelets Power Spectrum (GWPS) is produced, which is shown in \Figure{fig:surfaceRotation} in the right panel. The GWPS reveals significant power at periods of 131$\pm$11 and 69$\pm$4\,days (see\Table{tab:surfaceModulationFrequencies}), and is otherwise flat. The reported uncertainty corresponds to the width of the peak profile in the GWPS and is partly governed by the temporal variation or the lifetime of the spots.
By using the programme Period04 \citep{Lenz2005}, we performed a Fourier decomposition through prewhitening, and found four significant frequencies at S/N\,$\geq$\,4 \citep[][]{Breger1993}. The Fourier parameters of the extracted significant frequencies are reported in \Table{tab:surfaceModulationFrequencies} and the resulting fit is shown in \Figure{fig:LCcurve}.  A good agreement between the two different analysis techniques is found for the two dominant variation time scales.

The \new{$\sim$}130-day period of the main flux modulation is close to the binary orbital period of $\sim$120\,days. Formally, the uncertainty of the photometric period would agree with the orbital period. This measurement error however comprises many different effects such as temporal variation of spots or instrumental effects. \new{We note that GWPS and Fourier decomposition, two methods sensitive to different systematic noise sources,  agree well on the rotation period (\Table{tab:surfaceModulationFrequencies}). This period of $\sim$130\,days is also not a multiple of the quarter length of 90 days, as it would be expected for instrumental periods \citep[e.g.][]{Garcia2014b,Ceillier2017}. Therefore, we consider this period to originate from the stellar signal rather than being an instrumental artefact.}

Differential rotation is expected to have an effect on the measured rotation period, since spots are not expected to be located exactly on the equator and therefore will not deliver the actual equatorial rotation period. 
For red-giant stars, mainly solar-like profiles of differential surface rotation have been detected through Doppler-imaging. 
For example, \cite{Kuestler2015} discussed a solar-like latitudinal differential  rotation profile  at the surface about 10 times weaker than in the Sun. However, in a few cases also the detection of anti-solar differential rotation profiles were reported  \citep{Kovari2015}. From the shear-factors describing the differential rotation, the solar-like case would give a difference in the rotation period at the pole of +2 days with respect to the equator. In the anti-solar case, the rotation period of the pole would be 5 days shorter than the one of the equator. For both assumed types of rotation, the difference between surface rotation and orbital period does not lead to a satisfying solution.

From spectroscopy we know that the primary is substantially brighter than the secondary (see following Section\,\ref{sec:specAnalysis}). It can therefore be argued that the spots originate solely from the primary, as it outshines the secondary.
With periods at hand, we now can discuss the lightcurve variation in \Figure{fig:LCcurve} in a more precise way. The periodic and well-defined minima in the photometric variation occur quite regularly with the main photometric period of ($\sim$133, $f_1$). The shape of the maximum however is more complex, as it consists of a double-peak structure, whereby the two peaks change their absolute and relative amplitude over time. Only in cases where the first maximum per cycle is lower than the second, the photometric minimum is early with respect to the main photometric time scale $f_1$. Therefore, over the time, covered by the \Kepler photometry, no phase shift of the pattern is found, as it would be expected for spots on a differentially rotating surface. In a series of papers,  \citet[and references therein]{Jetsu2017} suggested that, spots in chromospherically active single or binary stars are located in long-lived active longitudes, which are separated by $\sim$180 degrees.  In active stars, the spot distribution differs with respect to the Sun, for which spots are evenly distributed in a band of latitude. The latitudinal differential rotation in these stars is probably weak and the change in the light curve comes from the shift of activity between the two regions. This explanation is in agreement with the  the double-peak structure, observed in the maximum. 
This picture is also fit by the 69-day modulation ($\sim$$f_1$/2), which is likely produced by the apparition of spots or active regions on opposite sides \citep{McQuillan2014}.

As the solar and anti-solar scenarios of differential surface rotation cannot explain the difference between surface rotation and orbital period, and the four years of \Kepler photometry do not indicate a strong level of differential surface rotation,
we consider that the system is not synchronised.  
A surface rotation period of 130\,d is also a normal value when compared to other red giants \citep{Ceillier2017}. 

The peak-to-peak amplitude of about 0.025\,mag (\Table{tab:surfaceModulationFrequencies}) is large enough to be easily detectable from ground-based multicolour photometry using small-sized telescopes utilising differential photometric techniques \new{to extend the lightcurve for spot modelling beyond the four years of data provided by the \Kepler mission.} \citep[e.g.][]{Breger2006,Saesen2010}. 


\begin{figure}[t!]
\centering
\includegraphics[width=0.49\textwidth]{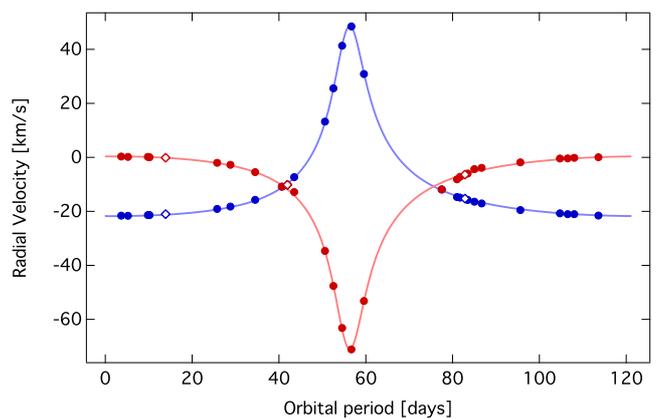}
\caption{\label{fig:sb2Curve}
Radial velocities and orbital solutions for the primary and secondary components of \Asterix, depicted in blue and red, respectively. Observations from 2013 are represented through dots, while observations from 2015 are marked as diamonds. The depicted orbital solution originates from spectral disentangling and was shifted by the systematic velocity to match with the measured RVs. 
}
\end{figure}




\section{Spectroscopic analysis \label{sec:specAnalysis}}

The visibility of both spectral components in the composite spectrum allows a full characterisation of their atmospheric properties, beside determination of their mass ratio. The spectroscopic observations are well sampled over the orbital cycle (\Table{tab:journalOfObservations} and \Figure{fig:sb2Curve}). Therefore, we have optimal conditions to investigate both components with spectroscopic tools. 
The radial velocities for both components were measured by fitting two Gaussians to the double-peaked cross-correlation response function of each spectrum. As error of the measurements, we adopt the 1-$\sigma$ level (70\,m/s) of the night-to-night stability of the \hermes high-resolution observing mode. The resulting RVs' are listed in \Table{tab:journalOfObservations} and depicted in \Figure{fig:sb2Curve}. The radial velocity amplitudes, visible in \Figure{fig:sb2Curve} from this simple approach show that the two components have relatively similar amplitudes, indicating a mass ratio close to unity.

\begin{figure*}[t!]
\centering
\includegraphics[width=\textwidth,height=70mm
]{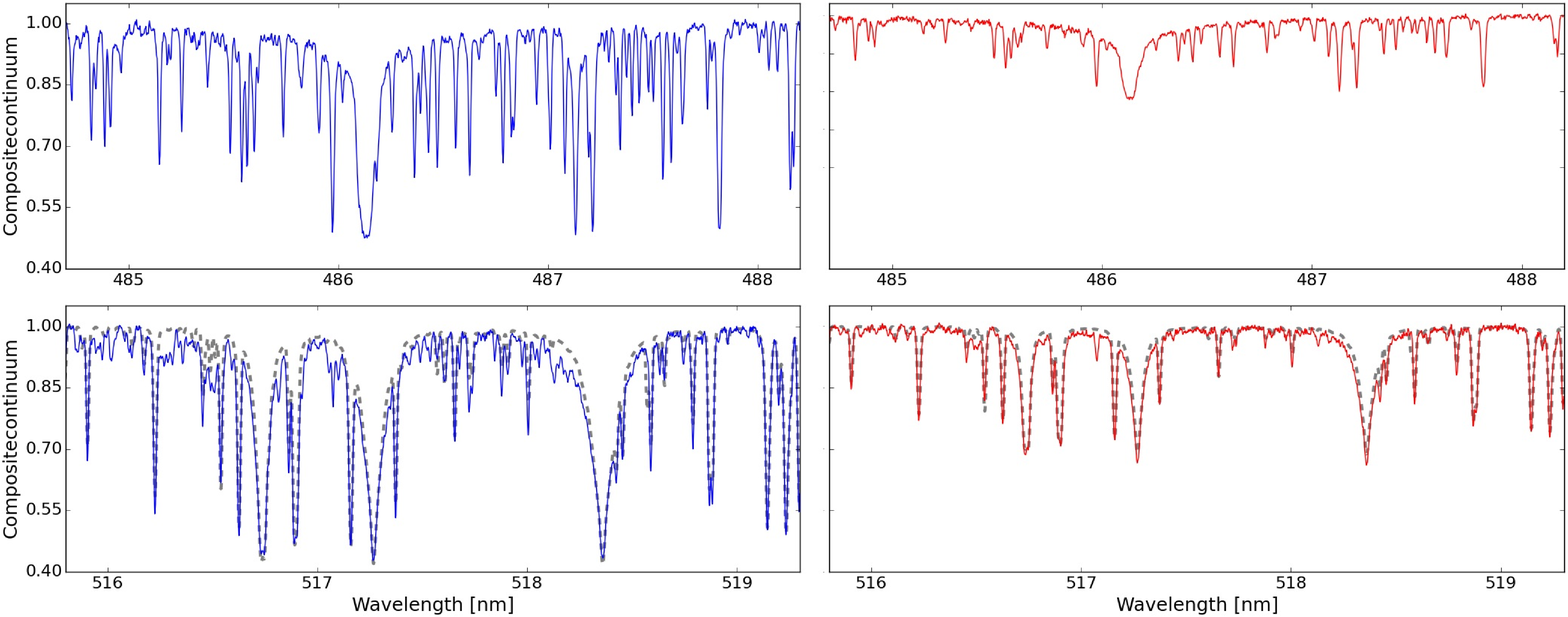}
\caption{Disentangled spectra of \Asterix. \new{The spectra of the primary in blue and secondary in red are shown in the left and right spectrum, respectively.} The \new{top} panels shows the region around the H$_\beta$ line, while the \new{bottom} panel depicts the magnesium triplet at 518\nm, normalised to the continuum flux of the composite spectrum. 
\new{The synthetic model of the best fit of the determination of fundamental parameters and metallicity is shown as dashed grey line is shown depicted in the region of the Mg triplet.}
\label{fig:spectralDisentangling} 
}
\end{figure*}


On first sight, it might look surprising that stars with such similar mass reveal quite different line depths for the two binary components \citep[see  \Figure{fig:spectralDisentangling}, or Figure\,2 in][]{Beck2016AzoresProceedings}. We therefore have good reasons to speculate at this point that both stars differ substantially in their fundamental parameters and therefore in their flux contribution. To determine the fundamental parameters of both stellar components, their individual contributions need to be extracted from the SB2-composite spectrum. Spectral disentangling (hereafter also \spd) is a powerful technique, which enables separation of individual spectral components from a series of composite spectra of a binary system, simultaneously optimising the orbital elements of a binary \citep{SimonSturm1994}. For \Asterix however, the clearly visible SB2 signature fortunately allows us to obtain reliable initial guesses on the orbital and fundamental parameters of both components a priori. 

\begin{table}[t!]
\caption{Orbital solutions for \Asterix.}
\centering
\tabcolsep=7pt
\begin{tabular}{lcrrr}
\hline\hline
Parameter & Unit & \\
\hline
$\Delta$T 			& [day] 	 	& 850\\
N				&[\#]			& 28\\
$P_{\rm orbit}$ (fixed)		& [day]		& 121.3\\
T$_{\rm peri}$		& [day]		& 2456485.53$\pm$0.03 \\
$e$ 				& []			& 0.692$\pm$0.002\\
$\omega$ 		& [rad] 		& 176.0$\pm$0.1\\ 
$K_1$ 			& [km/s]		& 35.25$\pm$0.12\\
$K_2$ 			& [km/s]		& 35.77$\pm$0.13\\
$q$				& []			& 1.015$\pm$0.005\\
\hline
\end{tabular}
\tablefoot{The set of optimal orbital elements found through spectral disentangling with FDBinary is reported. The timebase $\Delta$T and the number of data points $N$. The orbital period $P_{\rm orbit}$ from \Kepler photometry, the periastron passage $T_{\rm peri}$ and longitude $\Omega$ of periastron,  the  eccentricity $e$, the velocity of the system $\gamma$, the radial velocity amplitudes of both components $K_1$ \& $K_2$,  and the corresponding mass ratio $q$\,=\,M$_1$/M$_2$.
\label{tab:sb2Solution}}
\end{table}%

\begin{table}
\caption{Fundamental parameters derived from unconstrained fitting.}
\centering
\tabcolsep=3pt
\begin{tabular}{lcrr}
\hline\hline
Parameter & Unit &        Primary & Secondary \\
\hline
Effective temperature, \teff                 & [K]                 & 5020$\pm$100        & 5650$\pm$70 \\
Surface gravity, \logg                         & [dex]                 & 3.14$\pm$0.2        & 3.48$\pm$0.3\\
Microturbulence, $v_{\rm micro}$         & [km\,s${-1}$]                 & 1.2$\pm$0.3                & 1.43$\pm$0.2\\
Proj. surface rotation, \vsini         & [km\,s${-1}$]                 & 4.7$\pm$0.5                & 5.0$\pm$0.5 \\
Metallicity, $[$M/H]                                 & [dex]                 & -0.37$\pm$0.1        & -0.38$\pm$0.1\\
Lithium, A(Li)                                 & [dex]                 & 1.31$\pm$0.08        & 2.55$\pm$0.07\\\hline
Light factors                                                &                 & 0.63$\pm$0.03                & 0.37$\pm$0.02\\
Magnitude difference, $\Delta$m$_V$			& [mag]		&		\multicolumn{2}{c}{0.58$\pm$0.08}	\\
Apparent magnitude				& [mag]			&10.32$\pm$0.04				&10.90$\pm$0.09\\
\hline
\end{tabular}
\label{tab:fundamentalParameters}
\end{table}%

\subsection{Spectral disentangling \label{sec:SPD}}
We use the {\sc FDBinary} code \citep{Ilijic2004} implementing spectral disentangling in the Fourier domain \citep{Hadrava1995} to disentangle the composite spectra of \Asterix. 
Starting from the initial set of orbital parameters, this approach searches for the optimal set of orbital elements and individual spectra using a simplex algorithm. 
Due to possible imperfections in the normalisation and merging of \'echelle orders, \spd of a long spectral segment can produce undulations in the \hbox{disentangled spectra.} 

The orbital elements were determined from disentangling a $\sim$20\nm wide segment  at 520\nm. We kept the orbital period fixed to the value of 121.3\,d, reported by BHV14 from the autocorrelation of the \Kepler photometric light curve. After finding a consistent orbital solution, a major part of the optical spectrum is disentangled with the orbital elements fixed. The extracted segments are depicted in  \Figure{fig:spectralDisentangling} and the orbital solution is reported in \Table{tab:sb2Solution}.  Our final solution for \Asterix, which for the orbit of the primary is in good agreement with BHV14, leads to the RV semiamplitudes of $K_1$\,=\,35.25$\pm$0.12\kms and $K_2$\,=\,35.77$\pm$0.13\kms for the primary and secondary, respectively. The uncertainties were derived from a Gaussian fit to the parameter distributions from 5000 bootstrap simulations (Pavlovski et al. in prep.). The mass ratio found is $q$\,=\,$M_1/M_2$\,=\,1.015$\pm$0.005, i.e. a difference of 1.5\%.

The SB2 orbital model, derived from the set of optimal orbital elements is depicted in \Figure{fig:sb2Curve}. To meet with the measured RVs the model was shifted by the systematic velocity of the system. Using this orbital solution, also a 5\nm wide region of the Li\,\textsc{i} 670.78\,nm resonance transition doublet was extracted  (\Figure{fig:spectralDisentanglingLithium}).
We further note that the RV measurements from the year 2015 fit well the orbital solution based on the observations from the year 2013. Therefore, we conclude that no third massive body is present in the system.

\subsection{Fundamental parameters and abundances}

This analysis was performed through a grid search for the fundamental atmospheric parameters of a star, by comparing grids of synthetic spectra to the observed spectrum with the \textit{Grid Search in Stellar Parameters} (\gssp\footnote{The GSSP package is available for download at
https://fys.kuleuven.be/ster/meetings/binary-2015/gssp-software-package.}) software package by \cite*{Tkachenko2015} \citep[see also][]{Lehmann2011, Tkachenko2012}
Synthetic spectra were computed using the {\sc SynthV} radiative transfer code \citep{Tsymbal1996} based on a grid of model atmospheres precomputed with the \hbox{{\sc LLmodels} code \citep{Shulyak2004}.} 

As input we used the isolated spectra of the primary and secondary component in the region of the {Mg}\textsc{i} triplet at about 518\nm as shown in \Figure{fig:spectralDisentangling}. An unconstrained analysis was performed, treating the two stars as independent sources and the light factors are assumed to be wavelength independent. This allows us to determine all fundamental parameters as well as the light ratio. The fundamental parameters for both stellar components from the best fit are reported in \Table{tab:fundamentalParameters}.

The fundamental parameters show that the primary of \Asterix is approximately 600\,K cooler than the secondary component. A slightly  higher \logg~is found for the secondary component. The primary contributes about 63$\pm$3\% to the total flux of the system in Johnson\,V and the secondary contributes 37$\pm$3\%.  \cite{Tamajo2011} showed that it is possible to perform disentangling even if no information on the light ratio of both stellar components is available from external constraints (e.g. eclipses, interferometry). The unconstrained solution of \spd is stable if the light contributions add up to the 100\%, as it is the case for \Asterix. 
A well determined value of the light ratio is fundamental to obtain an accurate renormalisation of the two isolated spectra from their composite continuum to their  individual continuum. 

\new{The light ratios and the total flux of the system translate into a magnitude difference of 0.58$\pm$0.08\,mag. The average peak-to-peak amplitude of 0.025\,mag (see $f_1$ in\Table{tab:surfaceModulationFrequencies}) is small compared to and within the uncertainty of the obtained magnitude difference. Therefore, the long-periodic photometric flux modulations as described in \Section{sec:surfaceRotation} will not have an impact on the renormalisation and hence the spectroscopic results.
}


The rotational broadening of the absorption lines does not differ (within the uncertainties) between the two stars, which is not surprising, as it is known to be difficult to determine precise projected surface rotation velocities from lines, barely resolved by the spectrograph \citep{Gray2005,Hekker2007}. For red giants, the effects of rotation are also competing with the line broadening from macro turbulence. From the spectroscopically determined \logg~and \teff, the primary appears to be on the low-luminosity regime of the Red-Giant Branch (hereafter RGB). The temperature difference of $\sim$600\,K between the secondary and the primary is excluding the secondary ascending on the RGB. However, \teff, \logg~and the luminosity ratio are compatible with a late sub-giant or very early red-giant.

Comparing the derived metallicity of [M/H]$\simeq$-0.375 to large spectroscopic samples of red giants, such as the APOKASC survey of stars the \Kepler field of view \citep{Pinsonneault2014,Holtzman2015} shows that this system is in the sparsely populated under-abundant tail of the distribution of field stars. Such abundances are typical in stars in the galactic halo or bulge \citep{Epstein2014} but the relatively low luminosity of an early RGB (this identification of evolutionary state later confirmed in the following section) suggests that the system is close and cannot be located in the halo. We note that, although the metallicity for both components was allowed to vary freely in the fit, we find both values within 0.01\,dex and therefore basically the same metallicity. This is expected for stars born at the same time from the same primordial cloud. Substantially different metallicities, which would be confirmed from visual inspection of the composite spectra, could only be explained if binary system would have formed by one component capturing the other.

\begin{figure}[t!]
\centering
\includegraphics[width=\columnwidth,height=70mm]{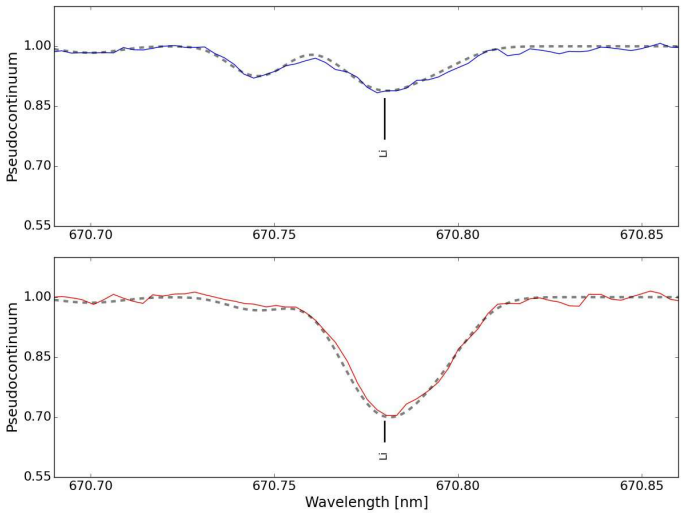}
\caption{Li\,\textsc{i} 670.78\,nm resonance doublet in both components of \Asterix. The spectra of the primary (\new{top} panel) and secondary (\new{bottom} panel) components have been renormalised to the individual continuum levels. The synthetic model of the best fit of the abundance determination is shown as dashed grey line. 
\label{fig:spectralDisentanglingLithium}
}
\end{figure}

The main abundance that is found to be significantly varying between the two components is the one of the fragile element of lithium, A(Li)\footnote{We use the standard abundance notation in which 
A(Li) = log[N(Li)/N(H)] + 12, where N(Li) and N(H) are the numbers of atoms per unit volume of element Li and of hydrogen} (\Figure{fig:spectralDisentanglingLithium}),  with A(Li) of 1.31$\pm$0.08 and 2.55$\pm$0.07\,dex for the primary and secondary, respectively. 
\new{The uncertainties in the Li abundances for both components were determined taking into account the uncertainties in the atmospheric parameters as they are listed in Table\,\ref{tab:fundamentalParameters}.}
 The less evolved component is found to be higher by 1.2\,dex than in the low-luminosity red-giant primary component of the system. Another case of a binary system with a strong difference in the Li abundance is \object{Capella}. \cite{Torres2015} recently redetermined the lithium abundance for the primary and secondary to be A(Li)$_{\rm 1}$\,=\,1.08$\pm$0.11 and A(Li)$_{\rm 2}$\,=\,3.28$\pm$0.13\,dex. This leads to a difference of 2.2\,dex between both components, being found on the secondary clump and the main sequence, whereby the more massive component has already undergone Li depletion. We therefore can assume a similar scenario also for \Asterix. Without a firm constraint on the mass and position of both stars in the \new{Hertzsprung-Russell Diagram}, it is difficult to distinguish between a lithium-rich giant or stars that have not yet undergone the FDU \citep{Charbonnel2000} as it depends on the Li abundance and the rotational history of the progenitor of the star on the main sequence \citep[e.g.][]{Talon1998}.

As discussed in the next Section, these stars exhibits solar-like oscillations, allowing to find a good mass estimates for the primary component that can serve as an input for stellar modelling. The astrophysical inference of this difference will be discussed in the modelling of the system in Section\,\ref{sec:lithium}. 



\section{Global seismic analysis of the power spectrum \label{sec:PSD}}

\begin{figure*}[t!]
\centering
\includegraphics[width=\columnwidth,height=55mm]{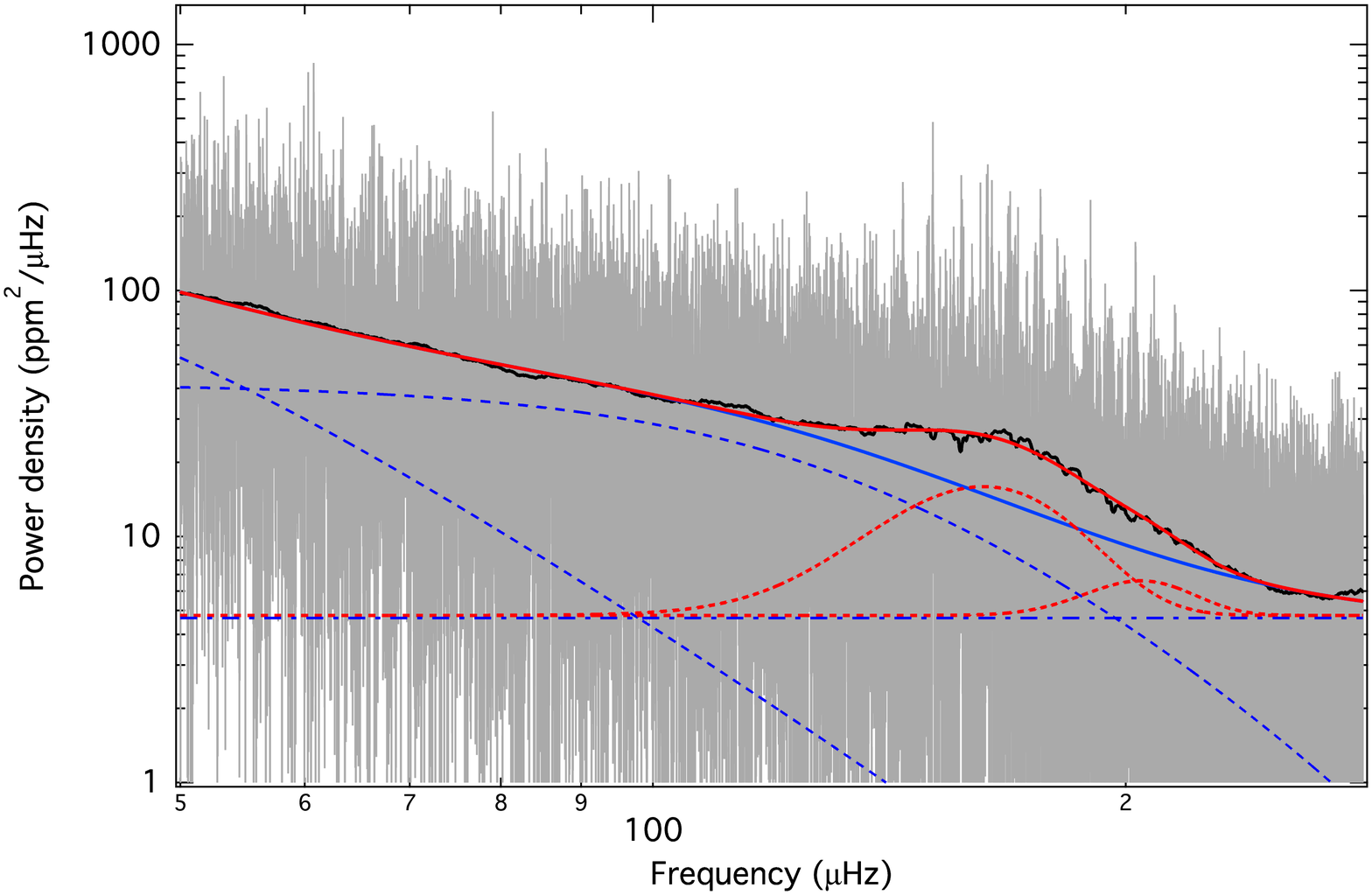}
\includegraphics[width=\columnwidth,height=55mm]{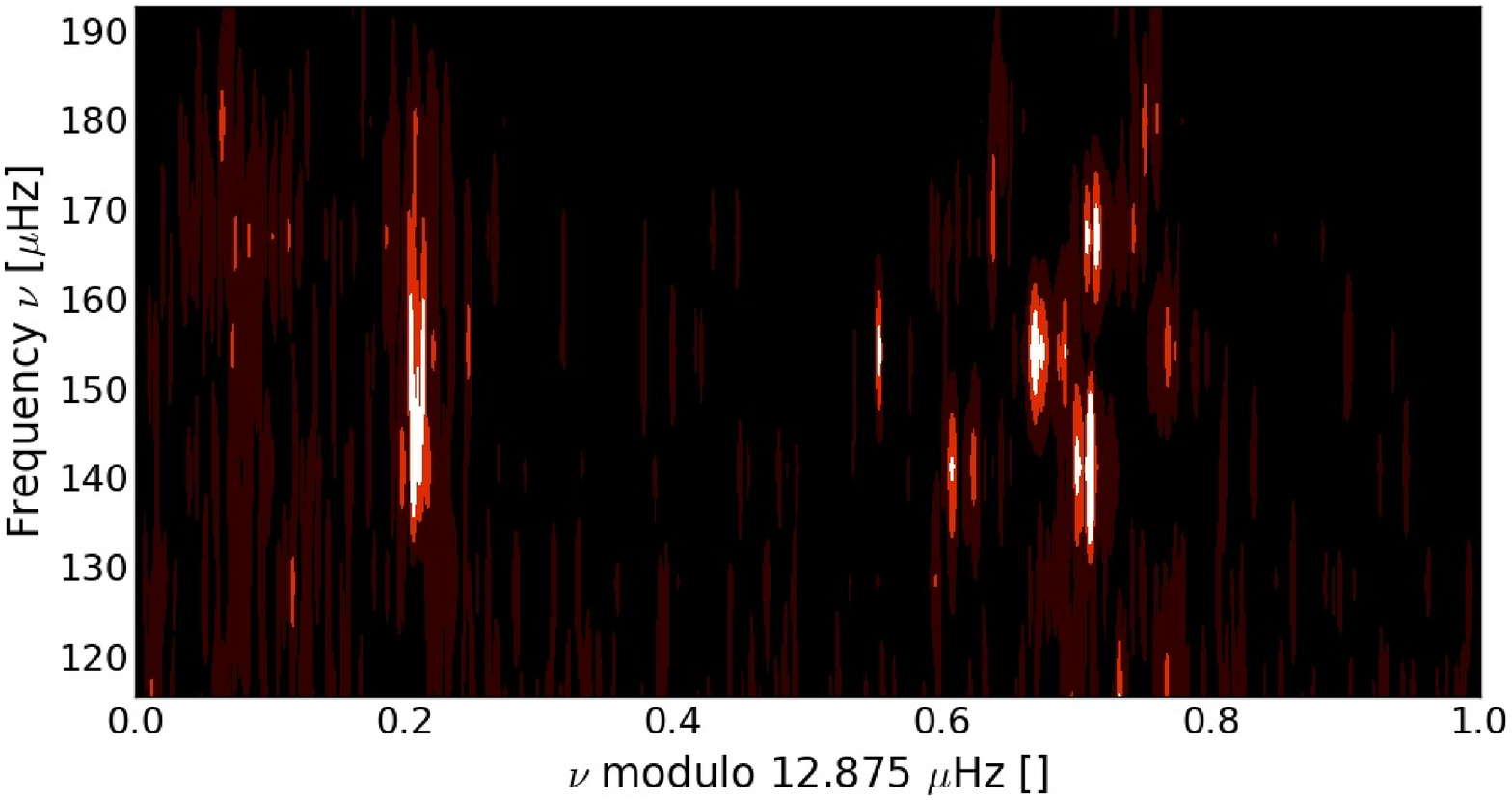}
\caption{Power spectral density (left) and \'echelle diagram (right) of \Asterix. 
In the left panel, the original and smoothed spectrum are shown as grey and black solid lines, respectively.  The different components that describe the granulation background are shown as blue dashed, while the white noise component as blue dash-dotted line. The two Gaussian envelopes to describe the power excess of the primary and reflection of the super-Nyquist power excess into the sub-Nyquist regime of the secondary are represented as red dotted lines. The combined fits with and without the oscillation components are depicted as red and blue solid lines.\newline
\label{fig:PSD}}
\end{figure*}

The power spectral density of \Asterix exhibits oscillation amplitudes that are substantially smaller than in normal, single-field red-giant stars (Figures\,\ref{fig:PSD}, \ref{fig:psdComparison}\,\&\,\ref{fig:comparingAmplitudes}). \new{The dominant cause of this effect is the photometric dilution of the seismic signal due to the bright companion, following
\begin{eqnarray}
\mathcal{A}_{{\rm diluted},i} = \mathcal{A}_{{\rm intrinsic},i}\cdot\mathcal{L}_i\,,
\label{eq:amplitude}
\end{eqnarray}
whereby for a given binary component $\mathcal{A}_{{\rm diluted},i}$, $\mathcal{A}_{{\rm diluted},i}$ are the measured and the intrinsic oscillation amplitude, respectively, and $\mathcal{L}_i$ is the spectroscopically light factor (see \Table{tab:fundamentalParameters} and Section\,\ref{sec:SPD}). Given} the spectroscopic light ratio between the primary and secondary star of about 3:2\new{, w}e can expect the mode heights (in power density) to be reduced by a factor of about 3 to 5 compared to a single red giant. Furthermore, several studies have suggested that stellar activity also reduces the amplitudes of solar-like oscillations \citep[e.g.][]{Mosser2009b,Garcia2010,Dall2010,Chaplin2011,Bonanno2014,Gaulme2014}. 
In addition to the low amplitudes, the frequency range of the oscillations ($\sim$120-250\muHz, see left panel of \Figure{fig:PSD}) is contaminated in the original power spectrum with power that leaks from high-amplitude low-frequency variations via the window function of the light curve.  
The analysis of the PSD and comparison to other stars is therefore not straightforward.

The resulting PSD reveals an excess of oscillation power at $\sim$160\muHz (\Figure{fig:PSD}). Stellar evolution is quite fast on the giant branch so that two stars of equal primordial chemical composition but slightly different mass \new{- as small as $\sim$1\% or even below -} can have significantly different radii. From the spectroscopic mass ratio we can therefore only constrain the location of the secondary power excess in a rather larger frequency range -- from overlapping the primary power excess to a frequency range typical for subgiants, well above the Nyquist frequency, $f_{Nq}\simeq$283\muHz.

\begin{table}[t!]
\caption{Seismic parameters derived for \Asterix.}
\centering
\tabcolsep=10pt
\begin{tabular}{crrr}
\hline\hline
Parameter & Unit &\multicolumn{1}{c}{\Asterix}&\multicolumn{1}{c}{KIC\,4586817}\\
\hline
\num				&	[$\mu$Hz]		& 165.3$\pm$1.3 		&161.4$\pm$0.4 \\
$\mathcal{A}_{\rm puls}^{\rm measured}$ & ppm 	& 29.1$\pm$1.9 		& \multirow{2}{*}{92.7$\pm$1.5} \\
$\mathcal{A}_{\rm puls}^{\rm intrinsic}$ 		& ppm & 46.2$\pm$4  \\
\dnu					&	[$\mu$Hz]		& 12.85$\pm$0.03 	& 12.759$\pm$0.009 \\
$\delta\nu_{02}$ 		&	[$\mu$Hz]		& 1.73$\pm$0.05		& 1.62$\pm$0.01 \\
$\Delta \Pi_1$			&	[s]				& 80.78 				& 79.05 \\
$\delta f_{\rm max}$	&	[$\mu$Hz]		& 0.24				& 0.22 \\ 
\hline
\end{tabular}
\tablefoot{The rows $\nu_{\rm max}$, $\mathcal{A}_{\rm puls}$, $\Delta\nu$ and $\delta\nu_{02}$ report the peak frequency, total amplitude of the oscillation power excess, and the large and small frequency separation of the central three radial orders for a given star.  $\Delta\Pi_1$ quantifies the true period spacing of dipole modes. The largest value of the detected rotational splitting \deltaf is also listed.\label{tab:seismicParameters}
}
\medskip
\medskip
\caption{Seismic fundamental parameters for \Asterix determined from seismic scaling relations and spectroscopic \Teff.}
\centering
\tabcolsep=10pt
\begin{tabular}{crrr}
\hline\hline
Parameter & Unit &\multicolumn{1}{c}{KIC\,9163796}&\multicolumn{1}{c}{KIC\,4586817}\\
\hline
Evol. state & & RGB & RGB\\ \hline
\Teff 	& 	[K]		& 4960$\pm$140 	& 4926$\pm$91\\
$M$		&	[M\sun]	& 1.39$\pm$0.06	& 1.36$\pm$0.03\\
$R$		&	[R\sun]	& 5.35$\pm$0.09	& 5.34$\pm$0.04\\
$L$ 		&      [L\sun] 	& 16$\pm$2 		& 15$\pm$1\\
$\log g$	&	[dex]		& 3.12$\pm$0.01 	& 3.117$\pm$0.004\\ 

\hline
\end{tabular}
\tablefoot{The evolutionary phase RGB describes a H-shell burning red giant on the ascending giant branch. Parameters for the primary star are calculated from grid modelling based on \num, $\Delta\nu$, and the spectroscopic estimate for \Teff. The secondary mass is estimated from the primary mass and $q$ and the other parameters are from grid modelling based on \num, \Teff, and the mass constraint. \label{tab:stellarParametersSeismology}}
\medskip
\end{table}%


\new{The stellar radius and mass, can be determined through the seismic scaling relations,  \citep[e.g.][]{Kjeldsen1995, Kallinger2010b,Kallinger2012,Kallinger2014, Chaplin2011}
\begin{eqnarray}
\frac{M}{M_\odot}&=&
\left(\frac{\nu_{\rm max}}{\nu_{\rm max}^\odot}\right)^{3}
\cdot\left(\frac{\Delta\nu}{\Delta\nu_\odot}\right)^{-4}
\cdot\left(\frac{T_{\rm eff}}{T_{\rm eff}^\odot}\right)^{3/2}\,,
\label{eq:mass}\\
\frac{R}{R_\odot}&=&
\left(\frac{\nu_{\rm max}}{\nu_{\rm max}^\odot}\right)
\cdot\left(\frac{\Delta\nu}{\Delta\nu_\odot}\right)^{-2}
\cdot\left(\frac{T_{\rm eff}}{T_{\rm eff}^\odot}\right)^{1/2}\,.
\label{eq:radius}
\end{eqnarray}
by comparing the} global seismic parameters, the central frequency of the oscillation power excess, \num, and  the large frequency separation between consecutive radial modes, \dnu \new{as well as the spectroscopic effective temperature, $T_{\rm eff}$ to the solar reference values}. 
Following the approach and \new{solar reference values} of \cite{Kallinger2010b,Kallinger2012,Kallinger2014} we derive $\nu_{\rm max}$\,=\,165$\pm$1\muHz and $\Delta\nu$\,=\,12.85$\pm$0.03\muHz, which translates into a mass and radius of 1.39$\pm$0.06\,M\sun and 5.35$\pm$0.09\,R\sun for the primary, respectively. We note, that if no index is given, we refer to parameters of the primary component \new{and} that \new{the reported} errors are internal uncertainties, which are likely to be too optimistic. 
\new{The result shows a discrepancy of the values of}  mass and radius of $\sim$0.9\,M\sun and $\sim$4.5\,R\sun, previously reported by BHV14\new{.} 
\new{The difference originates from the improved treatment of the light curve with gap-filling and inpainting techniques \citep[for details see, e.g.][]{Pires2009,Garcia2014}. Because approximately every three days one data point of the long-cadence observing mode (30\,min integrations) is rejected due to increased noise due to the on-board manoeuvre to withdraw the stored angular momentum of the spacecraft gyroscopes, a periodic gap is introduced to the data, which produces a high-frequency spectral window, which can hide the low-amplitude oscillation signal. 
The quality of the lightcurve is improved with respect to BHV14, by inpainting the gap through interpolation \citep[for details see][]{Garcia2014}, as well as by extending the length of the light curve by nearly a factor of two. Furthermore, the seismic data set was also corrected for the effects of spots.}

The \'echelle diagram is shown in the right panel of \Figure{fig:PSD} and the full list of all seismic global and fundamental parameters is given in \Table{tab:seismicParameters} and \ref{tab:stellarParametersSeismology}, respectively. The total pulsation amplitude $\mathcal{A}$ is defined as the square root of the integral over the power density excess defined by the Gaussian envelope. Photometric oscillation amplitudes quantify the light variation with respect to the mean brightness of the star. Because the total brightness of the binary system is larger than the mean brightness of the oscillating star, the amplitudes are affected by photometric dilution, but can be corrected if the light fractions are known (see \Table{tab:fundamentalParameters}). We refer to the amplitude which is corrected for the photometric dilution as the intrinsic one. For the primary component we find $\mathcal{A}_{\rm intrinsic}$\,=\,46$\pm$4\,ppm.



\subsection{Detection of the secondary \num}
Due to the low overall mode visibility, the oscillation spectrum of the primary star is difficult to analyse. Furthermore, \cite{Rawls2016} interpreted peaks which did not fit the general mode pattern as overlapping power excess of primary and secondary component.
To improve our understanding of the power excess of \Asterix and to exclude the possibility that modes from the secondary add to the individual mode pattern, we searched for a comparison star that follows  closely the frequency pattern of the radial and quadrupole modes of \Asterix's primary power excess. The best match is found with \KIC{4586817} ($\nu_{\rm max}$\,=\,161.4$\pm$0.4\muHz, $\Delta\nu$\,=\,12.76$\pm$0.01\muHz). To allow a better comparison,  the two oscillation spectra were normalised by their background signal \citep{Mathur2011,Kallinger2014}. The frequency axis is converted into radial orders via shifting the spectrum by the frequency of the central radial mode and by dividing by \dnu \citep[see also][]{Bedding2010CoAst}. From the representation of the power spectrum in terms of radial orders (\Figure{fig:psdComparison}), it can be seen that the frequency pattern of both stars is nearly identical. Therefore all significant peaks originate from the primary stellar component of \Asterix. If the secondary would have had a \num value very close to the primary, it should exhibit oscillations with about 40\% of the amplitudes of the primary.  
For \Asterix, we can exclude the scenario of overlapping power excesses as suggested for \KIC{9246715} by \cite{Rawls2016}. The power spectrum of \Asterix, as depicted in \Figure{fig:PSD} and \Figure{fig:psdComparison}, can therefore be treated as the one of a single star.

To search for the power excess of the secondary component, we included a second Gaussian component to the fitting standard approach of simultaneously fitting the granulation background and oscillation signal \citep{Kallinger2014}. Based on the Bayesian evidence we find that a model with two Gaussian components significantly better fits the observed PSD than a model with only one Gaussian. 
Tentatively, also the number of background components was allowed to vary to account for the contribution of the secondary to the background. In the standard approach, two components are used (from 0.1\num to $f_{Nq}$). Increasing the number of background components did, however, not improve the fit. 

According to this model the second power excess is located at 215$\pm$4\muHz with a measured total mode amplitude of 10$\pm$3\,ppm. Correcting for photometric dilution, we find $\sim$$30$\,ppm. The S/N is too low to be able to extract a value for \dnu from the PSD.  However, the measured \num of the secondary is very close to the value of the primary, which appears to contradict what was found from \Figure{fig:psdComparison} and the discussion above. It was shown and discussed in several papers, that power excesses, located above the Nyquist frequency for the \Kepler long cadence data of $\sim$283\muHz (also referred to as \textit{super Nyquist}) are reflected into the frequency regime below this value \citep[e.g.][BHV14]{Murphy2013,Chaplin2014SupNyq}. Because these stars are lower on the RGB or even subgiants, the intrinsic oscillation amplitudes are lower, which fits the picture for the detected \num of the secondary of \Asterix. In the super-Nyquist case,
 the true $\nu_{max,2}$ would be \new{located around $\sim$}350\,\muHz. To decide from seismology alone which scenario is correct we would need a clearly identified comb-like pattern of the pressure modes of the secondary power excess. Unfortunately, the mode visibility of the secondary is too low to determine a reliable estimate for \dnu, so that neither the subgiant nor the red-giant scenario can be rejected. However, the combination of asteroseismology and spectral disentangling offers an additional way of testing this hypothesis.

First, the spectroscopic fundamental parameters (derived in Section\,\ref{sec:specAnalysis}) are incompatible with $\nu_{max,2}$$\simeq$215\muHz as the temperature and fractional light ratio are indubitably  placing the secondary much further down, on the \new{very early} RGB, than the primary.
Furthermore, \new{Beck et al. (in prep)} propose a rewritten form of the seismic scaling relations, adapted to the output parameters of spectral disentangling \new{of the SB2 system} $-$ mass ratio $q$, the ratio of the light factors, $\mathcal{L}$, the effective temperatures \new{and the surface gravity $g$,} $-$ to gauge the ratio between the frequency of the two power excesses.
%
\new{Following the relation,}
\begin{eqnarray}
\frac{\nu_{\rm max,1}}{\nu_{\rm max,2}}\,&=&\,\frac{q}{\mathcal{L}}\cdot \left(\frac{T_{\rm eff,1}}{T_{\rm eff,2}}\right)^{3.5}\,,
\label{eq:disentanglingRelation}
\end{eqnarray}
\new{one finds $\nu_{\rm max,2}$=410$\pm50$\,$\mu$Hz, which gives an agreement within the reflected \num of the secondary ($\sim$350\,\muHz) within 1.2$\sigma$. If we assume a light ratio of 60:40, which is within the uncertainty of the spectroscopic solution, $\sim$360\,\muHz achieved. Using the spectroscopic information of the surface gravity,}
\begin{eqnarray}
\new{\frac{\nu_{\rm max,1}}{\nu_{\rm max,2}}\,}&\new{=}&\new{\,\frac{g_1}{g_2}\cdot \left(\frac{T_{\rm eff,1}}{T_{\rm eff,2}}\right)^{-0.5}\,,}
\label{eq:gRelation}
\end{eqnarray}
\new{we obtain $\nu_{\rm max,2}$=340$\pm$20\,$\mu$Hz, showing an even better agreement with the expected peak frequency. Therefore, we} find that KIC\,9163796 is a seismic binary with a sub and a super Nyquist oscillating components.



\subsection{Evolutionary stage \& dipole mode period spacing}

From the location of both power excesses clearly above 100\muHz and the seismic mass, it is evident that both components of the system are H-shell burning stars ascending the red-giant branch (RGB). Even members of the secondary clump would not reach such high values of $\nu_{\rm max,1}$ during their helium-core burning phase. This is confirmed by the observed period spacing of about 57\,s determined through autocorrelation of the mixed-dipole modes, which evidently corresponds to a H-shell burning star \citep{Bedding2011,Mosser2011a,Mosser2014}.

\section{Rotational gradient and envelope rotation\label{sec:peakbagging}
\label{sec:seismoRotation}}

\begin{figure*}[th!]
\centering
\includegraphics[width=\textwidth]{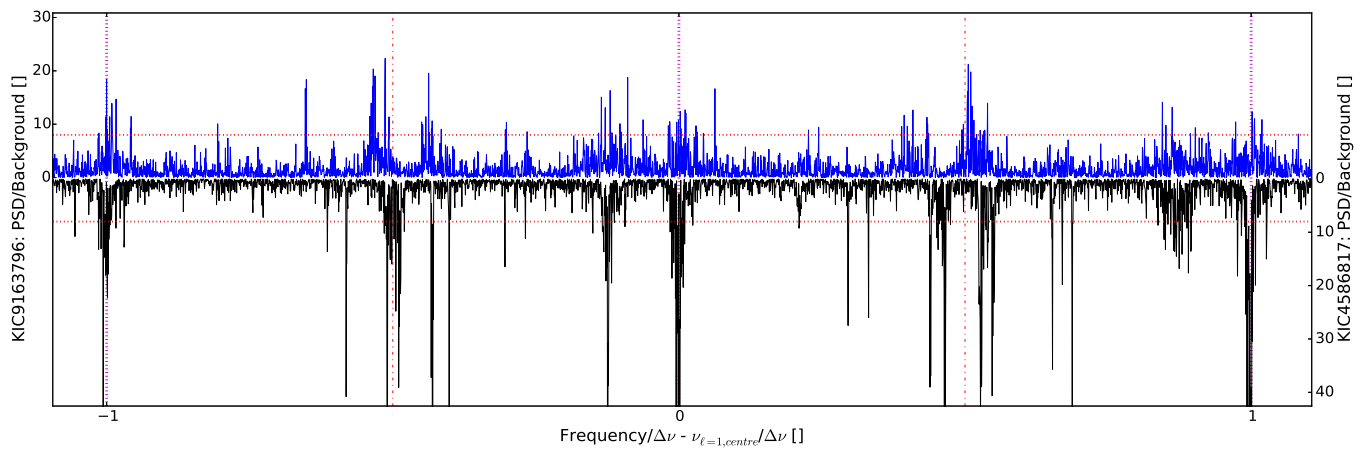}
\caption{  Normalised power spectral density of \Asterix (top) and \KIC{4586817} (bottom spectrum). The background contribution has been removed by division of the PSD through the background model. The frequency was divided by the respective large separation and centred on the central radial mode.
The frequencies of the $\ell$\,=\,0 and 1 modes are marked by blue dotted and red dash-dotted vertical lines. The dotted horizontal lines mark the traditional significance threshold of 8 times the background signal.
\label{fig:psdComparison}}
\end{figure*}

Seismic analysis relies upon the comparison of observed oscillation frequencies extracted from the PSD to theoretically computed frequencies to draw inference on stellar structure.
This is only possible with correctly identified pulsation modes, which is challenging for \Asterix due to the low S/N of the modes. 
Through the comparison with KIC\,4586817, the radial and quadrupole modes are, however, clearly identified (\Figure{fig:psdComparison}). 


For a rotating star, the Coriolis force as well as the coordinate transformation lift up the degeneracy of non-radial modes of the same radial order $n$ and spherical degree $\ell$, but different azimuthal order $m$, and thereby forms 2$\ell$+1 components. For slow rotators with a rotation profile depending on the radial coordinate only, $\Omega(r)$, the mode frequency can be written as,
\begin{equation}
f_{n,\ell,m}\,=\,f_{n,\ell,0} + m \cdot \delta f_{n,\ell}\,,\label{eq:rotSplit}
\end{equation} 
whereby $\delta f$ represent the rotational splitting. The dipole modes of the primary component of \Asterix show a clear signature of rotational splitting (Figures\,\ref{fig:psdComparison} and \ref{fig:MMID240}).

\subsection{Identification \& extraction of individual oscillation modes}
\begin{figure*}[t!]
\centering
\includegraphics[width=\textwidth,height=75mm]{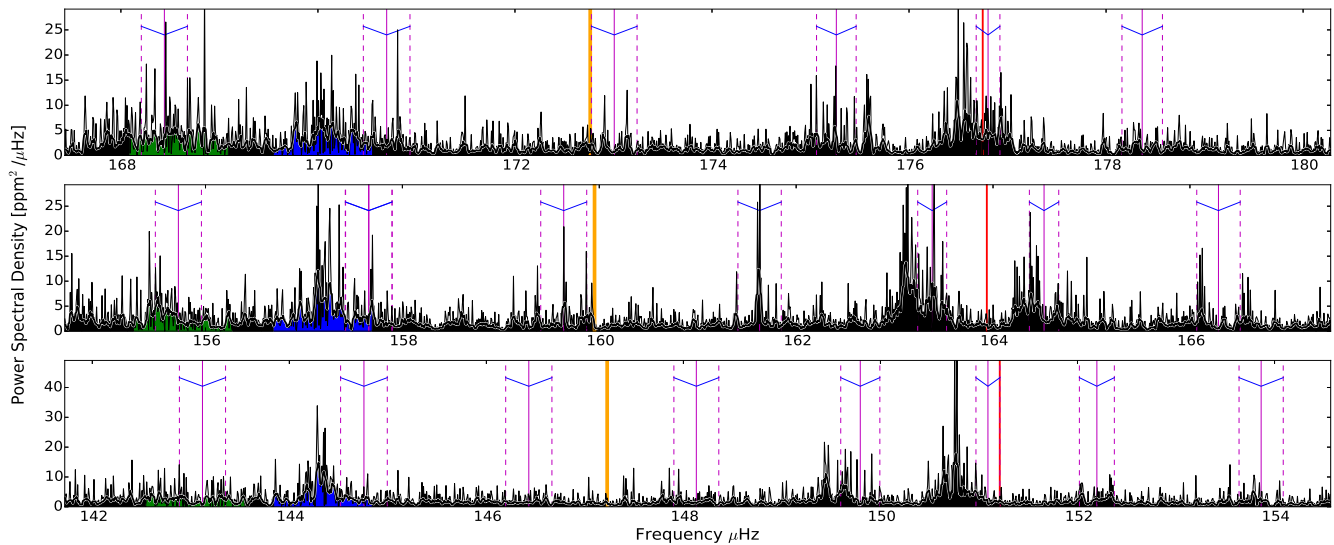}
\caption{Power-density spectrum of \Asterix. Each panel {contains} one radial order, depicting the original power spectrum as well as the spectrum smoothed through the 
Mode identifications of the pure (theoretical) $p$ modes for $\ell$\,=\,0, 1, 2 and 3 which are indicated with blue, red, green and yellow vertical bars, respectively. The effects of rotation are visible as the splitting of dipole modes (assumed $\delta f$\,=\,290nHz), located in the centre of each panel. The observed PSD is overlaid with the theoretical frequencies of mixed-dipole modes ($m$\,=\,0, solid thin lines) and the theoretical frequencies of the rotationally split components ($m$\,=\,$\pm$1, dashed thin lines). The components belonging to one rotationally split multiplet are indicated through $\rm V$-markers at the top of each panel.}
\label{fig:MMID240}
\end{figure*}

The resulting frequency pattern of rotationally split mixed-dipole modes in the power spectrum of a red giant can be described through the asymptotic expansion, developed by \cite{Mosser2012c,Mosser2012b}. In this approach, the adjustable parameters are the so called asymptotic period spacing of dipole modes, \Dpi, the coupling factor $\kappa$, between $p$- and $g$-modes \citep[see][]{Shibahashi1979,Unno1989}, and the rotational splitting of the gravity-dominated dipole modes, $\delta f_{max}$. We followed the grid search for the optimal combinations of \dpi and $\kappa$, as described by \cite{Buysschaert2016}, leading to  $\Delta\Pi_1$\,=\,80.78\,s. From the manual analysis, a $\delta f_{max}$\,$\simeq$\,0.29\muHz was found. These values were confirmed through the automated search of \cite{Mosser2015}. Therefore, the asymptotic expansion provides a firm mode identification of the individual peaks as shown in \Figure{fig:MMID240}. We note that the automated approach indicates the existence of a secondary solution of 81.15\,s and $\delta f_{max}$$\simeq$0.870\muHz. Such solution presents a special case, in which the rotational splittings are on the order of or even larger than the period spacing. In such, the rotational splitting is therefore not defined by the close pairs of peaks but by the complementary distance. After critically reviewing the solutions and from comparison of the PSD with KIC\,4586817 (\Figure{fig:psdComparison}) we can rule out the latter value.

For a full seismic analysis, the frequencies of the individual oscillation modes have to be extracted from the oscillation spectrum. Only in the central three radial orders modes have a sufficient S/N to be clearly distinguishable from noise peaks. The determination of the mode parameters of individual oscillation modes in the PSD of \Asterix has been performed with the \textsc{Diamonds} code of \cite{Corsaro2014}. The details of the Bayesian parameter estimation and model comparison by means of a nested sampling Monte Carlo algorithm are described in \cite{Corsaro2015a}. The detection probability $\mathcal{P}$ of an oscillation mode is defined as $\mathcal{E}_{\rm peak} / ( \mathcal{E}_{\rm no,peak} + \mathcal{E}_{\rm peak})$, where $\mathcal{E}$ represents the Bayesian evidence of the fitting models, one including and the other excluding the corresponding peak to be tested \citep[see][for more details]{Corsaro2014,Corsaro2015a}.
 
In total six pairs of radial and quadrupole modes have been measured (\Table{tab:individualFrequenciesPressure}) and 25 significant dipole modes were identified in the PSD (\Table{tab:individualFrequenciesDipole}). For ten dipole modes, the S/N allowed a clear identification of rotationally split components. For the rotationally split modes, an additional identification of the azimuthal order $m$, and \deltaf is provided in \Table{tab:individualFrequenciesDipole}. Rotational splittings of $g$- and $p$-dominated mixed modes carry information about the rotation rate weighted towards the core and envelope, respectively. Since the character of the g-dominated modes is governed by the properties of the core, the rotational splitting of the $g$-dominated modes is predominantly determined by the core rotation rate \citep{Beck2012,Mosser2012c,Goupil2013,Deheuvels2014,diMauro2016}. 
The situation is more complicated for the rotation rate of the envelope. Even though the core contribution to the rotational kernel of $p$-dominated modes is relatively small, the fast rotating core still contributes significantly to the rotational splitting of $p$-dominated modes. Therefore they need to be disentangled, which is presently only possible by using theoretical models. 

In principle, quadrupole modes are mixed modes as well but the observable modes contain a higher $p$-mode contribution than dipole modes. This makes them less sensitive to the core properties and more sensitive to the envelope rotation. For \Asterix, however, the noisy spectrum prevents us to see any split structure in quadrupole modes.

The value of the rotational splitting $\delta f_{n,\ell}$ in Eq\,(\ref{eq:rotSplit}) can be expressed as
\begin{equation} \label{eq:splittingKernel}
\delta f_{n,\ell}\,=\,\frac{1}{2\pi I_{n,\ell}} \int_{r=0}^{R} K_{n,\ell}(r) \Omega(r) dr,
\end{equation} 
where $I_{n,\ell}$ is the mode inertia and $K_{n,\ell}(r)$ gives the rotational kernel of a mode which depends on the mode eigenfunction \citep[e.g.][]{Cox1980}. $R$ is the radius of the model for the given star.

The visibility of the different $|m|$ components is independent of the rotation rate but is governed by the inclination between the rotation axis towards the line of sight \citep[e.g.][]{Gizon2003,Ballot2006}. In numerous works, \citep[e.g.][]{Ballot2006, Beck2013PhD,Deheuvels2015} it was shown that the actual geometry (i.e. the height ratio between the central and $|m|>1$ components) of a rotational multiplet in a solar-like oscillating star is strongly altered by lifetime effects of the stochastic modes. This is especially true for g-dominated mixed modes whose lifetime can be much longer than the time span covered by \Kepler observations. The inclination of the stellar rotation axis should therefore be extracted with care. In case of \Asterix,  the geometry of the splittings is extremely variable. Therefore we refrain from fitting the inclination but rather gauge the inclination to be between 40-70$^\circ$ from visual inspection. From the RV amplitude of the primary, we can estimate the mass function to be,
\begin{equation}
  M_1\cdot \sin^3 i = 0.85308 \pm 0.099  M_\odot.
  \end{equation} 
Using the seismic primary's mass, this function translates into an inclination of $\sim$58$^\circ$, which falls fairly well into the range constrained through seismology.  
Stronger constraints on the inclination could be provided from light curve fitting (e.g. BHV14) 
which is currently beyond the scope of this paper. 


\begin{table*}[t!]
\tabcolsep=20pt
\caption{Extracted radial and quadrupole oscillation modes and their parameters for \Asterix.}
\centering
\begin{tabular}{ccrrrcccc}
\hline\hline
\multicolumn{1}{c}{N}
& \multicolumn{1}{c}{$\ell$}
& \multicolumn{1}{c}{Frequency}
& \multicolumn{1}{c}{Amplitude}
& \multicolumn{1}{c}{FWHM }
& \multicolumn{1}{c}{$\mathcal{P}$}\\
& & 
\multicolumn{1}{c}{[$\mu$Hz]}
&\multicolumn{1}{c}{[ppm]} 
& \multicolumn{1}{c}{[$\mu$Hz]}
& \multicolumn{1}{c}{[]}\\
 
\hline 
 11 & 0  & $    144.337\pm0.009$ & $       7.44~_{-       0.84}^{+       0.78}$ & $       0.129~_{-       0.018}^{+       0.017}$   &     ...\\[2pt]
12 & 0 &   $    157.204\pm0.007$ & $      10.86~_{-       0.19}^{+       0.17}$ & $       0.239~_{-       0.005}^{+       0.005}$   &     ...\\[2pt]
 13 & 0 &   $    170.133\pm0.014$ & $       8.80~_{-       0.43}^{+       0.41}$ & $       0.378~_{-       0.015}^{+       0.015}$   &     ...\\[2pt]
14 & 0 &   $    182.996\pm0.021$ & $      10.92~_{-       0.30}^{+       0.39}$ & $       0.561~_{-       0.020}^{+       0.021}$   &     ...\\[2pt]
15 & 0  &  $    195.464\pm0.017$ & $       8.51~_{-       0.28}^{+       0.27}$ & $       0.589~_{-       0.018}^{+       0.019}$   &     ...\\[2pt]
16 & 0  &  $    209.046\pm0.028$ & $       3.41~_{-       0.34}^{+       0.33}$ & $       0.285~_{-       0.044}^{+       0.052}$   &       1.000\\[3pt] 
\hline
 11 & 2 &   $    142.824\pm0.086$ & $      17.19~_{-       0.45}^{+       0.55}$ & $       2.940~_{-       0.049}^{+       0.052}$   &     ...\\[2pt]
12 & 2 &   $    168.432~_{-      0.047}^{+      0.084}$ & $      17.09~_{-       0.25}^{+       0.22}$ & $       2.005~_{-       0.068}^{+       0.088}$   &     ...\\[2pt]
13 & 2 &   $    155.516\pm0.008$ & $       6.65~_{-       0.16}^{+       0.15}$ & $       0.185~_{-       0.005}^{+       0.005}$   &     ...\\[2pt]
14 & 2  &  $    181.071\pm0.017$ & $      11.19~_{-       0.23}^{+       0.21}$ & $       0.599~_{-       0.023}^{+       0.022}$   &     ...\\[2pt]
15 & 2 &   $    194.031\pm0.018$ & $       7.31~_{-       0.42}^{+       0.35}$ & $       0.484~_{-       0.029}^{+       0.032}$   &     ...\\[2pt]
16 & 2 &   $    207.743\pm0.047$ & $       7.71~_{-       0.35}^{+       0.39}$ & $       0.687~_{-       0.046}^{+       0.046}$   &     ...\\[3pt]
\hline
\end{tabular}
\tablefoot{The number of the radial order, spherical degree and azimuthal order $m$ of an oscillation mode. The parameters of the Lorentzian profile describing the mode are given, the median centre frequency, the mode amplitude and its full width at half maximum (FWHM). For the case of peaks with a lower S/N-ratio, the detection probability $\mathcal{P}$ is reported, which ranges between 0 and 1, whereby 1 corresponds to a detection probability of 100\%. A value '...' indicates modes with high S/N. }
\label{tab:individualFrequenciesPressure}
\end{table*}

\begin{table*}[t!]
\tabcolsep=12pt
\caption{Extracted dipole modes and their parameters for \Asterix. 
}
\centering
\begin{tabular}{cccccccr}
\hline\hline
\multicolumn{1}{c}{$\ell$}
& \multicolumn{1}{c}{$m$}
& \multicolumn{1}{c}{Frequency}
& \multicolumn{1}{c}{Amplitude}
& \multicolumn{1}{c}{FWHM or Height }
& \multicolumn{1}{c}{$\mathcal{P}$}
&\multicolumn{1}{c}{$\delta f$} \\
 & &  
\multicolumn{1}{c}{[$\mu$Hz]}
&\multicolumn{1}{c}{[ppm]} 
& \multicolumn{1}{c}{[$\mu$Hz] or [ppm$^2$/$\mu$Hz]}
& \multicolumn{1}{c}{[]}
&\multicolumn{1}{c}{[$\mu$Hz]}\\

\hline 
1 & ...& $    136.425~_{-      0.005}^{+      0.006}$ & $       2.91~_{-       0.28}^{+       0.26}$ & $       0.043~_{-       0.008}^{+       0.008}$ &        1.000 & $-$\\[1pt]
1 & ...& $    138.729~_{-      0.016}^{+      0.017}$ & $       4.16~_{-       0.45}^{+       0.49}$ & $       0.102~_{-       0.019}^{+       0.017}$ &        1.000& $-$\\[1pt]
1 & ...& $    147.901~_{-      0.032}^{+      0.035}$ & $       4.47~_{-       0.32}^{+       0.30}$ & $       0.388~_{-       0.049}^{+       0.049}$ &        0.674& $-$\\[1pt] 
1 & ...& $    152.318~_{-      0.005}^{+      0.005}$ & $       5.16~_{-       0.18}^{+       0.16}$ & $       0.232~_{-       0.012}^{+       0.016}$ &      ...& $-$\\[1pt]
1 & ...& $    154.706~_{-      0.006}^{+      0.007}$ & $       3.40~_{-       0.07}^{+       0.08}$ & $       0.075~_{-       0.002}^{+       0.002}$ &       1.000& $-$\\[1pt]
1 & ...& $    157.694~_{-      0.001}^{+      0.001}$ & 		[sinc] 					& $      198.33~_{-       10.35}^{+        2.82}$ &     ...& $-$\\[1pt]
1 & ...& $    164.389~_{-      0.013}^{+      0.017}$ & $       8.97~_{-       0.19}^{+       0.18}$ & $       0.298~_{-       0.019}^{+       0.019}$ &   ...& $-$\\[1pt]
1 & ...& $    176.627~_{-      0.013}^{+      0.014}$ & $      12.08~_{-       0.18}^{+       0.17}$ & $       0.318~_{-       0.009}^{+       0.009}$ &     ...& $-$\\[1pt]
1 & ...& $    177.971~_{-      0.001}^{+      0.002}$ &  		[sinc] 					& $       88.17~_{-        1.08}^{+        0.94}$ &     ...& $-$\\[1pt]
1 & ...& $    178.745~_{-      0.005}^{+      0.002}$ &  		[sinc]						& $       82.64~_{-        0.86}^{+        1.03}$ &     ...& $-$\\[1pt]
1 & ...& $    188.495~_{-      0.013}^{+      0.013}$ & $       9.71~_{-       0.43}^{+       0.50}$ & $       0.285~_{-       0.011}^{+       0.012}$ &      ...& $-$\\[1pt]
1 & ...& $    189.975~_{-      0.011}^{+      0.012}$ & $       8.29~_{-       0.21}^{+       0.21}$ & $       0.249~_{-       0.010}^{+       0.009}$ &      ...& $-$\\[1pt]
1 & ...& $    201.203~_{-      0.028}^{+      0.025}$ & $       6.27~_{-       0.34}^{+       0.34}$ & $       0.540~_{-       0.045}^{+       0.049}$ &      ...& $-$\\[1pt]
1 & ...& $    202.815~_{-      0.022}^{+      0.024}$ & $       8.27~_{-       0.43}^{+       0.48}$ & $       0.562~_{-       0.050}^{+       0.051}$ &      ...& $-$\\[1pt]
1 & ...& $    214.808~_{-      0.041}^{+      0.044}$ & $       6.67~_{-       0.43}^{+       0.46}$ & $       0.990~_{-       0.128}^{+       0.141}$ &      ...& $-$\\[1pt]
1 & ...& $    216.593~_{-      0.043}^{+      0.040}$ & $       4.91~_{-       0.36}^{+       0.38}$ & $       0.518~_{-       0.083}^{+       0.078}$  &     ...& $-$\\[5pt]
 \hline\hline

  \multirow{3}{*}{1} & -1& $    149.4228~_{-      0.0042}^{+      0.0051}$ & $       5.411~_{-       0.172}^{+       0.172}$ & $       0.077~_{-       0.002}^{+       0.002}$ &     ... &  \multirow{3}{*}{0.278$\pm$0.01}\\[1pt]
   & 0& $    149.6783~_{-      0.0040}^{+      0.0050}$ & $       5.311~_{-       0.161}^{+       0.155}$ & $       0.103~_{-       0.004}^{+       0.003}$ &    ...& & \\[1pt]
   & +1& $    149.9805~_{-      0.0095}^{+      0.0078}$ & $       3.891~_{-       0.116}^{+       0.118}$ & $       0.181~_{-       0.006}^{+       0.006}$ &     0.990\\[3pt]
 \hline
 
 \hline
  \multirow{3}{*}{1}  & -1& $    150.4854~_{-      0.0017}^{+      0.0017}$ & [sinc] & $      107.36~_{-       15.83}^{+        5.46}$ &       0.553& \multirow{3}{*}{0.157$\pm$0.005}\\[1pt]
  & 0& $    150.6321~_{-      0.0024}^{+      0.0020}$ & $       6.429~_{-       0.222}^{+       0.287}$ & $       0.139~_{-       0.004}^{+       0.005}$  &     ...& \\[1pt]
  & +1& $    150.8009~_{-      0.0050}^{+      0.0043}$ & $       9.702~_{-       0.399}^{+       0.246}$ & $       0.139~_{-       0.004}^{+       0.005}$ &     ...\\[1pt]

 \hline
 
  \multirow{2}{*}{1} & -1& $    153.5489~_{-      0.0009}^{+      0.0010}$ & [sinc] & $      138.02~_{-        1.88}^{+        3.39}$ &     ...&\multirow{3}{*}
  {0.245$\pm$0.003} \\[1pt]
   &0 & $    153.7943~_{-      0.0034}^{+      0.0033}$ & $       3.295~_{-       0.121}^{+       0.117}$ & $       0.081~_{-       0.002}^{+       0.002}$ &     1.000\\[3pt]
 \hline
 
  \multirow{3}{*}{1} & -1& $    159.3792~_{-      0.0125}^{+      0.0125}$ & $       2.943~_{-       0.170}^{+       0.203}$ & $       0.083~_{-       0.008}^{+       0.008}$ &      0.998& \multirow{3}{*}{0.23$\pm$0.02}\\[1pt]
   & 0& $    159.6479~_{-      0.0053}^{+      0.0056}$ & $       3.098~_{-       0.114}^{+       0.126}$ & $       0.075~_{-       0.004}^{+       0.004}$ &     0.986& \\[1pt]
   & +1& $    159.8589~_{-      0.0115}^{+      0.0109}$ & $       3.891~_{-       0.207}^{+       0.194}$ & $       0.188~_{-       0.022}^{+       0.016}$ &     1.000\\[3pt]
 \hline
 
  \multirow{2}{*}{1} & -1& $    161.3930~_{-      0.0007}^{+      0.0007}$ &[sinc] & $      119.15~_{-        9.10}^{+       14.91}$ &     ...&\multirow{3}{*}{0.237$\pm$0.002} \\[1pt]
   & 0& $    161.6303~_{-      0.0028}^{+      0.0029}$ & $       5.113~_{-       0.220}^{+       0.189}$ & $       0.069~_{-       0.005}^{+       0.004}$ &    ...\\[3pt]
 \hline
 
  \multirow{2}{*}{1} & -1& $    163.1311~_{-      0.0068}^{+      0.0072}$ & $       9.435~_{-       0.287}^{+       0.301}$ & $       0.169~_{-       0.008}^{+       0.009}$      &...&\multirow{3}{*}{0.15$\pm$0.02} \\[1pt]
 & +1& $    163.4220~_{-      0.0173}^{+      0.0156}$ & $       7.429~_{-       0.314}^{+       0.340}$ & $       0.242~_{-       0.028}^{+       0.031}$ &     ...\\[3pt]
 \hline
 
  \multirow{2}{*}{1} & -1& $    166.0994~_{-      0.0043}^{+      0.0054}$ & $       3.913~_{-       0.325}^{+       0.327}$ & $       0.070~_{-       0.004}^{+       0.004}$ &        1.000&\multirow{3}{*}{0.24$\pm$0.009} \\[1pt]
   & +1& $    166.5808~_{-      0.0089}^{+      0.0080}$ & $       3.383~_{-       0.129}^{+       0.142}$ & $       0.116~_{-       0.009}^{+       0.008}$ &        1.000\\[3pt]
 \hline
 
  \multirow{2}{*}{1} & 0& $    170.5549~_{-      0.0017}^{+      0.0012}$ & [sinc] & $      113.18~_{-        1.03}^{+        0.99}$ &     ...& \multirow{3}{*}{0.254$\pm$0.001}\\[1pt]
  & +1& $    170.8090~_{-      0.0005}^{+      0.0005}$ & [sinc] & $      258.74~_{-       20.37}^{+       33.16}$ &     ...\\[3pt]
 \hline
 
 \multirow{2}{*}{1} & 0& $    172.9106~_{-      0.0053}^{+      0.0048}$ & $       2.905~_{-       0.260}^{+       0.195}$ & $       0.064~_{-       0.003}^{+       0.003}$ &        0.985& \multirow{3}{*}{0.24$\pm$0.01}\\[1pt]
 & +1& $    173.1536~_{-      0.0092}^{+      0.0092}$ & $       3.015~_{-       0.119}^{+       0.100}$ & $       0.068~_{-       0.003}^{+       0.003}$ &    0.997\\[3pt]
 \hline
 
  \multirow{3}{*}{1} & -1& $    174.9999~_{-      0.0128}^{+      0.0105}$ & $       4.601~_{-       0.140}^{+       0.139}$ & $       0.139~_{-       0.007}^{+       0.008}$  &     ...& \multirow{3}{*}{0.29$\pm$0.01}\\[1pt]
  & 0& $    175.2731~_{-      0.0071}^{+      0.0046}$ & $       3.892~_{-       0.106}^{+       0.101}$ & $       0.086~_{-       0.009}^{+       0.006}$  &       1.000& \\[1pt]
  & +1& $    175.5968~_{-      0.0036}^{+      0.0032}$ & $       4.162~_{-       0.116}^{+       0.116}$ & $       0.052~_{-       0.003}^{+       0.002}$ &     ...\\[3pt]
\hline
\end{tabular}
\tablefoot{The first and second column list the spherical degree and azimuthal order $m$ of an identified dipole oscillation mode, respectively. For resolved dipole modes, described through a Lorentzian profile, the median centre frequency, the mode amplitude and its full width at half maximum (FWHM) are given. For unresolved peaks, described through a sinc function, the median frequency and the mode height is given. For the case of peaks with a low S/N-ratio, the detection probability $\mathcal{P}$ is reported, which ranges between 0 and 1, whereby 1 corresponds to a detection probability of 100\%. In case of rotationally split dipole modes the average separation of the $m$\,=\,$\pm$1 modes is given for the individual component frequencies. 
}
\label{tab:individualFrequenciesDipole}
\vspace{-3.5mm}
\end{table*}

\subsection{Core rotation and surface rotation}

To compute the rotational kernels for the modes listed in Tab.\,\ref{tab:individualFrequenciesDipole} we use a representative model with 1.4\,M\sun and 5.4\,R\sun and Guenther's non-adiabatic and non-radial pulsation code \citep{guenther1994}. The model was calculated with the Yale Stellar Evolution Code \citep[YREC;][]{demarque2008, guenther1992} for near-solar composition ($Z\,=\,0.02, Y\,=\,0.28$) assuming the solar mixture by \citet{grevesse1996} and a mixing length parameter $\alpha_\mathrm{MLT}\,=\,1.8$. It has a He-core mass fraction of about 0.14.

For a more detailed picture of the rotational properties, Eq.\,(\ref{eq:splittingKernel}) needs to be inverted. To solve this equation, several inversion techniques have been developed in the past aiming to determine the internal rotation profile of the Sun \citep[e.g][and references therein]{Howe2009}. The inversion of this integral is, however, a highly ill-conditioned problem that requires, for example, numerical regularisation. BHV14 found that classical approaches like the RLS method \citep[e.g.][]{Dalsgaard1990} or the SOLA technique \citep[e.g.][]{Schou1998} are not well-suited for red giants.  They easily become numerically unstable and it is often impossible to evaluate the reliability of the result.  Recently, in a comprehensive study of rotational inversion techniques \cite{diMauro2016} found that both methods, SOLA and OLA fail to fit the averaging kernels for fractional radii $r/R_\star$\,$<$\,0.01. These methods agree in general on the overall rotational gradient between the surface and the core. However, the methods fail to agree on the rotation law in the lower region of the convective envelope and are dependent on the chosen stellar model.

We therefore follow the forward modelling approach from \cite{Kallinger2017}, which has proven its ability to accurately reveal the rotational behaviour of a red giant (BHV14). The algorithm computes synthetic rotational splittings for a model of rigidly rotating shells and compares them to the observed splittings. To fit the individual rotation rates a Bayesian nested sampling algorithm is used. The advantage of this method is that it provides reliable parameters and their uncertainties as well as a comparison of different models (with, e.g. a different number of shells) to evaluate which model reproduces the observation best. As for KIC\,5006817 (BHV14) we find for \Asterix that a 2-zone model (core and envelope) gives the most reliable result and that the exact position of transition between core and envelope rotation cannot be determined with the available observations. We therefore fix the border between the two shells to a fractional mass of 0.14 (i.e. the He-core mass fraction and approximate position of the  H-burning shell that separates the contracting core from the expanding envelope.).

The core and envelope of \Asterix are found to rotate with an average rate of 545$\pm$9 and 79$\pm$14\,nHz, respectively. The measured rotation rates translate into a core-to-envelope rotational gradient of $6.9^{+2.0}_{-1.0}$. The core-rotation rate is a typical value on the RGB, compared to the large sample studies of \cite{Mosser2012c}. It is evident that the uncertainty of this result is dominated by the relative error ($\sim$18\%) of the envelope rotation rate. For \Asterix we have, however, an estimate for the surface rotation rate from the dominant period of the light curve modulation ($f_1$\,=\,87$\pm$8\,nHz). 
The core-surface rotation gradient is typical for RGB stars \citep[$\Omega_{\rm core}$/$\Omega_{\rm envelope}$\,$\simeq$\,6 to 30 times, ][]{Beck2012,Beck2014a,Deheuvels2012,Deheuvels2014,diMauro2016,Goupil2013,Triana2017} but appears a the lower end of the values found so far. Yet we note that this is small number statistics. 


\subsection{Rotation in the convective envelope \& angular momentum transfer}
Using surface rotation rate derived from \kepler photometry, we find a core-to-surface rotation rate of  $6.3^{+0.7}_{-0.6}$, which is in good agreement with the seismic value. 
The consistency between the surface and seismic envelope rotation rate (which represents the average rotation rate of the external 86\% and 99\% of the mass and radius, respectively, of a representative model of \Asterix) provides the observational evidence that the rotational gradient in the convective envelope of red giants is in fact very small or even zero. A similar result was found by \cite{diMauro2016} from asteroseismic inversions. 
\new{This is different from the findings of \cite{BrunPalacios2009} in the case of a lower-mass upper RGB star, where they showed from three-dimensional global non-linear simulations that, for very low rotation rates, shellular radial differential rotation is expected with a low latitudinal dependence. It is however in fair agreement with similar 3-D simulations of the star Pollux by \cite{PB2014}, which resembles more the primary of KIC9165796, and for which they obtain a 40\% radial contrast between core and envelope ratio.}
As a consequence the still unknown physical process that decelerates the core of red giants during stellar evolution is likely to act in the radiative zone or at the transition regions between the core and the outer convective envelope of subgiants and red giants. 


 \begin{figure}[t!]
\centering
\includegraphics[width=0.81\columnwidth]{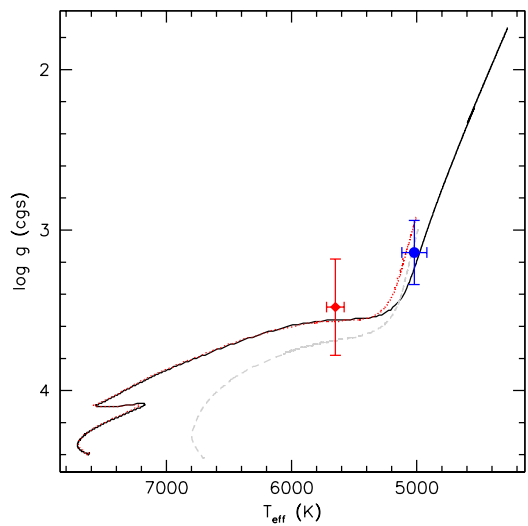}
\caption{Evolutionary track of representative models of the red-giant stars with [M/H]\,=\,-0.37 star in the \teff-\logg~plane. The black solid and red dotted lines represent the evolution of a 1.39\,M$_\odot$ star computed assuming a grey atmosphere or a fit to realistic {\sc PHOENIX} model atmospheres as an outer boundary condition to the stellar structure equations, respectively. The grey dashed line is also computed with {\sc PHOENIX} atmospheres and assuming a reduced mass of 1.18\,M$_\odot$.
The positions of the primary and the secondary components of \Asterix along this evolutionary track from spectroscopic parameters are marked in blue and red, respectively.}
\label{fig:evolutionaryTrack}
\end{figure}

\section{Lithium abundance \& rotational mixing \label{sec:lithium}}


The binary system of \Asterix provides a highly constrained set of stars, with all spectroscopic fundamental parameters well known, both stars located at the same distance from the observer and the same primordial metallicity. As shown by \cite{Lagarde2015}, that the initial conditions of the rotation rate of stars on the main sequence is a critical parameter and governs the Li abundance in the more advanced stages. Not knowing the initial conditions is complicating the analysis. By having two nearly equal mass stars, born and evolved under the same conditions, we can ignore these differences in the rotation history. 
{As discussed in \Section{sec:specAnalysis}, an A(Li) of 1.31$\pm$0.08 and 2.55$\pm$0.07\,dex as well as a difference of about 600\,K was found between the primary and the secondary component, respectively (also see \Table{tab:fundamentalParameters} and \Figure{fig:spectralDisentanglingLithium}).}
Although the mass of a star cannot be determined with better precision than a few percent, an accurate estimate of the \textit{ratio in mass} is provided from the ratio of radial velocity amplitudes. This allows us to study the history of the system, based on the differences in abundances of lithium. 

\begin{figure}[t!]
\centering
\includegraphics[width=0.81\columnwidth]{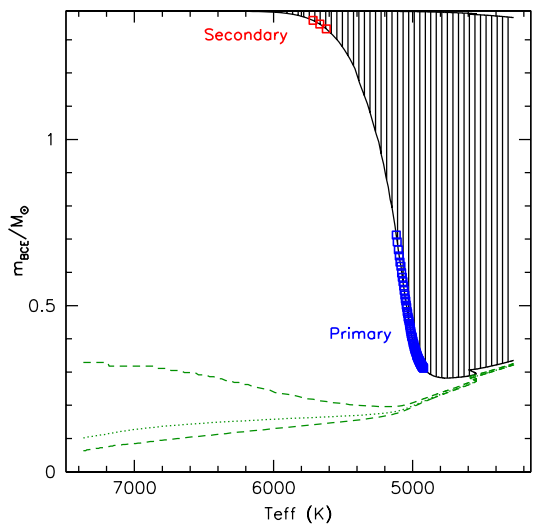}
\caption{Kippenhahn-like diagram in mass for the representative model of \Asterix. The shaded areas represent convective regions, and the green dashed lines mark out the H burning shell. The internal structure of the primary and secondary is indicated in this diagram using the same symbols as in Fig.\,\ref{fig:evolutionaryTrack}. }
\label{fig:kippenhahnDiagram}
\end{figure}

Representative models were calculated for stars in the mass range of 1.36\,M\sun to 1.41\,M\sun at  a metallicity of Z\,=\,0.005869 corresponding to [Fe/H]\,=\,-0.37\,dex when using the \cite{Asplund2009} solar abundances as a reference.  These models are computed with the latest version of the \textsc{starevol} code \citep[see][]{Amard2016}. 
The impact of rotation on the effective gravity is accounted for following the formalism by \cite{EndalSofia1978}, and the transport of \new{angular momentum and} chemicals is treated using the \cite{MaederZahn1998} and \cite{MathisZahn2004} formalisms with the \cite{Mathis2004} prescription for the horizontal viscosity, extensively described in \cite{Amard2016}. 
The mass loss is accounted for beyond the main sequence using the Reimer's formula with $\eta_R$\,=\,0.5. 
The nuclear reaction rates for the pp-chains and CNO cycle have been updated using the NACRE\,II compilation \citep{Xu2013}.

\begin{figure}[t!]
\centering
\includegraphics[width=0.81\columnwidth]{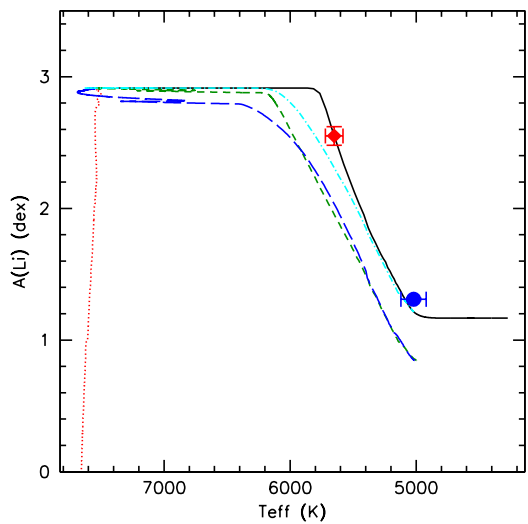}
\caption{Surface lithium abundance for the representative model of \Asterix, with different scenarios of rotation history as function of the stellar effective temperature. The primary and secondary are marked as dot and diamond respectively. {The solid (black) line represent the non-rotating model and the solid-body rotating ones, the dotted (red) line represents a differentially rotating model with very efficient turbulent mixing \citep[$\nu_h$ prescription from][]{Zahn1992}, the short-dashed (green) line represents a similar model using a different prescription for turbulent shear  \citep[$\nu_h$ prescription from][]{Mathis2004}, the long-dashed (blue) line represents a model similar to the previous one but slowly rotating on the ZAMS and the dashed-dotted (cyan) line represents a model computed assuming solid-body rotation on the main sequence and differential rotation beyond the TAMS. The two components of \Asterix are shown using the same symbols as in previous figures}.
}
\label{fig:LithiumModelling}
\end{figure}

The \new{non-rotating} evolutionary tracks of components in the \teff-\logg~plane as well as the  corresponding Kippenhahn diagram are depicted in \Figure{fig:evolutionaryTrack} and \ref{fig:kippenhahnDiagram}, respectively. \new{These are representative of all our models, including the rotating ones, because the slow rotation does not noticeably modify the stellar structure and evolution in our models.}
Because the mass used for the models come from scaling relations, we check if the position of the stars is compatible with the main seismic mass as well as with the mass corrected for the 15\% mass reduction as proposed by \cite{Gaulme2016}. Figure 11 shows three evolutionary tracks: the solid black line is our standard model, computed assuming a grey atmosphere as outer boundary condition to the stellar structure equations. The lines represent models with 1.39\,M$_\odot$ and a 1.18\,M$_\odot$, computing assuming a more realistic outer boundary condition. These two masses correspond to the original stellar mass and the value, reduced by the proposed 15\%. For these models, the thermal structure of the uppermost shells is fitted to a grid of detailed {\sc
 PHOENIX} model atmospheres \citep{Allard2001,Allard2012} in order to retrieve realistic \teff~and \logg~values.
The secondary is well fitted by both models of 1.39\,M$_\odot$ assuming either grey or realistic {\sc PHOENIX} atmosphere, while the agreement with the primary is less good. For the primary, the 1.18\,M$_\odot$ model with {\sc PHOENIX} atmosphere achieves a better fit but it shows a worse fit of the secondary.
Judged from the stellar mass and the models, the main-sequence progenitors of the red giants in \Asterix were late F-type stars and the best model fits  for an age of about 2.7\,Gyr. With a mass of $\sim$1.4\,M\sun, the progenitor stars were located on the blue, hotter edge of the so called Li-dip on the main sequence \citep{Wallerstein1965,Boesgaard1987,Talon1998}. In these stars, the microscopic settling of chemical elements is counter acted by macroscopic effects and therefore, the measured abundance is typically close to the  cosmic A(Li), known from the analysis from meteorites \citep{Talon1998,Castro2016}. 
\new{However, it was shown by \cite{Guiglion2016}
the initial Li abundance should be lower than the 3.3\,dex for the solar value
for lower stellar metallicity.} We therefore assume an initial A(Li) for the main sequence-progenitor stars of 2.7\,dex, which is a typical value for low-metallicity stars .

\new{Because tidal interaction alternates the angular momentum transport inside stars, \cite{Zahn1994} showed that synchronised binary systems have substantially differing lithium abundances. However, the system of \Asterix is wide, suggesting that tidal interaction has only started recently and, as described in the following section, is in fact weak. Therefore, we can treat \Asterix as a system where tidal induced transport of angular moment and chemical species plays no significant role in the previous rotation and activity history of the system and the alternation of the  A(Li)-evolution  can be ignored.}
 
As it can be seen from the Kippenhahn-diagram in \Figure{fig:kippenhahnDiagram}, the secondary and the primary component are in the early and late phase of the FDU phase on the RGB, respectively. Therefore, the components of \Asterix are framing the first-dredge-up event. Since lithium is consumed through thermonuclear reactions in the stellar interior at temperatures above $\sim$3 million Kelvin, its surface abundance is a strong tracer of mixing processes, rotation and loss of angular momentum inside a star. This makes it an interesting system in the context of stellar evolution as we can assume that the observed difference in surface Li abundance between the two stellar components, described in Section\,\ref{sec:specAnalysis} and \Table{tab:fundamentalParameters}, is solely governed  by the difference in mass. Therefore, also the difference in the stellar evolution relates directly to the difference in mass between the two components.

Motivated by the need to fit the Li surface abundance of the two stars in the system, we tested various descriptions of the rotational history. Starting with a typical initial lithium abundance expected for the metallicity of KIC 9163796, we look for the rotational history that could best explain the determined A(Li) in both components. The models discussed below are depicted in \Figure{fig:LithiumModelling}. We note that all scenarios lead to a basic agreement with the observed surface rotation velocity estimated from the spot modulation but vary drastically in the internal rotational gradient. 

The first, simplified approach of assuming \citep[e.g.][]{TalonZahn1997} solid-body rotation ensures no transport of chemicals by the vertical turbulent shear instability since the efficiency of this instability entirely depends on the angular velocity gradient. This approach produces a good agreement with the spectroscopic Li abundances (\Figure{fig:LithiumModelling}, black solid line), which is fully compatible with standard models first dredge-up dilution, thus indicating that lithium has been preserved from destruction via additional transport processes during the main-sequence evolution, such as anisotropic turbulent transport (Mathis et\,al.\ 2017 submitted), internal gravity waves \citep{Talon2005} or magnetic fields \citep{Strugarek2011}.  
The presence of rotationally split dipole modes found in the power spectrum of the primary component, analysed in Section\,\ref{sec:seismoRotation} however provides firm evidence that the star is now rotating non-rigidly, with the typical rotation gradient,  found in other field and binary stars (see discussion in \Section{sec:seismoRotation}). This may induce that the internal transport of angular momentum induced by tidal waves \citep[e.g.][]{GoldreichNicholson1989,TalonKumar1998,MathisRemus2013} is weak in comparison with other transport mechanisms. 
Therefore, several scenarios of non-rigid rotation were investigated.

Allowing the rotational gradient in the models to evolve freely, we  obtain a huge rotational gradient of a core-to-surface rotation rate ratio of 800, which is by far too steep when compared to the observational results, suggesting a factor of 10 for stars at the low luminosity edge of the RGB. Also, the evolution of the lithium abundance in \Figure{fig:LithiumModelling} (red dotted and green dashed line) strongly indicates that {fast} non-rigid rotation throughout all phase of \Asterix can be excluded. 
In models experiencing non-rigid rotation during their entire evolution, the surface lithium abundance is decreased during the main-sequence evolution due to the vertical turbulent shear instability, and it is not possible to reconcile the predicted abundances with the ones derived from high resolution spectra.
In an additional test scenario, the solid-body rotation hypothesis was only forced on the model during the main sequence (cyan dashed-dotted line). This is justified, following the indications given by helio- and asteroseismology that low-mass and intermediate stars could be solid-body rotators on the main sequence \citep[e.g.][]{Kurtz2014,Benomar2015,Murphy2016}. In this case, the rotational mixing that develops beyond the Terminating Age Main Sequence (TAMS) in the radiative interior leads to a decrease of the surface lithium abundance prior to the FDU, and the predicted A(Li) for the secondary is smaller than the observed one. On the other hand, the surface lithium abundance is fully controlled by the deepening of the convective envelope once the FDU starts and the model prediction is fully compatible with the A(Li) of the primary.

Therefore, the lithium abundance of the primary is best compatible with the scenarios of rigid rotation throughout the entire evolution. This finding is independent of the chosen mass for the two stars. In both cases (1.39 and 1.18\,M\sun), the Li abundance is consistent with predictions of standard stellar evolution or solid-body rotation in the radiative region.
Rigid rotation in this phase however is quite unusual for stars of this mass and evolutionary status, as also typical sub-giant progenitors of \Asterix appear to be rotating non-rigidly in the Hertzsprung gap \cite[e.g.][]{CantoMartins2011,Deheuvels2014}. 
However, the gradient between the core and surface rotation was found to be low.  It deviates by one to two orders of magnitude from the rotational gradient predicted by the differentially rotating models computed here \citep[][]{Palaciosetal2006,Marquesetal2013,Ceillieretal2013}. Such a small rotational gradient would be accompanied by weak shear and the expected effect on the surface lithium abundance would be negligible, so that such a small core to surface rotation gradient should be compatible with the predictions of the rigid case scenario depicted in \Figure{fig:LithiumModelling}.



\section{Stellar activity in context \label{sec:activity}}
A modulation of the stellar activity could be expected due to several intrinsic or causes, such as tidal interaction or to spots. The latter was found for \object{EK\,Eri} by \cite{Strassmeier1999}.

The signature of spots, shown in \Figure{fig:LCcurve} indicates that at least one component of the binary system is very active. Another well known indicator of stellar, chromospheric activity is the emission in the cores of the Ca\textsc{ii}\,H\&K lines in the near ultra violet (394 and 393nm, respectively). A visual inspection of \Hermes spectra of red giants in binaries, observed with \Kepler (BHV14) or giants in the Hyades, shows that \Asterix has by far the strongest emission in those lines among this sample. \Figure{fig:CaHKAsterixHyades} compares the emission at the core of the Ca\,K-line in the average spectrum \Asterix to the strength of this emission in two Hyades giants \ttau and $\epsilon$\,Tau, which exhibit typical red-giant activity. 

\subsection{Chromospheric activity}

The classical way of exploiting information on the chromospheric activity contained in these Ca lines is the Mount Wilson Observatory (MWO) \sindex \citep[][and references therein]{Duncan1991}. The multiplicative factor to scale from the instrumental \Sindex, measured in red-giant stars to meet the MWO-reference frame was determined to be 18.9 by \cite{Beck2017theta}, from comparison of the measured \Ssymbol for \ttau \new{(=\object{77\,Tau})} by \cite{Auriere2015} and \Hermes observations. When calculating the \Sindex of the individual frames, corrected for the individual radial velocity, a clear modulation with the orbital phase is found, with a peak during the periastron passage. This is an artefact of the formalism of the \Sindex due to the changing fractional light contribution in the centre of the 0.109\nm wide triangular windows, centred on the cores of the H and K line and the alternated number of lines in the normalising windows. Measuring \Ssymbol in an SB2-system will always lead to an underestimation as a function of the fractional light. 

\begin{figure}
\includegraphics[width=\columnwidth,height=50mm]{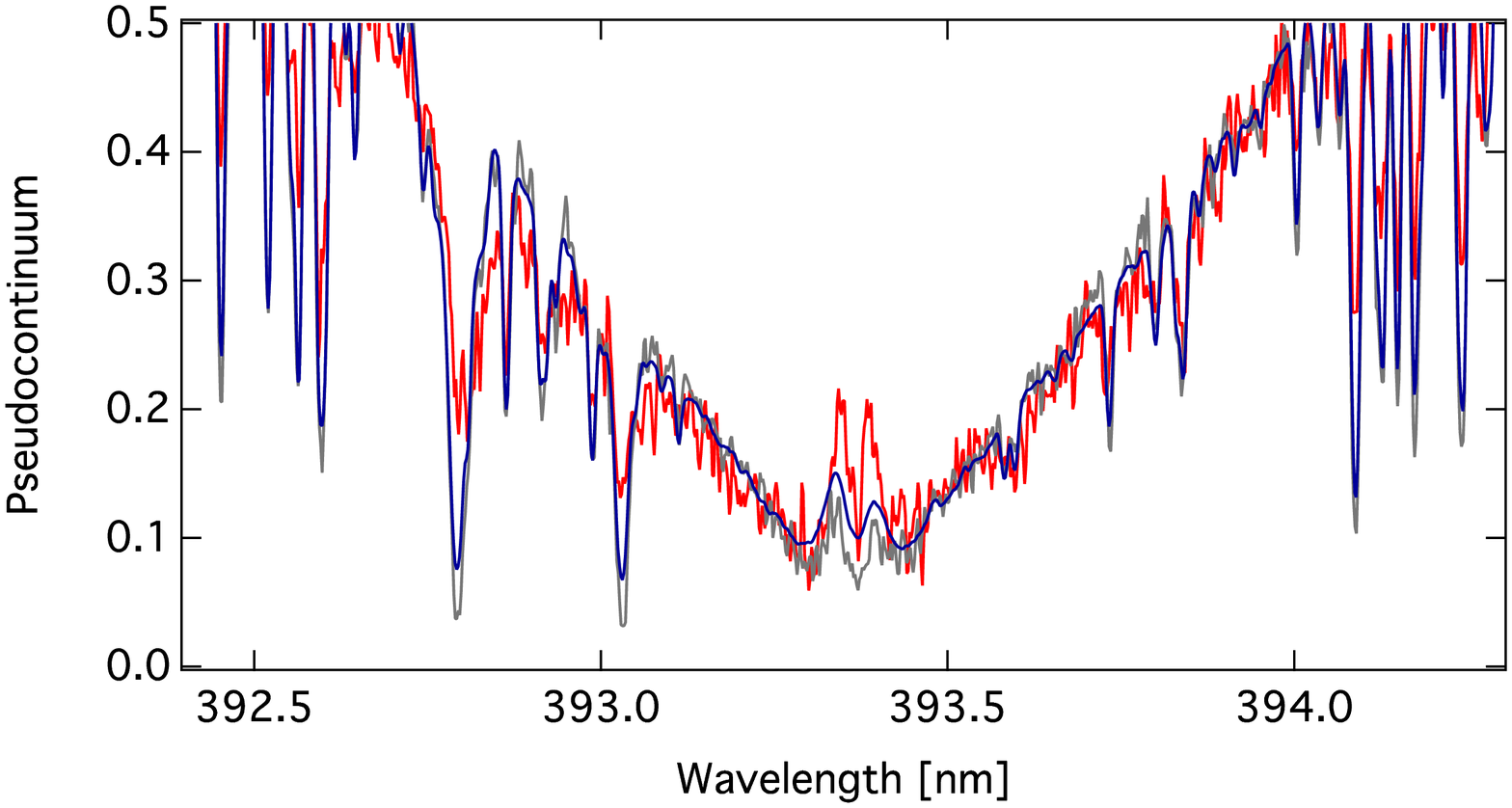}
\caption{\label{fig:CaHKAsterixHyades} The emission in the Ca\,K line in the median spectrum of the primary of \Asterix (red), compared to the emission of the Hyades giants \ttau (middle curve, blue) and $\epsilon$\,Tauri (bottom curve, grey). }
\end{figure}

To test if both stellar components of the system show signs of chromospheric activity, we inspected the four observations during periastron, when the difference in RV is larger than 50\kms. In those observations, the signature of emission is moving in phase with the spectrum of the primary component. We therefore argue that the most active component is the primary. From these four spectra, we measure an \Sindex of 0.219, which translates into an \Ssymbol\,$\simeq$\,0.36 through the correction for the differential light contribution. Such value is in good agreement with the \sindex of stars with detected Zeeman-splittings. \cite{Auriere2015} report that rather strong values (\Ssymbol$\gtrsim$\,0.2) were found for most magnetically active stars, while stars with non-detections are typically found below this value. 

\new{Comparing \Asterix with the sample of active stars in single red-giant stars and those in wide binaries in the \Ssymbol-P$_{\rm rot}$ plane \cite[Fig.\,9 in][]{Auriere2015}, we find that this system is above the fit they present to their sample but note, that there is a large scatter. One of the stars closest to \Asterix in terms of rotational period is the secondary clump primary in the wide binary system of \ttau (P$_{\rm rot}$$\simeq$140\,days, also \Figure{fig:CaHKAsterixHyades}) which varies between 0.16$\lesssim$\Ssymbol$\lesssim$0.2. over the campaign of \cite{Auriere2015}. Beck et al. (in prep) confirms similar variations. Given the large scatter in the \Sindex of active stars, which can also be modulated by long term activity cycles \citep{Saar1999} we find that \Asterix is an active red giant.} 
\new{Furthermore, we can compare \Asterix also via the Rossby number, \Ro, describing the ratio between the observed rotation period and the maximum convective turnover timescale within the convective envelope.
Based on the models, presented in \Section{sec:lithium}, we calculated the \Ro for both component, following the semi-emperical definition used by \cite{Auriere2015}. Because the rotation rate of the secondary component cannot be gauged either from the light curve nor from the unresolved rotational broadening in spectroscopy, we used the surface rotation of  the rigidly rotating model to calculate \Ro for both components. The rotation period of this model sufficiently reproduces the surface rotation rate found for the primary from the \Kepler light curve (Tab.\ref{tab:surfaceModulationFrequencies} \& Fig.\ref{fig:LCcurve}).
For the primary and secondary we obtain $\mathcal{R}_{\rm O,1}$=2.63 and $\mathcal{R}_{\rm O,2}$=0.74, respectively.}

The majority of red-giant stars with detected magnetic signatures are \new{found} either at the beginning of core-He burning or in the first dredge-up phase \citep{Auriere2015}. A theoretical interpretation for this behaviour has been proposed by \cite{Charbonnel2017} in the frame work of convective dynamos. \new{\Asterix is located in this strip of magnetic activity.} 
\new{Therefore, we can use} the relation \new{between the measured \Sindex and the large scale unsigned longitudinal magnetic field $|B_l|_{\rm max}$,} \new{depicted} in Figure\,10 \new{of} \cite{Auriere2015}, 0.3\,$<$\,\Ssymbol$<\,$0.4 would indicate a magnetic field strength 5$\lesssim$$|B_l|_{\rm max}$$\lesssim$8\,G.
\new{Using the Rossby number instead of the \Sindex leads to similar results. For the primary and secondary we find $\sim$4\,G and $\sim$7G for $\lesssim$$|B_l|_{\rm max}$, respectively. Given the large scatter and unknown systematics in the measurements of the large scale unsigned longitudinal magnetic field, the two methods are in good agreement.}

\subsection{Effect of activity on the oscillation amplitude \label{sec:oscAmplitude}}
\begin{figure}[t!]
\centering
\includegraphics[width=\columnwidth
]{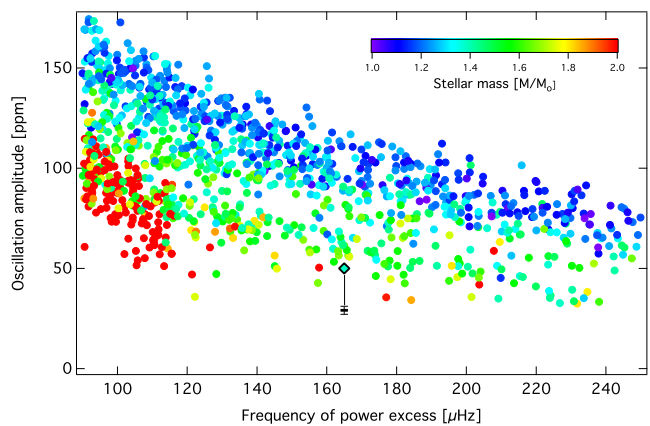}
\caption{Total oscillation amplitudes $\mathcal{A}_{\rm puls}$ for the extended  sample of red giants \citep{Kallinger2014}, with their seismic mass colour-coded. The bulk of red dots on the left hand side mark secondary-clump stars. The measured and intrinsic amplitude of the primary component of \Asterix are shown as the marker with error bars and the black framed diamond symbol, respectively.} 
\label{fig:comparingAmplitudes}
\end{figure}

To test if the interaction of the two stellar components in \Asterix has an overall influence  on the oscillation amplitude, we compare in \Figure{fig:comparingAmplitudes} the amplitude of the primary to those of a large sample of red giants observed with \Kepler \citep{Kallinger2014}. We find that the corrected amplitude $\mathcal{A}_{\rm puls,1}$ is slightly lower (by about 20\%) than what is expected for a 1.4\,M\sun star, but it is not an extreme outlier. 


Furthermore, we tested wether the overall oscillation amplitude is modulated by the orbital phase.  From the full light curve, which was folded into a phase diagram, we described the flux modulation at periastron with a high-order Legendre-polynomial fit. When using the polynomial fit, to remove the flux modulation through normalisation, the residual flux does not show any significant amplitude modulation a as function of the orbital phase. 
We therefore conclude that the tidal interaction has no immediate effects on the mode parameters, but will act throughout all phases of the orbit. 


\subsection{Effect of activity on the mode linewidth}
The oscillation amplitude depends on numerous factors and is subject to the constant interaction between the driving and damping mechanisms \citep[e.g.][]{BelkacemSamadi2013}. 
From the analysis of activity in eclipsing binary stars, \cite{Gaulme2014} suggested that such effects lead to additional damping of solar-like oscillations. From the precise measurements of the mode parameters, we can now  test the influence of activity on the mode parameters

Unlike the mode amplitude, the full width at half maximum of a mode, $\Gamma$, is independent of photometric dilution. Therefore, the measured values of $\Gamma$ in \Asterix can be directly compared to the typical values of the mode width in larger samples of stars. In this analysis, we only compare the width of radial modes, as $\ell$\,=\,2 might be affected by rotation and the presence of quadrupole mixed modes, which would lead to an overestimate. \cite{Corsaro2015a} have performed a detailed analysis of the individual mode parameters in 19 red-giant stars, observed by \Kepler for more than four years. For the effective temperature of the primary component of about 5000\,K (see \Table{tab:fundamentalParameters}), they report an average value for $\Gamma_{\ell\,=\,0}$$\simeq$0.1\muHz. In \Asterix, we measure much broader peaks. From the six radial modes that we measure (see \Table{tab:individualFrequenciesPressure}), we find an average of $\Gamma$\,=\,0.4$\pm$0.2\muHz. Such a broad mode width can only be explained due to strong damping, as expected for active stars as it is the case for \Asterix\ from spots (\Figure{fig:LCcurve}) and chromospheric activity (\Figure{fig:CaHKAsterixHyades}).



\section{Tidal interaction \label{sec:modelingTides}}
In eccentric close binary systems such as KIC 9163796, tidal interactions modify the orbits of the stars and their rotation. 
The pace of the evolution towards an equilibrium state where the orbits are circularised and the rotational and orbital spins are aligned and synchronised is determined by the distance between the two stellar components, their mass and radius, and the strength of the dissipative processes applied on tidal flows in their interiors \citep[e.g.][]{Zahn1989,MathisRemus2013}. In red giant stars, tidal flows are constituted by the large-scale equilibrium tide \citep[][]{Zahn1966,Remus2012}, generated by the ellipsoidal hydrostatic adjustment of the star because of the presence of the companion, and the dynamical tide, meaning tidal inertial waves propagating in their large convective envelope when $P_{\rm orb}>1/2P_{\rm star}$ and tidal gravity waves propagating in their stably stratified radiative core \citep[][]{Zahn1975,OgilvieLin2007}. We recall here that inertial waves have the Coriolis acceleration as a restoring force. For stars with large convective envelopes like red giant stars, turbulent friction applied by convection on tidal flows is the main dissipation process that converts their kinetic energy into heat \citep{Zahn1966,OgilvieLin2007,ADMLP2015}. In stellar radiation zones, radiative damping acting on tidal gravity waves is the dominant mechanism \citep[][]{Zahn1975,ADMLP2015}.
Having good representative stellar models of the primary and secondary (see \Figure{fig:kippenhahnDiagram}) {allows us to} discuss the dissipation of tidal energy in the system studied here to understand its orbital state and explore which tidal wave can potentially be detected.


\subsection{Equilibrium tide and circularisation state}

The observed modulation of the brightness of heartbeat stars is caused by the distorted stellar shape associated to the equilibrium tide \citep[for a more extensive description of the observational aspects of tides in red-giant binary systems we refer the reader to][]{Beck2017Circularization}. Moreover, \cite{Gallet2017b} demonstrated that the dissipation of tidal inertial waves in the deep convective envelope of red giant stars is weak.
\new{Indeed, \cite{Beck2017TideLetter} have shown from an ensemble study of red-giant binary systems, observed with \Kepler, that the contribution of the time dependent component of tidal interaction to the full budget of dissipation is negligible for all systems.}
 Therefore, we focus here on the dissipation of the equilibrium tide by the turbulent friction in the deep convective envelope to understand the observed non-circularised state of the system.

In this framework, \cite{Verbunt1995} developed a formalism to predict the expected circularisation state in red giant binary stars based on the theory derived in \cite{Zahn1966,Zahn1989}. They obtained the rate of circularisation of a system for evolving red giant stars as the KIC 9163796 system 
\begin{equation}
\frac{\Delta \ln e}{f} = -1.7\cdot10^{-5}\,\left(\frac{M}{M_\odot}\right)^{-11/3}\,q^{-1}\left(1+q^{-1}\right)^{-5/3}\,I(t)\,\left(\frac{P_{\rm orb}}{\rm day}\right)^{-16/3}.~~
\end{equation} 
In this expression $e$ is the orbital eccentricity, $M$ the mass of the primary, $q$ the mass ratio where $m$ is the mass of the secondary, and $P_{\rm orb}$ the orbital period. The circularisation strength $I\left(t\right)$, where $t$ is the time, is a monotonic function of the stellar radius and $f$ is a dimensionless parameter of order unity that quantifies the details of the convective tidal turbulent friction applied on the equilibrium tide. We see from the Kippenhahn diagram reported in \Figure{fig:kippenhahnDiagram}, that only the primary is nearly fully convective. The dissipation of the equilibrium tide depending on the thickness of the convective envelope, we therefore assume that the primary is likely to be the main seat of tidal dissipation in the KIC 9163796 system.

Following this formalism, we have computed the expected reduction of the orbital eccentricity, based on the stellar parameters of the primary star deduced from seismology. From the analysis of their sample, \cite{Verbunt1995} concluded that binary systems with a value of $\frac{\Delta \ln e}{f}$$\gtrsim$3 should be quasi-circularised. For \Asterix, we find $\frac{\Delta \ln e}{f}$$\simeq$$10^{-3.3}$. Therefore, the pace of the circularisation is currently slow with a low level of dissipation of tidal kinetic energy. Therefore, the system is not expected to be circularised yet as observed with the eccentric orbit. This finding is in good agreement with systems in the sample of \cite{Verbunt1995} with similar eccentricities. A slightly stronger tidal dissipation is found in KIC\,5006817 with $\frac{\Delta \ln e}{f}$$\simeq$$10^{-2.9}$. This system has nearly the same eccentricity and the orbital period is only 30 days shorter than \Asterix while its primary component have similar mass and radius. The main difference of KIC\,5006817 with the system discussed in this work is the mass ratio, which was determined by \hbox{BHV14 to be 1/5}. 

From this comparison, it was found that both systems are thus far from circularisation. The huge spread of eccentricity in the sample of \cite{Verbunt1995} indicates that, at such low values of $\frac{\Delta \ln e}{f}\lesssim10^{-2}$, circularisation and synchronisation are determined by the previous history of the systems. The low values of $\frac{\Delta \ln e}{f}$ indicate that the evolution of the systems studied here is currently dominated by the time scales of stellar evolution, as the stars ascend the RGB, until the separation between the two components is getting small enough to allow stronger tidal forces to act. 

\subsection{Search for the dynamical  tide}
In numerous eccentric binary systems with main-sequence components, low-frequency oscillations, that are not excited through opacity variations \citep[i.e. by the $\kappa$-mechanism, e.g.][]{kippenhahn2013} but by tidal interactions were found \citep[e.g.][]{Welsh2011,Thompson2012,Hambleton2013}. Their frequencies are in resonance with the orbital period and the modes correspond to the dynamical tide introduced above. In red giant stars, the two families of candidates are tidal gravity waves propagating in the radiative core and tidal inertial waves propagating in the convective envelope. As in the case of main-sequence solar-type stars, tidal gravity waves would be difficult to detect because of their screening by the convective envelope \citep[except if they are mixed gravito-acoustic modes,][]{Appourchauxetal2010}. Therefore, in the case of \Asterix, we focus on tidal inertial waves that can be excited when \hbox{$P_{\rm orb}$$\,>\,$1/2\,$\cdot$\,$ P_{\rm star}$} and propagate in the convective envelopes of the components.

\begin{figure}[t!]
\centering
\includegraphics[width=\columnwidth, height=45mm]{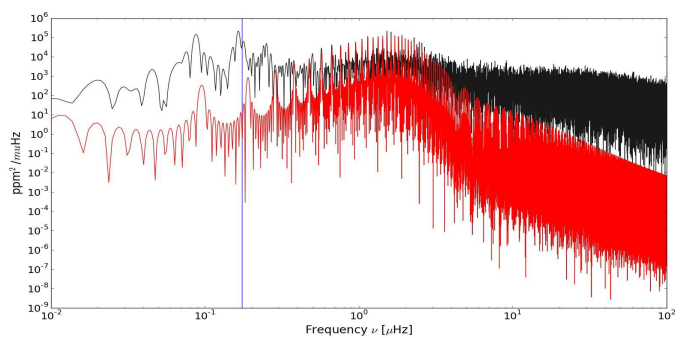}
\caption{Comparison of PSD from synthetic light curve without oscillation and the spot-free light curve in red and black, respectively. The vertical line indicates the frequency corresponding to twice the rotational frequency, $\Omega_{\rm star}$.} 
\label{fig:fourierOfHBLC}
\end{figure}

To guide the search for tidal inertial modes in the low-frequency regime of the power spectrum of \Asterix, we computed the PSD of a synthetic light curve (\Figure{fig:fourierOfHBLC} , also see \Section{sec:oscAmplitude}). Analysing a detrended light curve is problematic as imperfections of the detrending routine lead to side peaks that mimic these frequencies.  Therefore, the high-order Legendre polynomial fit of the flux modulation, described in the preceding subsection, was used to construct a synthetic light curve of the pure ellipsoidal modulation. The vertical line in \Figure{fig:fourierOfHBLC} indicates the frequency corresponding to twice the stellar rotation frequency $\Omega_{\rm star}$ 
Only signal below this value could originate from tidal inertial modes. 
The two highest peaks in the real  spectrum (black spectrum in \Figure{fig:fourierOfHBLC}) with frequencies below this frequency limit  correspond to $f_1$ and $f_2$ of the spot modulation. 

The non-detection of tidally induced modes does not mean that they do not exist in systems like \Asterix. To reach a detectable signal, these modes would however need to produce large photometric signals in order to compete with the high-level of the granulation background (see \Figure{fig:fourierOfHBLC}), typical for solar-like oscillators in advanced stellar evolution. Because RV obtained through high-resolution spectroscopy are less affected by the convective signal and the heartbeat event, a high-quality time series of radial velocity measurements might be better suited for the search of the signature of the dynamical tide, i.e. tidal inertial and gravity waves, in both components in the case of well separated SB2 systems.

\section{Summary \& conclusions \label{sec:Conclusions}}

In this work we have combined combined space photometry, ground-based high-resolution spectroscopy, and a seismic analysis, as well as theoretical modelling to study the eccentric binary system, \Asterix. 
From the solution of the spectral disentangling of both components, we find that the mass ratio between both components is 1.015$\pm$0.005. Therefore, we know the mass ratio between the two components by far better than the stellar mass of the primary alone which $\sim$1.4\,M\sun from asteroseismology. The temperature difference between the two components of $\sim$600K places the secondary and the primary into the early and late phase of the first dredge-up event at the bottom of the RGB.

The main seismic analysis focused on the power excess originating from the primary component.  By comparing the oscillation spectrum to the spectrum of a comparison star with similar values of \num and \dnu, an unambiguous mode identification could be achieved. Furthermore, we identify the power excess of the secondary. This detection support this by reconstructing the expected $\nu_{\rm max}^{\rm secondary}$ by using only spectroscopic inputs from the spectral disentangling solution.

\new{In fair agreement with three -dimensional numerical MHD simulations of a slow rotating low-mass star at the base of the RGB by \cite[][and Palacios private communication]{PB2014},} we show that there is no clear signature of differential surface rotation in the light curve. The main variation of the light curve could originate from the varying activity of two large active regions.  
From the analysis of the rotationally split modes in the oscillation spectrum of the primary, the rotational gradient between the core and the surface could be determined to be $6.9^{+2.0}_{-1.0}$. The value of the seismically determined rotation gradient in the radial direction is compatible within the error bars with the value of the rotational gradient estimated by using the surface rotation rate from photometry ($6.3^{+0.7}_{-0.6}$). This strongly suggests that the rotational gradient inside the convective envelope is  flat or even rigid. 
Such flat rotational gradient in the convective envelope of a red-giant branch star shows the validity of the usually chosen two-zone model for straight-forward testings. It also confirms the findings of \cite{diMauro2016}, who were not able to detect a variation of the rotational gradient inside the convective envelope but accounted this to the very limited radial resolution of the rotational kernels in the convective envelope. 
The core-to-surface rotation gradient is compatible to single field stars, which could indicate that the angular momentum transport through tides is negligible in such systems.
As a consequence the still unknown physical process that decelerates the core during stellar evolution is likely to act in the transition region between the core and radiative inner envelope of subgiants and red giants, such as internal gravity waves \citep{Talon2008}.
A flat rotational gradient also allows a simplified treatment of the propagation of the dynamical tide in stars, such as inertial waves.  \cite{Guenel2016} has shown that differential rotation is imposing strong selection effects on the possible propagation paths. In this context, although we see power in the PSD which could originate from the inertial waves propagating in the convective envelope, we cannot rule out that this is an effect of the segmentation of the \Kepler data.

Such a well-constrained binary system allows us to study the interactions between the two stellar components. Besides the surface rotation, we measure in the optical spectrum a high level of chromospheric activity. Also the line width of the oscillation modes is indicating a much stronger damping than found in other red giants in the same range of effective temperature. This could be related to magnetic fields induced through the strong interaction at periastron. However, no immediate effect on the measured overall oscillation amplitude was found as a function of the orbital phase. 
Additionally, we studied the tidal interactions in the system. We confirm the result of \cite{Verbunt1995}, that the dissipation of tidal energy is low in systems with low-luminosity red-giant components. 
No low-frequent modes, induced by the dynamical tide were found. Judged by the dimension of the convective envelope and the stratification of the stellar structure, the amplitude of these modes is expected to be minor, compared to the dominant granulation noise of the primary. 
Therefore, the observed flux modulation are the sole effect of the equilibrium tide.

From the disentangled spectra of both stellar components, we find that the abundance of the fragile element lithium is differing by 1.2\,dex between the components, while the overall metallicity is the same for both components. From comparison of the predicted lithium value from rotating and non-rotating models, we find that the measured lithium abundance in both components is compatible with a  rigid rotating along stellar evolution. It can be argued that this scenario is justified by the low level of the rotational gradient, found from the asteroseismic analysis.

\Asterix thus provides us with a unique and impressive example on how strong the effect of a difference in mass of about 1-2\% can be on stellar evolution and consequently in lithium abundance. This system in particular allows us to study a binary system, whose components are in the early and late stages of the first dredge-up event $-$ one of the most interesting evolutionary phases in red-giant star evolution. Binary systems like \Asterix provide a robust and well tested benchmark for testing seismic scaling relations with observations as well as confronting models of chemical mixing and angular momentum transport. In last consequence, binary systems like \Asterix will allow us to improve the determination of stellar ages.
 \\

\begin{acknowledgements}
We acknowledge the work of the team behind \textit{Kepler}. Funding for the \textit{Kepler} Mission is provided by NASA's Science Mission Directorate. 
We thank the technical team as well as the observers of the \Hermes spectrograph and Mercator Telescope, operated on the island of La Palma by the Flemish Community, at the Spanish Observatorio del Roque de los Muchachos of the Instituto de Astrof{\'i}sica de Canarias. 
PGB acknowledges the support of the Spanish Ministry of Economy and Competitiveness (MINECO) under the programme 'Juan de la Cierva \new{Incorporacion}' (IJCI-2015-26034). PGB, RAG and BM acknowledge the ANR (Agence Nationale de la Recherche, France) programme IDEE (n$^\circ$ANR-12-BS05-0008) 'Interaction Des Etoiles et des Exoplanetes'.  PGB also received funding from the CNES grants at CEA. KP was supported by the Croatian Science Foundation grant 2014-09-8656. E.C. is funded by the European Community's Seventh Framework Programme (FP7/2007-2013) under grant agreement N$^\circ$312844 (SPACEINN).
The research leading to these results has received funding from the European Community's Seventh Framework Programme ([FP7/2007-2013]) under grant agreement No. 312844 (SPACEINN) and under grant agreement No. 269194 (IRSES/ASK). SM acknowledges support support from the European Research Council  (ERC) through the grant 647383 (SPIRE). \new{AT acknowledges the support of the Fonds Wetenschappelijk Onderzoek - Vlaanderen (FWO) under the grant agreement G0H5416N (ERC Opvangproject).  The research leading to these results has partially received funding from the European Research Council (ERC) under the European Union's Horizon 2020 research and innovation programme (grant agreement N$^\circ$670519: MAMSIE) and from the Belgian Science Policy Office 
(Belspo) under ESA/PRODEX grant ``PLATO mission development''.}
\end{acknowledgements}


\bibliographystyle{aa}
\bibliography{bibliographyAuO.bib}

\begin{thebibliography}{165}
\expandafter\ifx\csname natexlab\endcsname\relax\def\natexlab#1{#1}\fi

\bibitem[{{Allard} {et~al.}(2001){Allard}, {Hauschildt}, {Alexander},
  {Tamanai}, \& {Schweitzer}}]{Allard2001}
{Allard}, F., {Hauschildt}, P.~H., {Alexander}, D.~R., {Tamanai}, A., \&
  {Schweitzer}, A. 2001, \apj, 556, 357

\bibitem[{{Allard} {et~al.}(2012){Allard}, {Homeier}, \&
  {Freytag}}]{Allard2012}
{Allard}, F., {Homeier}, D., \& {Freytag}, B. 2012, Philosophical Transactions
  of the Royal Society of London Series A, 370, 2765

\bibitem[{{Amard} {et~al.}(2016){Amard}, {Palacios}, {Charbonnel}, {Gallet}, \&
  {Bouvier}}]{Amard2016}
{Amard}, L., {Palacios}, A., {Charbonnel}, C., {Gallet}, F., \& {Bouvier}, J.
  2016, \aap, 587, A105

\bibitem[{{Appourchaux} {et~al.}(2015){Appourchaux}, {Antia}, {Ball},
  {Creevey}, {Lebreton}, {Verma}, {Vorontsov}, {Campante}, {Davies}, {Gaulme},
  {R{\'e}gulo}, {Horch}, {Howell}, {Everett}, {Ciardi}, {Fossati}, {Miglio},
  {Montalb{\'a}n}, {Chaplin}, {Garc{\'{\i}}a}, \& {Gizon}}]{Appourchaux2015}
{Appourchaux}, T., {Antia}, H.~M., {Ball}, W., {et~al.} 2015, A\&A, 582, A25

\bibitem[{{Appourchaux} {et~al.}(2010){Appourchaux}, {Belkacem}, {Broomhall},
  {Chaplin}, {Gough}, {Houdek}, {Provost}, {Baudin}, {Boumier}, {Elsworth},
  {Garc{\'{\i}}a}, {Andersen}, {Finsterle}, {Fr{\"o}hlich}, {Gabriel}, {Grec},
  {Jim{\'e}nez}, {Kosovichev}, {Sekii}, {Toutain}, \&
  {Turck-Chi{\`e}ze}}]{Appourchauxetal2010}
{Appourchaux}, T., {Belkacem}, K., {Broomhall}, A.-M., {et~al.} 2010, \aapr,
  18, 197

\bibitem[{{Arentoft} {et~al.}(2008){Arentoft}, {Kjeldsen}, {Bedding}, {Bazot},
  {Christensen-Dalsgaard}, {Dall}, {Karoff}, {Carrier}, {Eggenberger},
  {Sosnowska}, {Wittenmyer}, {Endl}, {Metcalfe}, {Hekker}, {Reffert}, {Butler},
  {Bruntt}, {Kiss}, {O'Toole}, {Kambe}, {Ando}, {Izumiura}, {Sato}, {Hartmann},
  {Hatzes}, {Bouchy}, {Mosser}, {Appourchaux}, {Barban}, {Berthomieu},
  {Garcia}, {Michel}, {Provost}, {Turck-Chi{\`e}ze}, {Marti{\'c}}, {Lebrun},
  {Schmitt}, {Bertaux}, {Bonanno}, {Benatti}, {Claudi}, {Cosentino}, {Leccia},
  {Frandsen}, {Brogaard}, {Glowienka}, {Grundahl}, \&
  {Stempels}}]{Arentoft2008}
{Arentoft}, T., {Kjeldsen}, H., {Bedding}, T.~R., {et~al.} 2008, ApJ, 687, 1180

\bibitem[{{Asplund} {et~al.}(2009){Asplund}, {Grevesse}, {Sauval}, \&
  {Scott}}]{Asplund2009}
{Asplund}, M., {Grevesse}, N., {Sauval}, A.~J., \& {Scott}, P. 2009, \araa, 47,
  481

\bibitem[{{Auclair Desrotour} {et~al.}(2015){Auclair Desrotour}, {Mathis}, \&
  {Le Poncin-Lafitte}}]{ADMLP2015}
{Auclair Desrotour}, P., {Mathis}, S., \& {Le Poncin-Lafitte}, C. 2015, \aap,
  581, A118

\bibitem[{{Auri{\`e}re} {et~al.}(2015){Auri{\`e}re}, {Konstantinova-Antova},
  {Charbonnel}, {Wade}, {Tsvetkova}, {Petit}, {Dintrans}, {Drake}, {Decressin},
  {Lagarde}, {Donati}, {Roudier}, {Ligni{\`e}res}, {Schr{\"o}der},
  {Landstreet}, {L{\`e}bre}, {Weiss}, \& {Zahn}}]{Auriere2015}
{Auri{\`e}re}, M., {Konstantinova-Antova}, R., {Charbonnel}, C., {et~al.} 2015,
  A\&A, 574, A90

\bibitem[{{Ballot} {et~al.}(2006){Ballot}, {Garc{\'{\i}}a}, \&
  {Lambert}}]{Ballot2006}
{Ballot}, J., {Garc{\'{\i}}a}, R.~A., \& {Lambert}, P. 2006, MNRAS, 369, 1281

\bibitem[{{Beck} {et~al.}(2017{\natexlab{a}}){Beck}, {Mathis}, {Gallet},
  {Charbonnel}, {Garcia}, \& {Benbakoura}}]{Beck2017TideLetter}
{Beck}, P.\, G., {Mathis}, S., {Gallet}, F., {et~al.} 2017{\natexlab{a}}, MNRAS
  (submitted)

\bibitem[{{Beck} {et~al.}(2017{\natexlab{b}}){Beck}, {Mathis}, {Kallinger},
  {Garcia}, {Benbakoura}, \& {Benbakoura}}]{Beck2017Circularization}
{Beck}, P.\, G., {Mathis}, S., {Kallinger}, T., {et~al.} 2017{\natexlab{b}},
  proceedings of the J.P. Zahn Symposium 2016 (accepted)

\bibitem[{{Beck} {et~al.}(2017{\natexlab{c}}){Beck}, {Pavlovski}, {Tkachenko},
  {Johnston}, {Vos}, {Torres}, \& {Garc{\'i}a}}]{Beck2017theta}
{Beck}, P.\, G., {Pavlovski}, K., {Tkachenko}, A., {et~al.} 2017{\natexlab{c}},
  A\&A (in prep)

\bibitem[{{Beck}(2013)}]{Beck2013PhD}
{Beck}, P.~G. 2013, PhD thesis, University of Leuven, Belgium.

\bibitem[{{Beck} {et~al.}(2017{\natexlab{d}}){Beck}, {do Nascimento}, {Duarte},
  {Salabert}, {Tkachenko}, {Mathis}, {Mathur}, {Garc{\'{\i}}a}, {Castro},
  {Pall{\'e}}, {Egeland}, {Montes}, {Creevey}, {Andersen}, {Kamath}, \& {van
  Winckel}}]{Beck2017Li}
{Beck}, P.~G., {do Nascimento}, Jr., J.-D., {Duarte}, T., {et~al.}
  2017{\natexlab{d}}, \aap, 602, A63

\bibitem[{{Beck} {et~al.}(2014){Beck}, {Hambleton}, {Vos}, {Kallinger},
  {Bloemen}, {Tkachenko}, {Garc{\'{\i}}a}, {{\O}stensen}, {Aerts}, {Kurtz}, {De
  Ridder}, {Hekker}, {Pavlovski}, {Mathur}, {De Smedt}, {Derekas}, {Corsaro},
  {Mosser}, {Van Winckel}, {Huber}, {Degroote}, {Davies}, {Pr{\v s}a},
  {Debosscher}, {Elsworth}, {Nemeth}, {Siess}, {Schmid}, {P{\'a}pics}, {de
  Vries}, {van Marle}, {Marcos-Arenal}, \& {Lobel}}]{Beck2014a}
{Beck}, P.~G., {Hambleton}, K., {Vos}, J., {et~al.} 2014, A\&A, 564, A36

\bibitem[{{Beck} {et~al.}(2015{\natexlab{a}}){Beck}, {Hambleton}, {Vos},
  {Kallinger}, {Garcia}, {Mathur}, \& {Houmani}}]{Beck2015Toulouse}
{Beck}, P.~G., {Hambleton}, K., {Vos}, J., {et~al.} 2015{\natexlab{a}}, in
  European Physical Journal Web of Conferences, Vol. 101, EPJWoC, 06004

\bibitem[{{Beck} {et~al.}(2017{\natexlab{e}}){Beck}, {Kallinger}, {Pavlovski},
  {Palacios}, {Tkachenko}, {Garc{\'{\i}}a}, {Mathis}, {Corsaro}, {Johnston},
  {Mosser}, {Ceillier}, {do Nascimento}, \&
  {Raskin}}]{Beck2016AzoresProceedings}
{Beck}, P.~G., {Kallinger}, T., {Pavlovski}, K., {et~al.} 2017{\natexlab{e}},
  ArXiv: 1611.01402

\bibitem[{{Beck} {et~al.}(2015{\natexlab{b}}){Beck}, {Kambe}, {Hillen},
  {Corsaro}, {Van Winckel}, {Moravveji}, {De Ridder}, {Bloemen}, {Saesen},
  {Mathias}, {Degroote}, {Kallinger}, {Verhoelst}, {Ando}, {Carrier}, {Acke},
  {Oreiro}, {Miglio}, {Eggenberger}, {Sato}, {Zwintz}, {P{\'a}pics},
  {Marcos-Arenal}, {Sans Fuentes}, {Schmid}, {Waelkens}, {{\O}stensen},
  {Matthews}, {Yoshida}, {Izumiura}, {Koyano}, {Nagayama}, {Shimizu}, {Okada},
  {Okita}, {Sakamoto}, {Yamamuro}, \& {Aerts}}]{Beck2015a}
{Beck}, P.~G., {Kambe}, E., {Hillen}, M., {et~al.} 2015{\natexlab{b}}, A\&A,
  573, A138

\bibitem[{{Beck} {et~al.}(2012){Beck}, {Montalban}, {Kallinger}, {De Ridder},
  {Aerts}, {Garc{\'{\i}}a}, {Hekker}, {Dupret}, {Mosser}, {Eggenberger},
  {Stello}, {Elsworth}, {Frandsen}, {Carrier}, {Hillen}, {Gruberbauer},
  {Christensen-Dalsgaard}, {Miglio}, {Valentini}, {Bedding}, {Kjeldsen},
  {Girouard}, {Hall}, \& {Ibrahim}}]{Beck2012}
{Beck}, P.~G., {Montalban}, J., {Kallinger}, T., {et~al.} 2012, Nature, 481, 55

\bibitem[{{Bedding} \& {Kjeldsen}(2010)}]{Bedding2010CoAst}
{Bedding}, T.~R. \& {Kjeldsen}, H. 2010, Comm. in Asteroseismology, 161, 3

\bibitem[{{Bedding} {et~al.}(2011){Bedding}, {Mosser}, {Huber},
  {Montalb{\'a}n}, {Beck}, {Christensen-Dalsgaard}, {Elsworth},
  {Garc{\'{\i}}a}, {Miglio}, {Stello}, {White}, {De Ridder}, {Hekker}, {Aerts},
  {Barban}, {Belkacem}, {Broomhall}, {Brown}, {Buzasi}, {Carrier}, {Chaplin},
  {di Mauro}, {Dupret}, {Frandsen}, {Gilliland}, {Goupil}, {Jenkins},
  {Kallinger}, {Kawaler}, {Kjeldsen}, {Mathur}, {Noels}, {Aguirre}, \&
  {Ventura}}]{Bedding2011}
{Bedding}, T.~R., {Mosser}, B., {Huber}, D., {et~al.} 2011, Nature, 471, 608

\bibitem[{{Belkacem} \& {Samadi}(2013)}]{BelkacemSamadi2013}
{Belkacem}, K. \& {Samadi}, R. 2013, in Lecture Notes in Physics, Berlin
  Springer Verlag, Vol. 865, Lecture Notes in Physics, Berlin Springer Verlag,
  ed. M.~{Goupil}, K.~{Belkacem}, C.~{Neiner}, F.~{Ligni{\`e}res}, \& J.~J.
  {Green}, 179

\bibitem[{{Benomar} {et~al.}(2015){Benomar}, {Takata}, {Shibahashi},
  {Ceillier}, \& {Garc{\'{\i}}a}}]{Benomar2015}
{Benomar}, O., {Takata}, M., {Shibahashi}, H., {Ceillier}, T., \&
  {Garc{\'{\i}}a}, R.~A. 2015, \mnras, 452, 2654

\bibitem[{{Bloemen}(2013)}]{Bloemen2013}
{Bloemen}, S. 2013, PhD thesis, University of Leuven, Belgium.

\bibitem[{{Boesgaard}(1987)}]{Boesgaard1987}
{Boesgaard}, A.~M. 1987, \pasp, 99, 1067

\bibitem[{{Bonanno} {et~al.}(2014){Bonanno}, {Corsaro}, \&
  {Karoff}}]{Bonanno2014}
{Bonanno}, A., {Corsaro}, E., \& {Karoff}, C. 2014, A\&A, 571, A35

\bibitem[{{Borucki} {et~al.}(2010){Borucki}, {Koch}, {Basri}, {Batalha},
  {Brown}, {Caldwell}, {Caldwell}, {Christensen-Dalsgaard}, {Cochran},
  {DeVore}, {Dunham}, {Dupree}, {Gautier}, {Geary}, {Gilliland}, {Gould},
  {Howell}, {Jenkins}, {Kondo}, {Latham}, {Marcy}, {Meibom}, {Kjeldsen},
  {Lissauer}, {Monet}, {Morrison}, {Sasselov}, {Tarter}, {Boss}, {Brownlee},
  {Owen}, {Buzasi}, {Charbonneau}, {Doyle}, {Fortney}, {Ford}, {Holman},
  {Seager}, {Steffen}, {Welsh}, {Rowe}, {Anderson}, {Buchhave}, {Ciardi},
  {Walkowicz}, {Sherry}, {Horch}, {Isaacson}, {Everett}, {Fischer}, {Torres},
  {Johnson}, {Endl}, {MacQueen}, {Bryson}, {Dotson}, {Haas}, {Kolodziejczak},
  {Van Cleve}, {Chandrasekaran}, {Twicken}, {Quintana}, {Clarke}, {Allen},
  {Li}, {Wu}, {Tenenbaum}, {Verner}, {Bruhweiler}, {Barnes}, \&
  {Prsa}}]{Borucki2010}
{Borucki}, W.~J., {Koch}, D., {Basri}, G., {et~al.} 2010, Science, 327, 977

\bibitem[{{Breger} {et~al.}(2006){Breger}, {Beck}, {Lenz}, {Schmitzberger},
  {Guggenberger}, \& {Shobbrook}}]{Breger2006}
{Breger}, M., {Beck}, P., {Lenz}, P., {et~al.} 2006, A\&A, 455, 673

\bibitem[{{Breger} {et~al.}(1993){Breger}, {Stich}, {Garrido}, {Martin},
  {Jiang}, {Li}, {Hube}, {Ostermann}, {Paparo}, \& {Scheck}}]{Breger1993}
{Breger}, M., {Stich}, J., {Garrido}, R., {et~al.} 1993, A\&A, 271, 482

\bibitem[{{Brown} {et~al.}(1989){Brown}, {Sneden}, {Lambert}, \&
  {Dutchover}}]{Brown1989}
{Brown}, J.~A., {Sneden}, C., {Lambert}, D.~L., \& {Dutchover}, Jr., E. 1989,
  \apjs, 71, 293

\bibitem[{{Brun} \& {Palacios}(2009)}]{BrunPalacios2009}
{Brun}, A.~S. \& {Palacios}, A. 2009, \apj, 702, 1078

\bibitem[{{Buysschaert} {et~al.}(2016){Buysschaert}, {Beck}, {Corsaro},
  {Christensen-Dalsgaard}, {Aerts}, {Arentoft}, {Kjeldsen}, {Garc{\'{\i}}a},
  {Silva Aguirre}, \& {Degroote}}]{Buysschaert2016}
{Buysschaert}, B., {Beck}, P.~G., {Corsaro}, E., {et~al.} 2016, \aap, 588, A82

\bibitem[{{Canto Martins} {et~al.}(2011){Canto Martins}, {L{\`e}bre},
  {Palacios}, {de Laverny}, {Richard}, {Melo}, {Do Nascimento}, \& {de
  Medeiros}}]{CantoMartins2011}
{Canto Martins}, B.~L., {L{\`e}bre}, A., {Palacios}, A., {et~al.} 2011, A\&A,
  527, A94

\bibitem[{{Casagrande} {et~al.}(2016){Casagrande}, {Silva Aguirre},
  {Schlesinger}, {Stello}, {Huber}, {Serenelli}, {Sch{\"o}nrich}, {Cassisi},
  {Pietrinferni}, {Hodgkin}, {Milone}, {Feltzing}, \&
  {Asplund}}]{Casagrande2016}
{Casagrande}, L., {Silva Aguirre}, V., {Schlesinger}, K.~J., {et~al.} 2016,
  \mnras, 455, 987

\bibitem[{{Castro} {et~al.}(2016){Castro}, {Duarte}, {Pace}, \& {do
  Nascimento}}]{Castro2016}
{Castro}, M., {Duarte}, T., {Pace}, G., \& {do Nascimento}, J.-D. 2016, \aap,
  590, A94

\bibitem[{{Ceillier} {et~al.}(2013){Ceillier}, {Eggenberger}, {Garc{\'{\i}}a},
  \& {Mathis}}]{Ceillieretal2013}
{Ceillier}, T., {Eggenberger}, P., {Garc{\'{\i}}a}, R.~A., \& {Mathis}, S.
  2013, \aap, 555, A54

\bibitem[{{Ceillier} {et~al.}(2017){Ceillier}, {Tayar}, {Mathur}, {Salabert},
  {Garc{\'{\i}}a}, {Stello}, {Pinsonneault}, {van Saders}, {Beck}, \&
  {Bloemen}}]{Ceillier2017}
{Ceillier}, T., {Tayar}, J., {Mathur}, S., {et~al.} 2017, \aap, 605, A111

\bibitem[{{Chaplin} {et~al.}(2014){Chaplin}, {Elsworth}, {Davies}, {Campante},
  {Handberg}, {Miglio}, \& {Basu}}]{Chaplin2014SupNyq}
{Chaplin}, W.~J., {Elsworth}, Y., {Davies}, G.~R., {et~al.} 2014, \mnras, 445,
  946

\bibitem[{{Chaplin} {et~al.}(2011){Chaplin}, {Kjeldsen},
  {Christensen-Dalsgaard}, {Basu}, {Miglio}, {Appourchaux}, {Bedding},
  {Elsworth}, {Garc{\'{\i}}a}, {Gilliland}, {Girardi}, {Houdek}, {Karoff},
  {Kawaler}, {Metcalfe}, {Molenda-{\.Z}akowicz}, {Monteiro}, {Thompson},
  {Verner}, {Ballot}, {Bonanno}, {Brand{\~a}o}, {Broomhall}, {Bruntt},
  {Campante}, {Corsaro}, {Creevey}, {Do{\u g}an}, {Esch}, {Gai}, {Gaulme},
  {Hale}, {Handberg}, {Hekker}, {Huber}, {Jim{\'e}nez}, {Mathur}, {Mazumdar},
  {Mosser}, {New}, {Pinsonneault}, {Pricopi}, {Quirion}, {R{\'e}gulo},
  {Salabert}, {Serenelli}, {Silva Aguirre}, {Sousa}, {Stello}, {Stevens},
  {Suran}, {Uytterhoeven}, {White}, {Borucki}, {Brown}, {Jenkins}, {Kinemuchi},
  {Van Cleve}, \& {Klaus}}]{Chaplin2011}
{Chaplin}, W.~J., {Kjeldsen}, H., {Christensen-Dalsgaard}, J., {et~al.} 2011,
  Science, 332, 213

\bibitem[{{Charbonnel} \& {Balachandran}(2000)}]{Charbonnel2000}
{Charbonnel}, C. \& {Balachandran}, S.~C. 2000, \aap, 359, 563

\bibitem[{{Charbonnel} {et~al.}(2017){Charbonnel}, {Decressin}, {Lagarde},
  {Gallet}, {Palacios}, {Auri{\`e}re}, {Konstantinova-Antova}, {Mathis},
  {Anderson}, \& {Dintrans}}]{Charbonnel2017}
{Charbonnel}, C., {Decressin}, T., {Lagarde}, N., {et~al.} 2017, \aap, 605,
  A102

\bibitem[{{Charbonnel} {et~al.}(1994){Charbonnel}, {Vauclair}, {Maeder},
  {Meynet}, \& {Schaller}}]{Charbonnel1994}
{Charbonnel}, C., {Vauclair}, S., {Maeder}, A., {Meynet}, G., \& {Schaller}, G.
  1994, \aap, 283, 155

\bibitem[{{Christensen-Dalsgaard} {et~al.}(1990){Christensen-Dalsgaard},
  {Schou}, \& {Thompson}}]{Dalsgaard1990}
{Christensen-Dalsgaard}, J., {Schou}, J., \& {Thompson}, M.~J. 1990, \mnras,
  242, 353

\bibitem[{{Corsaro} \& {De Ridder}(2014)}]{Corsaro2014}
{Corsaro}, E. \& {De Ridder}, J. 2014, A\&A, 571, A71

\bibitem[{{Corsaro} {et~al.}(2015){Corsaro}, {De Ridder}, \&
  {Garc{\'{\i}}a}}]{Corsaro2015a}
{Corsaro}, E., {De Ridder}, J., \& {Garc{\'{\i}}a}, R.~A. 2015, A\&A, 579, A83

\bibitem[{{Cox}(1980)}]{Cox1980}
{Cox}, J.~P. 1980, {Theory of stellar pulsation}

\bibitem[{{Dall} {et~al.}(2010){Dall}, {Bruntt}, {Stello}, \&
  {Strassmeier}}]{Dall2010}
{Dall}, T.~H., {Bruntt}, H., {Stello}, D., \& {Strassmeier}, K.~G. 2010, A\&A,
  514, A25

\bibitem[{{Davies} {et~al.}(2015){Davies}, {Chaplin}, {Farr}, {Garc{\'{\i}}a},
  {Lund}, {Mathis}, {Metcalfe}, {Appourchaux}, {Basu}, {Benomar}, {Campante},
  {Ceillier}, {Elsworth}, {Handberg}, {Salabert}, \& {Stello}}]{Davies2015}
{Davies}, G.~R., {Chaplin}, W.~J., {Farr}, W.~M., {et~al.} 2015, \mnras, 446,
  2959

\bibitem[{{Deheuvels} {et~al.}(2015){Deheuvels}, {Ballot}, {Beck}, {Mosser},
  {{\O}stensen}, {Garc{\'{\i}}a}, \& {Goupil}}]{Deheuvels2015}
{Deheuvels}, S., {Ballot}, J., {Beck}, P.~G., {et~al.} 2015, A\&A, 580, A96

\bibitem[{{Deheuvels} {et~al.}(2014){Deheuvels}, {Do{\u g}an}, {Goupil},
  {Appourchaux}, {Benomar}, {Bruntt}, {Campante}, {Casagrande}, {Ceillier},
  {Davies}, {De Cat}, {Fu}, {Garc{\'{\i}}a}, {Lobel}, {Mosser}, {Reese},
  {Regulo}, {Schou}, {Stahn}, {Thygesen}, {Yang}, {Chaplin},
  {Christensen-Dalsgaard}, {Eggenberger}, {Gizon}, {Mathis},
  {Molenda-{\.Z}akowicz}, \& {Pinsonneault}}]{Deheuvels2014}
{Deheuvels}, S., {Do{\u g}an}, G., {Goupil}, M.~J., {et~al.} 2014, A\&A, 564,
  A27

\bibitem[{{Deheuvels} {et~al.}(2012){Deheuvels}, {Garc{\'{\i}}a}, {Chaplin},
  {Basu}, {Antia}, {Appourchaux}, {Benomar}, {Davies}, {Elsworth}, {Gizon},
  {Goupil}, {Reese}, {Regulo}, {Schou}, {Stahn}, {Casagrande},
  {Christensen-Dalsgaard}, {Fischer}, {Hekker}, {Kjeldsen}, {Mathur}, {Mosser},
  {Pinsonneault}, {Valenti}, {Christiansen}, {Kinemuchi}, \&
  {Mullally}}]{Deheuvels2012}
{Deheuvels}, S., {Garc{\'{\i}}a}, R.~A., {Chaplin}, W.~J., {et~al.} 2012, ApJ,
  756, 19

\bibitem[{{Demarque} {et~al.}(2008){Demarque}, {Guenther}, {Li}, {Mazumdar}, \&
  {Straka}}]{demarque2008}
{Demarque}, P., {Guenther}, D.~B., {Li}, L.~H., {Mazumdar}, A., \& {Straka},
  C.~W. 2008, ApSS, 316, 31

\bibitem[{{Di Mauro} {et~al.}(2016){Di Mauro}, {Ventura}, {Cardini}, {Stello},
  {Christensen-Dalsgaard}, {Dziembowski}, {Patern{\`o}}, {Beck}, {Bloemen},
  {Davies}, {De Smedt}, {Elsworth}, {Garc{\'{\i}}a}, {Hekker}, {Mosser}, \&
  {Tkachenko}}]{diMauro2016}
{Di Mauro}, M.~P., {Ventura}, R., {Cardini}, D., {et~al.} 2016, ApJ, 817, 65

\bibitem[{{Duncan} {et~al.}(1991){Duncan}, {Vaughan}, {Wilson}, {Preston},
  {Frazer}, {Lanning}, {Misch}, {Mueller}, {Soyumer}, {Woodard}, {Baliunas},
  {Noyes}, {Hartmann}, {Porter}, {Zwaan}, {Middelkoop}, {Rutten}, \&
  {Mihalas}}]{Duncan1991}
{Duncan}, D.~K., {Vaughan}, A.~H., {Wilson}, O.~C., {et~al.} 1991, ApJS, 76,
  383

\bibitem[{{Endal} \& {Sofia}(1978)}]{EndalSofia1978}
{Endal}, A.~S. \& {Sofia}, S. 1978, ApJ, 220, 279

\bibitem[{{Epstein} {et~al.}(2014){Epstein}, {Elsworth}, {Johnson}, {Shetrone},
  {Mosser}, {Hekker}, {Tayar}, {Harding}, {Pinsonneault}, {Silva Aguirre},
  {Basu}, {Beers}, {Bizyaev}, {Bedding}, {Chaplin}, {Frinchaboy},
  {Garc{\'{\i}}a}, {Garc{\'{\i}}a P{\'e}rez}, {Hearty}, {Huber}, {Ivans},
  {Majewski}, {Mathur}, {Nidever}, {Serenelli}, {Schiavon}, {Schneider},
  {Sch{\"o}nrich}, {Sobeck}, {Stassun}, {Stello}, \& {Zasowski}}]{Epstein2014}
{Epstein}, C.~R., {Elsworth}, Y.~P., {Johnson}, J.~A., {et~al.} 2014, \apjl,
  785, L28

\bibitem[{{Frandsen} {et~al.}(2013){Frandsen}, {Lehmann}, {Hekker},
  {Southworth}, {Debosscher}, {Beck}, {Hartmann}, {Pigulski}, {Kopacki},
  {Ko{\l}aczkowski}, {St{\c e}{\'s}licki}, {Thygesen}, {Brogaard}, \&
  {Elsworth}}]{Frandsen2013}
{Frandsen}, S., {Lehmann}, H., {Hekker}, S., {et~al.} 2013, A\&A, 556, A138

\bibitem[{{Gallet} {et~al.}(2017){Gallet}, {Bolmont}, {Mathis}, {Charbonnel},
  {Amard}, {Amard}, \& {Amard}}]{Gallet2017b}
{Gallet}, F., {Bolmont}, E., {Mathis}, S., {et~al.} 2017, ArXiv: 1705.10164

\bibitem[{{Garc{\'{\i}}a} {et~al.}(2014{\natexlab{a}}){Garc{\'{\i}}a},
  {Ceillier}, {Salabert}, {Mathur}, {van Saders}, {Pinsonneault}, {Ballot},
  {Beck}, {Bloemen}, {Campante}, {Davies}, {do Nascimento}, {Mathis},
  {Metcalfe}, {Nielsen}, {Su{\'a}rez}, {Chaplin}, {Jim{\'e}nez}, \&
  {Karoff}}]{Garcia2014b}
{Garc{\'{\i}}a}, R.~A., {Ceillier}, T., {Salabert}, D., {et~al.}
  2014{\natexlab{a}}, A\&A, 572, A34

\bibitem[{{Garc{\'{\i}}a} {et~al.}(2011){Garc{\'{\i}}a}, {Hekker}, {Stello},
  {Guti{\'e}rrez-Soto}, {Handberg}, {Huber}, {Karoff}, {Uytterhoeven},
  {Appourchaux}, {Chaplin}, {Elsworth}, {Mathur}, {Ballot},
  {Christensen-Dalsgaard}, {Gilliland}, {Houdek}, {Jenkins}, {Kjeldsen},
  {McCauliff}, {Metcalfe}, {Middour}, {Molenda-Zakowicz}, {Monteiro}, {Smith},
  \& {Thompson}}]{Garcia2011}
{Garc{\'{\i}}a}, R.~A., {Hekker}, S., {Stello}, D., {et~al.} 2011, MNRAS, 414,
  L6

\bibitem[{{Garc{\'{\i}}a} {et~al.}(2014{\natexlab{b}}){Garc{\'{\i}}a},
  {Mathur}, {Pires}, {R{\'e}gulo}, {Bellamy}, {Pall{\'e}}, {Ballot},
  {Barcel{\'o} Forteza}, {Beck}, {Bedding}, {Ceillier}, {Roca Cort{\'e}s},
  {Salabert}, \& {Stello}}]{Garcia2014}
{Garc{\'{\i}}a}, R.~A., {Mathur}, S., {Pires}, S., {et~al.} 2014{\natexlab{b}},
  A\&A, 568, A10

\bibitem[{{Garc{\'{\i}}a} {et~al.}(2010){Garc{\'{\i}}a}, {Mathur}, {Salabert},
  {Ballot}, {R{\'e}gulo}, {Metcalfe}, \& {Baglin}}]{Garcia2010}
{Garc{\'{\i}}a}, R.~A., {Mathur}, S., {Salabert}, D., {et~al.} 2010, Science,
  329, 1032

\bibitem[{{Gaulme} {et~al.}(2014){Gaulme}, {Jackiewicz}, {Appourchaux}, \&
  {Mosser}}]{Gaulme2014}
{Gaulme}, P., {Jackiewicz}, J., {Appourchaux}, T., \& {Mosser}, B. 2014, 785, 5

\bibitem[{{Gaulme} {et~al.}(2016){Gaulme}, {McKeever}, {Jackiewicz}, {Rawls},
  {Corsaro}, {Mosser}, {Southworth}, {Mahadevan}, {Bender}, \&
  {Deshpande}}]{Gaulme2016}
{Gaulme}, P., {McKeever}, J., {Jackiewicz}, J., {et~al.} 2016, \apj, 832, 121

\bibitem[{{Gaulme} {et~al.}(2013){Gaulme}, {McKeever}, {Rawls}, {Jackiewicz},
  {Mosser}, \& {Guzik}}]{Gaulme2013}
{Gaulme}, P., {McKeever}, J., {Rawls}, M.~L., {et~al.} 2013, ApJ, 767, 82

\bibitem[{{Gizon} \& {Solanki}(2003)}]{Gizon2003}
{Gizon}, L. \& {Solanki}, S.~K. 2003, ApJ, 589, 1009

\bibitem[{{Goldreich} \& {Nicholson}(1989)}]{GoldreichNicholson1989}
{Goldreich}, P. \& {Nicholson}, P.~D. 1989, \apj, 342, 1079

\bibitem[{{Goupil} {et~al.}(2013){Goupil}, {Mosser}, {Marques}, {Ouazzani},
  {Belkacem}, {Lebreton}, \& {Samadi}}]{Goupil2013}
{Goupil}, M.~J., {Mosser}, B., {Marques}, J.~P., {et~al.} 2013, A\&A, 549, A75

\bibitem[{{Gray}(2005)}]{Gray2005}
{Gray}, D.~F. 2005, {The Observation and Analysis of Stellar Photospheres}

\bibitem[{{Grevesse} {et~al.}(1996){Grevesse}, {Noels}, \&
  {Sauval}}]{grevesse1996}
{Grevesse}, N., {Noels}, A., \& {Sauval}, A.~J. 1996, in Astronomical Society
  of the Pacific Conference Series, Vol.~99, Cosmic Abundances, ed. {S.~S.~Holt
  \& G.~Sonneborn}, 117

\bibitem[{{Grundahl} {et~al.}(2008){Grundahl}, {Clausen}, {Hardis}, \&
  {Frandsen}}]{Grundahl2008}
{Grundahl}, F., {Clausen}, J.~V., {Hardis}, S., \& {Frandsen}, S. 2008, \aap,
  492, 171

\bibitem[{{Grundahl} {et~al.}(2017){Grundahl}, {Fredslund Andersen},
  {Christensen-Dalsgaard}, {Antoci}, {Kjeldsen}, {Handberg}, {Houdek},
  {Bedding}, {Pall{\'e}}, {Jessen-Hansen}, {Silva Aguirre}, {White},
  {Frandsen}, {Albrecht}, {Andersen}, {Arentoft}, {Brogaard}, {Chaplin},
  {Harps{\o}e}, {J{\o}rgensen}, {Karovicova}, {Karoff}, {Kj{\ae}rgaard
  Rasmussen}, {Lund}, {Sloth Lundkvist}, {Skottfelt}, {Norup S{\o}rensen},
  {Tronsgaard}, \& {Weiss}}]{Grundahl2017}
{Grundahl}, F., {Fredslund Andersen}, M., {Christensen-Dalsgaard}, J., {et~al.}
  2017, \apj, 836, 142

\bibitem[{{Guenel} {et~al.}(2016){Guenel}, {Baruteau}, {Mathis}, \&
  {Rieutord}}]{Guenel2016}
{Guenel}, M., {Baruteau}, C., {Mathis}, S., \& {Rieutord}, M. 2016, \aap, 589,
  A22

\bibitem[{{Guenther}(1994)}]{guenther1994}
{Guenther}, D.~B. 1994, ApJ, 422, 400

\bibitem[{{Guenther} {et~al.}(1992){Guenther}, {Demarque}, {Kim}, \&
  {Pinsonneault}}]{guenther1992}
{Guenther}, D.~B., {Demarque}, P., {Kim}, Y.-C., \& {Pinsonneault}, M.~H. 1992,
  ApJ, 387, 372

\bibitem[{{Guiglion} {et~al.}(2016){Guiglion}, {de Laverny}, {Recio-Blanco},
  {Worley}, {De Pascale}, {Masseron}, {Prantzos}, \&
  {Mikolaitis}}]{Guiglion2016}
{Guiglion}, G., {de Laverny}, P., {Recio-Blanco}, A., {et~al.} 2016, \aap, 595,
  A18

\bibitem[{{Hadrava}(1995)}]{Hadrava1995}
{Hadrava}, P. 1995, AApS, 114, 393

\bibitem[{{Hambleton} {et~al.}(2013){Hambleton}, {Kurtz}, {Pr{\v s}a}, {Guzik},
  {Pavlovski}, {Bloemen}, {Southworth}, {Conroy}, {Littlefair}, \&
  {Fuller}}]{Hambleton2013}
{Hambleton}, K.~M., {Kurtz}, D.~W., {Pr{\v s}a}, A., {et~al.} 2013, \mnras,
  434, 925

\bibitem[{{Hekker} {et~al.}(2010){Hekker}, {Debosscher}, {Huber}, {Hidas}, {De
  Ridder}, {Aerts}, {Stello}, {Bedding}, {Gilliland}, {Christensen-Dalsgaard},
  {Brown}, {Kjeldsen}, {Borucki}, {Koch}, {Jenkins}, {Van Winckel}, {Beck},
  {Blomme}, {Southworth}, {Pigulski}, {Chaplin}, {Elsworth}, {Stevens},
  {Dreizler}, {Kurtz}, {Maceroni}, {Cardini}, {Derekas}, \&
  {Suran}}]{Hekker2010}
{Hekker}, S., {Debosscher}, J., {Huber}, D., {et~al.} 2010, ApJ, 713, L187

\bibitem[{{Hekker} \& {Mel{\'e}ndez}(2007)}]{Hekker2007}
{Hekker}, S. \& {Mel{\'e}ndez}, J. 2007, A\&A, 475, 1003

\bibitem[{{Holtzman} {et~al.}(2015){Holtzman}, {Shetrone}, {Johnson}, {Allende
  Prieto}, {Anders}, {Andrews}, {Beers}, {Bizyaev}, {Blanton}, {Bovy},
  {Carrera}, {Chojnowski}, {Cunha}, {Eisenstein}, {Feuillet}, {Frinchaboy},
  {Galbraith-Frew}, {Garc{\'{\i}}a P{\'e}rez}, {Garc{\'{\i}}a-Hern{\'a}ndez},
  {Hasselquist}, {Hayden}, {Hearty}, {Ivans}, {Majewski}, {Martell},
  {Meszaros}, {Muna}, {Nidever}, {Nguyen}, {O'Connell}, {Pan}, {Pinsonneault},
  {Robin}, {Schiavon}, {Shane}, {Sobeck}, {Smith}, {Troup}, {Weinberg},
  {Wilson}, {Wood-Vasey}, {Zamora}, \& {Zasowski}}]{Holtzman2015}
{Holtzman}, J.~A., {Shetrone}, M., {Johnson}, J.~A., {et~al.} 2015, \aj, 150,
  148

\bibitem[{Howe(2009)}]{Howe2009}
Howe, R. 2009, Living Reviews in Solar Physics, 6

\bibitem[{{Ilijic} {et~al.}(2004){Ilijic}, {Hensberge}, {Pavlovski}, \&
  {Freyhammer}}]{Ilijic2004}
{Ilijic}, S., {Hensberge}, H., {Pavlovski}, K., \& {Freyhammer}, L.~M. 2004, in
  ASPCS, Vol. 318, Spectroscopically and Spatially Resolving the Components of
  the Close Binary Stars, ed. R.~W. {Hilditch}, H.~{Hensberge}, \&
  K.~{Pavlovski}, 111--113

\bibitem[{{Jetsu} {et~al.}(2017){Jetsu}, {Henry}, \& {Lehtinen}}]{Jetsu2017}
{Jetsu}, L., {Henry}, G.~W., \& {Lehtinen}, J. 2017, \apj, 838, 122

\bibitem[{{Jofr{\'e}} {et~al.}(2015){Jofr{\'e}}, {Petrucci}, {Garc{\'{\i}}a},
  \& {G{\'o}mez}}]{Jofre2015}
{Jofr{\'e}}, E., {Petrucci}, R., {Garc{\'{\i}}a}, L., \& {G{\'o}mez}, M. 2015,
  \aap, 584, L3

\bibitem[{{Kallinger} {et~al.}(2014){Kallinger}, {De Ridder}, {Hekker},
  {Mathur}, {Mosser}, {Gruberbauer}, {Garc{\'{\i}}a}, {Karoff}, \&
  {Ballot}}]{Kallinger2014}
{Kallinger}, T., {De Ridder}, J., {Hekker}, S., {et~al.} 2014, A\&A, 570, A41

\bibitem[{{Kallinger} {et~al.}(2010){Kallinger}, {Gruberbauer}, {Guenther},
  {Fossati}, \& {Weiss}}]{Kallinger2010b}
{Kallinger}, T., {Gruberbauer}, M., {Guenther}, D.~B., {Fossati}, L., \&
  {Weiss}, W.~W. 2010, A\&A, 510, A106

\bibitem[{{Kallinger} {et~al.}(2012){Kallinger}, {Hekker}, {Mosser}, {De
  Ridder}, {Bedding}, {Elsworth}, {Gruberbauer}, {Guenther}, {Stello}, {Basu},
  {Garc{\'{\i}}a}, {Chaplin}, {Mullally}, {Still}, \&
  {Thompson}}]{Kallinger2012}
{Kallinger}, T., {Hekker}, S., {Mosser}, B., {et~al.} 2012, A\&A, 541, A51

\bibitem[{{Kallinger} {et~al.}(2017){Kallinger}, {Weiss}, {Beck}, {Pigulski},
  {Kuschnig}, {Tkachenko}, {Pakhomov}, {Ryabchikova}, {L{\"u}ftinger}, {Palle},
  {Semenko}, {Handler}, {Koudelka}, {Matthews}, {Moffat}, {Pablo}, {Popowicz},
  {Rucinski}, {Wade}, \& {Zwintz}}]{Kallinger2017}
{Kallinger}, T., {Weiss}, W.~W., {Beck}, P.~G., {et~al.} 2017, \aap, 603, A13

\bibitem[{{K{\H o}v{\'a}ri} {et~al.}(2015){K{\H o}v{\'a}ri}, {Kriskovics},
  {K{\"u}nstler}, {Carroll}, {Strassmeier}, {Vida}, {Ol{\'a}h}, {Bartus}, \&
  {Weber}}]{Kovari2015}
{K{\H o}v{\'a}ri}, Z., {Kriskovics}, L., {K{\"u}nstler}, A., {et~al.} 2015,
  \aap, 573, A98

\bibitem[{{Kippenhahn} {et~al.}(2013){Kippenhahn}, {Weigert}, \&
  {Weiss}}]{kippenhahn2013}
{Kippenhahn}, R., {Weigert}, A., \& {Weiss}, A. 2013, {Stellar Structure and
  Evolution}

\bibitem[{{Kjeldsen} \& {Bedding}(1995)}]{Kjeldsen1995}
{Kjeldsen}, H. \& {Bedding}, T.~R. 1995, A\&A, 293, 87

\bibitem[{{K{\"u}nstler} {et~al.}(2015){K{\"u}nstler}, {Carroll}, \&
  {Strassmeier}}]{Kuestler2015}
{K{\"u}nstler}, A., {Carroll}, T.~A., \& {Strassmeier}, K.~G. 2015, \aap, 578,
  A101

\bibitem[{{Kurtz} {et~al.}(2014){Kurtz}, {Saio}, {Takata}, {Shibahashi},
  {Murphy}, \& {Sekii}}]{Kurtz2014}
{Kurtz}, D.~W., {Saio}, H., {Takata}, M., {et~al.} 2014, \mnras, 444, 102

\bibitem[{{Lagarde} {et~al.}(2015){Lagarde}, {Miglio}, {Eggenberger}, {Morel},
  {Montalb{\'a}n}, {Mosser}, {Rodrigues}, {Girardi}, {Rainer}, {Poretti},
  {Barban}, {Hekker}, {Kallinger}, {Valentini}, {Carrier}, {Hareter},
  {Mantegazza}, {Elsworth}, {Michel}, \& {Baglin}}]{Lagarde2015}
{Lagarde}, N., {Miglio}, A., {Eggenberger}, P., {et~al.} 2015, \aap, 580, A141

\bibitem[{{Lambert} \& {Reddy}(2004)}]{Lambert2004}
{Lambert}, D.~L. \& {Reddy}, B.~E. 2004, \mnras, 349, 757

\bibitem[{{Lebreton} {et~al.}(2001){Lebreton}, {Fernandes}, \&
  {Lejeune}}]{Lebreton2001}
{Lebreton}, Y., {Fernandes}, J., \& {Lejeune}, T. 2001, A\&A, 374, 540

\bibitem[{{Lehmann} {et~al.}(2011){Lehmann}, {Tkachenko}, {Semaan},
  {Guti{\'e}rrez-Soto}, {Smalley}, {Briquet}, {Shulyak}, {Tsymbal}, \& {De
  Cat}}]{Lehmann2011}
{Lehmann}, H., {Tkachenko}, A., {Semaan}, T., {et~al.} 2011, A\&A, 526, A124

\bibitem[{{Lenz} \& {Breger}(2005)}]{Lenz2005}
{Lenz}, P. \& {Breger}, M. 2005, Communications in Asteroseismology, 146, 53

\bibitem[{{Liu} {et~al.}(2014){Liu}, {Tan}, {Wang}, {Zhao}, {Sato}, {Takeda},
  \& {Li}}]{Liu2014}
{Liu}, Y.~J., {Tan}, K.~F., {Wang}, L., {et~al.} 2014, ApJ, 785, 94

\bibitem[{{Maeder} \& {Zahn}(1998)}]{MaederZahn1998}
{Maeder}, A. \& {Zahn}, J.-P. 1998, A\&A, 334, 1000

\bibitem[{{Marques} {et~al.}(2013){Marques}, {Goupil}, {Lebreton}, {Talon},
  {Palacios}, {Belkacem}, {Ouazzani}, {Mosser}, {Moya}, {Morel}, {Pichon},
  {Mathis}, {Zahn}, {Turck-Chi{\`e}ze}, \& {Nghiem}}]{Marquesetal2013}
{Marques}, J.~P., {Goupil}, M.~J., {Lebreton}, Y., {et~al.} 2013, \aap, 549,
  A74

\bibitem[{{Mathis} {et~al.}(2004){Mathis}, {Palacios}, \& {Zahn}}]{Mathis2004}
{Mathis}, S., {Palacios}, A., \& {Zahn}, J.-P. 2004, A\&A, 425, 243

\bibitem[{{Mathis} \& {Remus}(2013)}]{MathisRemus2013}
{Mathis}, S. \& {Remus}, F. 2013, in Lecture Notes in Physics, Berlin Springer
  Verlag, Vol. 857, Lecture Notes in Physics, Berlin Springer Verlag, ed. J.-P.
  {Rozelot} \& C.~. {Neiner}, 111--147

\bibitem[{{Mathis} \& {Zahn}(2004)}]{MathisZahn2004}
{Mathis}, S. \& {Zahn}, J.-P. 2004, A\&A, 425, 229

\bibitem[{{Mathur} {et~al.}(2011){Mathur}, {Hekker}, {Trampedach}, {Ballot},
  {Kallinger}, {Buzasi}, {Garc{\'{\i}}a}, {Huber}, {Jim{\'e}nez}, {Mosser},
  {Bedding}, {Elsworth}, {R{\'e}gulo}, {Stello}, {Chaplin}, {De Ridder},
  {Hale}, {Kinemuchi}, {Kjeldsen}, {Mullally}, \& {Thompson}}]{Mathur2011}
{Mathur}, S., {Hekker}, S., {Trampedach}, R., {et~al.} 2011, \apj, 741, 119

\bibitem[{{McQuillan} {et~al.}(2014){McQuillan}, {Mazeh}, \&
  {Aigrain}}]{McQuillan2014}
{McQuillan}, A., {Mazeh}, T., \& {Aigrain}, S. 2014, ApJS, 211, 24

\bibitem[{{Metcalfe} {et~al.}(2015){Metcalfe}, {Creevey}, \&
  {Davies}}]{Metcalfe2015}
{Metcalfe}, T.~S., {Creevey}, O.~L., \& {Davies}, G.~R. 2015, ApJl, 811, L37

\bibitem[{{Mosser} {et~al.}(2011){Mosser}, {Barban}, {Montalb{\'a}n}, {Beck},
  {Miglio}, {Belkacem}, {Goupil}, {Hekker}, {De Ridder}, {Dupret}, {Elsworth},
  {Noels}, {Baudin}, {Michel}, {Samadi}, {Auvergne}, {Baglin}, \&
  {Catala}}]{Mosser2011a}
{Mosser}, B., {Barban}, C., {Montalb{\'a}n}, J., {et~al.} 2011, A\&A, 532, A86

\bibitem[{{Mosser} {et~al.}(2014){Mosser}, {Benomar}, {Belkacem}, {Goupil},
  {Lagarde}, {Michel}, {Lebreton}, {Stello}, {Vrard}, {Barban}, {Bedding},
  {Deheuvels}, {Chaplin}, {De Ridder}, {Elsworth}, {Montalban}, {Noels},
  {Ouazzani}, {Samadi}, {White}, \& {Kjeldsen}}]{Mosser2014}
{Mosser}, B., {Benomar}, O., {Belkacem}, K., {et~al.} 2014, A\&A, 572, L5

\bibitem[{{Mosser} {et~al.}(2012{\natexlab{a}}){Mosser}, {Goupil}, {Belkacem},
  {Marques}, {Beck}, {Bloemen}, {De Ridder}, {Barban}, {Deheuvels}, {Elsworth},
  {Hekker}, {Kallinger}, {Ouazzani}, {Pinsonneault}, {Samadi}, {Stello},
  {Garc{\'{\i}}a}, {Klaus}, {Li}, {Mathur}, \& {Morris}}]{Mosser2012c}
{Mosser}, B., {Goupil}, M.~J., {Belkacem}, K., {et~al.} 2012{\natexlab{a}},
  A\&A, 548, A10

\bibitem[{{Mosser} {et~al.}(2012{\natexlab{b}}){Mosser}, {Goupil}, {Belkacem},
  {Michel}, {Stello}, {Marques}, {Elsworth}, {Barban}, {Beck}, {Bedding}, {De
  Ridder}, {Garc{\'{\i}}a}, {Hekker}, {Kallinger}, {Samadi}, {Stumpe},
  {Barclay}, \& {Burke}}]{Mosser2012b}
{Mosser}, B., {Goupil}, M.~J., {Belkacem}, K., {et~al.} 2012{\natexlab{b}},
  A\&A, 540, A143

\bibitem[{{Mosser} {et~al.}(2009){Mosser}, {Michel}, {Appourchaux}, {Barban},
  {Baudin}, {Boumier}, {Bruntt}, {Catala}, {Deheuvels}, {Garc{\'{\i}}a},
  {Gaulme}, {Regulo}, {Roxburgh}, {Samadi}, {Verner}, {Auvergne}, {Baglin},
  {Ballot}, {Benomar}, \& {Mathur}}]{Mosser2009b}
{Mosser}, B., {Michel}, E., {Appourchaux}, T., {et~al.} 2009, \aap, 506, 33

\bibitem[{{Mosser} {et~al.}(2013){Mosser}, {Michel}, {Belkacem}, {Goupil},
  {Baglin}, {Barban}, {Provost}, {Samadi}, {Auvergne}, \&
  {Catala}}]{Mosser2013}
{Mosser}, B., {Michel}, E., {Belkacem}, K., {et~al.} 2013, A\&A, 550, A126

\bibitem[{{Mosser} {et~al.}(2015){Mosser}, {Vrard}, {Belkacem}, {Deheuvels}, \&
  {Goupil}}]{Mosser2015}
{Mosser}, B., {Vrard}, M., {Belkacem}, K., {Deheuvels}, S., \& {Goupil}, M.~J.
  2015, A\&A, 584, A50

\bibitem[{{Murphy} {et~al.}(2016){Murphy}, {Fossati}, {Bedding}, {Saio},
  {Kurtz}, {Grassitelli}, \& {Wang}}]{Murphy2016}
{Murphy}, S.~J., {Fossati}, L., {Bedding}, T.~R., {et~al.} 2016, \mnras, 459,
  1201

\bibitem[{{Murphy} {et~al.}(2013){Murphy}, {Shibahashi}, \&
  {Kurtz}}]{Murphy2013}
{Murphy}, S.~J., {Shibahashi}, H., \& {Kurtz}, D.~W. 2013, MNRAS, 430, 2986

\bibitem[{{Ogilvie} \& {Lin}(2007)}]{OgilvieLin2007}
{Ogilvie}, G.~I. \& {Lin}, D.~N.~C. 2007, \apj, 661, 1180

\bibitem[{{Palacios} \& {Brun}(2014)}]{PB2014}
{Palacios}, A. \& {Brun}, A.~S. 2014, in IAU Symposium, Vol. 302, Magnetic
  Fields throughout Stellar Evolution, ed. P.~{Petit}, M.~{Jardine}, \& H.~C.
  {Spruit}, 363--364

\bibitem[{{Palacios} {et~al.}(2006){Palacios}, {Charbonnel}, {Talon}, \&
  {Siess}}]{Palaciosetal2006}
{Palacios}, A., {Charbonnel}, C., {Talon}, S., \& {Siess}, L. 2006, \aap, 453,
  261

\bibitem[{{Pinsonneault} {et~al.}(2014){Pinsonneault}, {Elsworth}, {Epstein},
  {Hekker}, {M{\'e}sz{\'a}ros}, {Chaplin}, {Johnson}, {Garc{\'{\i}}a},
  {Holtzman}, {Mathur}, {Garc{\'{\i}}a P{\'e}rez}, {Silva Aguirre}, {Girardi},
  {Basu}, {Shetrone}, {Stello}, {Allende Prieto}, {An}, {Beck}, {Beers},
  {Bizyaev}, {Bloemen}, {Bovy}, {Cunha}, {De Ridder}, {Frinchaboy},
  {Garc{\'{\i}}a-Hern{\'a}ndez}, {Gilliland}, {Harding}, {Hearty}, {Huber},
  {Ivans}, {Kallinger}, {Majewski}, {Metcalfe}, {Miglio}, {Mosser}, {Muna},
  {Nidever}, {Schneider}, {Serenelli}, {Smith}, {Tayar}, {Zamora}, \&
  {Zasowski}}]{Pinsonneault2014}
{Pinsonneault}, M.~H., {Elsworth}, Y., {Epstein}, C., {et~al.} 2014, ApJS, 215,
  19

\bibitem[{{Pires} {et~al.}(2009){Pires}, {Starck}, {Amara}, {Teyssier},
  {R{\'e}fr{\'e}gier}, \& {Fadili}}]{Pires2009}
{Pires}, S., {Starck}, J.-L., {Amara}, A., {et~al.} 2009, \mnras, 395, 1265

\bibitem[{{Raskin}(2011)}]{RaskinPhD}
{Raskin}, G. 2011, PhD thesis, Institute of Astronomy, Katholieke Universiteit
  Leuven, Belgium

\bibitem[{{Raskin} {et~al.}(2011){Raskin}, {van Winckel}, {Hensberge},
  {Jorissen}, {Lehmann}, {Waelkens}, {Avila}, {de Cuyper}, {Degroote},
  {Dubosson}, {Dumortier}, {Fr{\'e}mat}, {Laux}, {Michaud}, {Morren}, {Perez
  Padilla}, {Pessemier}, {Prins}, {Smolders}, {van Eck}, \&
  {Winkler}}]{Raskin2011}
{Raskin}, G., {van Winckel}, H., {Hensberge}, H., {et~al.} 2011, A\&A, 526, A69

\bibitem[{{Rawls} {et~al.}(2016){Rawls}, {Gaulme}, {McKeever}, {Jackiewicz},
  {Orosz}, {Corsaro}, {Beck}, {Mosser}, {Latham}, \& {Latham}}]{Rawls2016}
{Rawls}, M.~L., {Gaulme}, P., {McKeever}, J., {et~al.} 2016, \apj, 818, 108

\bibitem[{{Remus} {et~al.}(2012){Remus}, {Mathis}, \& {Zahn}}]{Remus2012}
{Remus}, F., {Mathis}, S., \& {Zahn}, J.-P. 2012, A\&A, 544, A132

\bibitem[{{Saar} \& {Brandenburg}(1999)}]{Saar1999}
{Saar}, S.~H. \& {Brandenburg}, A. 1999, ApJ, 524, 295

\bibitem[{{Saesen} {et~al.}(2010){Saesen}, {Carrier}, {Pigulski}, {Aerts},
  {Handler}, {Narwid}, {Fu}, {Zhang}, {Jiang}, {Vanautgaerden}, {Kopacki},
  {St{\c e}{\'s}licki}, {Acke}, {Poretti}, {Uytterhoeven}, {Gielen},
  {{\O}stensen}, {De Meester}, {Reed}, {Ko{\l}aczkowski}, {Michalska},
  {Schmidt}, {Yakut}, {Leitner}, {Kalomeni}, {Cherix}, {Spano}, {Prins}, {van
  Helshoecht}, {Zima}, {Huygen}, {Vandenbussche}, {Lenz}, {Ladjal}, {Puga
  Antol{\'{\i}}n}, {Verhoelst}, {De Ridder}, {Niarchos}, {Liakos}, {Lorenz},
  {Dehaes}, {Reyniers}, {Davignon}, {Kim}, {Kim}, {Lee}, {Lee}, {Kwon},
  {Broeders}, {van Winckel}, {Vanhollebeke}, {Waelkens}, {Raskin}, {Blom},
  {Eggen}, {Degroote}, {Beck}, {Puschnig}, {Schmitzberger}, {Gelven},
  {Steininger}, {Blommaert}, {Drummond}, {Briquet}, \&
  {Debosscher}}]{Saesen2010}
{Saesen}, S., {Carrier}, F., {Pigulski}, A., {et~al.} 2010, A\&A, 515, A16

\bibitem[{{Salaris} \& {Cassisi}(2005)}]{Salaris2005}
{Salaris}, M. \& {Cassisi}, S. 2005, {Evolution of Stars and Stellar
  Populations}

\bibitem[{{Satsuka} {et~al.}(2017){Satsuka}, {Tsuribe}, {Tanaka}, \&
  {Nagamine}}]{Satsuka2017}
{Satsuka}, T., {Tsuribe}, T., {Tanaka}, S., \& {Nagamine}, K. 2017, \mnras,
  465, 986

\bibitem[{{Schmid} \& {Aerts}(2016)}]{SchmidAerts2016}
{Schmid}, V.~S. \& {Aerts}, C. 2016, \aap, 592, A116

\bibitem[{{Schou} {et~al.}(1998){Schou}, {Antia}, {Basu}, {Bogart}, {Bush},
  {Chitre}, {Christensen-Dalsgaard}, {di Mauro}, {Dziembowski}, {Eff-Darwich},
  {Gough}, {Haber}, {Hoeksema}, {Howe}, {Korzennik}, {Kosovichev}, {Larsen},
  {Pijpers}, {Scherrer}, {Sekii}, {Tarbell}, {Title}, {Thompson}, \&
  {Toomre}}]{Schou1998}
{Schou}, J., {Antia}, H.~M., {Basu}, S., {et~al.} 1998, ApJ, 505, 390

\bibitem[{{Shibahashi}(1979)}]{Shibahashi1979}
{Shibahashi}, H. 1979, \pasj, 31, 87

\bibitem[{{Shulyak} {et~al.}(2004){Shulyak}, {Tsymbal}, {Ryabchikova},
  {St{\"u}tz}, \& {Weiss}}]{Shulyak2004}
{Shulyak}, D., {Tsymbal}, V., {Ryabchikova}, T., {St{\"u}tz}, C., \& {Weiss},
  W.~W. 2004, A\&A, 428, 993

\bibitem[{{Silva Aguirre} {et~al.}(2014){Silva Aguirre}, {Ruchti}, {Hekker},
  {Cassisi}, {Christensen-Dalsgaard}, {Datta}, {Jendreieck}, {Jessen-Hansen},
  {Mazumdar}, {Mosser}, {Stello}, {Beck}, \& {de Ridder}}]{SilvaAguirre2014}
{Silva Aguirre}, V., {Ruchti}, G.~R., {Hekker}, S., {et~al.} 2014, ApJL, 784,
  L16

\bibitem[{{Silva Aguirre} \& {Serenelli}(2016)}]{SilvaAguirre2016}
{Silva Aguirre}, V. \& {Serenelli}, A.~M. 2016, Astronomische Nachrichten, 337,
  823

\bibitem[{{Simon} \& {Sturm}(1994)}]{SimonSturm1994}
{Simon}, K.~P. \& {Sturm}, E. 1994, A\&A, 281, 286

\bibitem[{{Skumanich}(1972)}]{Skumanich1972}
{Skumanich}, A. 1972, ApJ, 171, 565

\bibitem[{{Strassmeier} {et~al.}(1999){Strassmeier}, {St{\c e}pie{\'n} },
  {Henry}, \& {Hall}}]{Strassmeier1999}
{Strassmeier}, K.~G., {St{\c e}pie{\'n} }, K., {Henry}, G.~W., \& {Hall}, D.~S.
  1999, \aap, 343, 175

\bibitem[{{Strugarek} {et~al.}(2011){Strugarek}, {Brun}, \&
  {Zahn}}]{Strugarek2011}
{Strugarek}, A., {Brun}, A.~S., \& {Zahn}, J.-P. 2011, \aap, 532, A34

\bibitem[{{Talon} \& {Charbonnel}(1998)}]{Talon1998}
{Talon}, S. \& {Charbonnel}, C. 1998, \aap, 335, 959

\bibitem[{{Talon} \& {Charbonnel}(2005)}]{Talon2005}
{Talon}, S. \& {Charbonnel}, C. 2005, \aap, 440, 981

\bibitem[{{Talon} \& {Charbonnel}(2008)}]{Talon2008}
{Talon}, S. \& {Charbonnel}, C. 2008, \aap, 482, 597

\bibitem[{{Talon} \& {Kumar}(1998)}]{TalonKumar1998}
{Talon}, S. \& {Kumar}, P. 1998, \apj, 503, 387

\bibitem[{{Talon} \& {Zahn}(1997)}]{TalonZahn1997}
{Talon}, S. \& {Zahn}, J.-P. 1997, \aap, 317, 749

\bibitem[{{Tamajo} {et~al.}(2011){Tamajo}, {Pavlovski}, \&
  {Southworth}}]{Tamajo2011}
{Tamajo}, E., {Pavlovski}, K., \& {Southworth}, J. 2011, A\&A, 526, A76

\bibitem[{{Theme{\ss}l} {et~al.}(2017){Theme{\ss}l}, {Hekker}, {Southworth},
  {Beck}, {Pavlovski}, \& {Tkachenko}}]{Themessl2017}
{Theme{\ss}l}, N., {Hekker}, S., {Southworth}, J., {et~al.} 2017, MNRAS
  (submitted)

\bibitem[{{Thompson} {et~al.}(2012){Thompson}, {Everett}, {Mullally},
  {Barclay}, {Howell}, {Still}, {Rowe}, {Christiansen}, {Kurtz}, {Hambleton},
  {Twicken}, {Ibrahim}, \& {Clarke}}]{Thompson2012}
{Thompson}, S.~E., {Everett}, M., {Mullally}, F., {et~al.} 2012, ApJ, 753, 86

\bibitem[{{Tkachenko}(2015)}]{Tkachenko2015}
{Tkachenko}, A. 2015, A\&A, 581, A129

\bibitem[{{Tkachenko} {et~al.}(2012){Tkachenko}, {Lehmann}, {Smalley},
  {Debosscher}, \& {Aerts}}]{Tkachenko2012}
{Tkachenko}, A., {Lehmann}, H., {Smalley}, B., {Debosscher}, J., \& {Aerts}, C.
  2012, MNRAS, 422, 2960

\bibitem[{{Torres} {et~al.}(2015){Torres}, {Claret}, {Pavlovski}, \&
  {Dotter}}]{Torres2015}
{Torres}, G., {Claret}, A., {Pavlovski}, K., \& {Dotter}, A. 2015, ApJ, 807, 26

\bibitem[{{Triana} {et~al.}(2017){Triana}, {Corsaro}, {De Ridder}, {Bonanno},
  {P{\'e}rez Hern{\'a}ndez}, \& {Garc{\'{\i}}a}}]{Triana2017}
{Triana}, S.~A., {Corsaro}, E., {De Ridder}, J., {et~al.} 2017, \aap, 602, A62

\bibitem[{{Tsymbal}(1996)}]{Tsymbal1996}
{Tsymbal}, V. 1996, in Astronomical Society of the Pacific Conference Series,
  Vol. 108, M.A.S.S., Model Atmospheres and Spectrum Synthesis, ed. S.~J.
  {Adelman}, F.~{Kupka}, \& W.~W. {Weiss}, 198

\bibitem[{{Unno} {et~al.}(1989){Unno}, {Osaki}, {Ando}, {Saio}, \&
  {Shibahashi}}]{Unno1989}
{Unno}, W., {Osaki}, Y., {Ando}, H., {Saio}, H., \& {Shibahashi}, H. 1989,
  {Nonradial oscillations of stars}

\bibitem[{{Verbunt} \& {Phinney}(1995)}]{Verbunt1995}
{Verbunt}, F. \& {Phinney}, E.~S. 1995, A\&A, 296, 709

\bibitem[{{Wallerstein} {et~al.}(1965){Wallerstein}, {Herbig}, \&
  {Conti}}]{Wallerstein1965}
{Wallerstein}, G., {Herbig}, G.~H., \& {Conti}, P.~S. 1965, \apj, 141, 610

\bibitem[{{Welsh} {et~al.}(2011){Welsh}, {Orosz}, {Aerts}, {Brown},
  {Brugamyer}, {Cochran}, {Gilliland}, {Guzik}, {Kurtz}, {Latham}, {Marcy},
  {Quinn}, {Zima}, {Allen}, {Batalha}, {Bryson}, {Buchhave}, {Caldwell},
  {Gautier}, {Howell}, {Kinemuchi}, {Ibrahim}, {Isaacson}, {Jenkins}, {Prsa},
  {Still}, {Street}, {Wohler}, {Koch}, \& {Borucki}}]{Welsh2011}
{Welsh}, W.~F., {Orosz}, J.~A., {Aerts}, C., {et~al.} 2011, ApJS, 197, 4

\bibitem[{{White} {et~al.}(2017){White}, {Benomar}, {Silva Aguirre}, {Ball},
  {Bedding}, {Chaplin}, {Christensen-Dalsgaard}, {Garcia}, {Gizon}, {Stello},
  {Aigrain}, {Antia}, {Appourchaux}, {Bazot}, {Campante}, {Creevey}, {Davies},
  {Elsworth}, {Gaulme}, {Handberg}, {Hekker}, {Houdek}, {Howe}, {Huber},
  {Karoff}, {Marques}, {Mathur}, {McQuillan}, {Metcalfe}, {Mosser}, {Nielsen},
  {R{\'e}gulo}, {Salabert}, \& {Stahn}}]{White2017}
{White}, T.~R., {Benomar}, O., {Silva Aguirre}, V., {et~al.} 2017, \aap, 601,
  A82

\bibitem[{{Xu} {et~al.}(2013){Xu}, {Takahashi}, {Goriely}, {Arnould}, {Ohta},
  \& {Utsunomiya}}]{Xu2013}
{Xu}, Y., {Takahashi}, K., {Goriely}, S., {et~al.} 2013, Nuclear Physics A,
  918, 61

\bibitem[{{Zahn}(1966)}]{Zahn1966}
{Zahn}, J.~P. 1966, Annales d'Astrophysique, 29, 489

\bibitem[{{Zahn}(1975)}]{Zahn1975}
{Zahn}, J.-P. 1975, \aap, 41, 329

\bibitem[{{Zahn}(1989)}]{Zahn1989}
{Zahn}, J.-P. 1989, A\&A, 220, 112

\bibitem[{{Zahn}(1992)}]{Zahn1992}
{Zahn}, J.-P. 1992, A\&A, 265, 115

\bibitem[{{Zahn}(1994)}]{Zahn1994}
{Zahn}, J.-P. 1994, A\&A, 288, 829

\end{thebibliography}
\newpage




\end{document}